\documentclass[11pt]{article}
\pdfoutput=1

\usepackage{fullpage}
\usepackage{amsmath}
\usepackage{amssymb}
\usepackage{setspace}
\usepackage{bm}
\usepackage{dsfont}
\usepackage{graphics}
\usepackage{tabu}
\usepackage{mathrsfs}
\usepackage{empheq}
\usepackage{tikz-cd}
\usepackage{cite}
\usepackage{multirow}
\usepackage[font=footnotesize,labelfont=bf,justification=centerlast,width=.94\textwidth]{caption}
\usepackage[citebordercolor={.8 .8 1},urlbordercolor={.8 .8 1},pdfstartview=FitH]{hyperref}

\usepackage{color}

\usepackage[normalem]{ulem}

\newcommand{\be}{\begin{equation}}
\newcommand{\ee}{\end{equation}}
\newcommand{\bes}{\begin{equation*}}
\newcommand{\ees}{\end{equation*}}
\newcommand{\bea}{\begin{eqnarray}}
\newcommand{\eea}{\end{eqnarray}}
\newcommand{\beas}{\begin{eqnarray*}}
\newcommand{\eeas}{\end{eqnarray*}}

\newcommand{\eps}{\epsilon}
\newcommand{\epsJ}{\varepsilon}

\newcommand{\cN}{\mathcal{N}}
\newcommand{\cL}{\mathcal{L}}
\newcommand{\cW}{\mathcal{W}}
\newcommand{\cA}[1]{\mathscr{A}^{[#1]}}
\newcommand{\cB}[1]{\mathscr{B}^{[#1]}}
\newcommand{\cC}[1]{\mathscr{C}^{[#1]}}
\newcommand{\cD}[1]{\mathscr{D}^{[#1]}}
\newcommand{\cE}[2]{\mathscr{E}^{[#1]}_{#2}}
\newcommand{\cPhi}[2]{\Phi^{[#1]}_{#2}}
\newcommand{\cPsi}[2]{\Psi^{[#1]}_{#2}}

\newcommand{\lG}{\eps_{-\frac{1}{2}}^{-}}
\newcommand{\lH}{\eps_{-\frac{1}{2}}^{+}}
\newcommand{\blH}{\bar\eps_{-\frac{1}{2}}^{+}}
\newcommand{\lS}{\eps_{-\frac{3}{2}}^{-}}
\newcommand{\lT}{\eps_{-\frac{3}{2}}^{+}}
\newcommand{\blT}{\bar\eps_{-\frac{3}{2}}^{+}}
\newcommand{\nJ}{\nu_{0}}

\newcommand{\nW}{\nu_{-2}}
\newcommand{\nA}{\tilde{\nu}_{-1}}

\newcommand{\qv}{q_{2}}
\newcommand{\qw}{q_{3}}
\newcommand{\hwr}{\left| \text{h.w.}\right\rangle}

\begin{document}
\numberwithin{equation}{section}
{
\begin{titlepage}
\begin{center}

\hfill \\
\hfill \\
\vskip 0.35in

{\Large \bf Extremal Higher Spin Black Holes}\\

\vskip 0.4in

{\large M\'aximo Ba\~nados$^a$, Alejandra Castro$^b$, Alberto Faraggi$^a$ and Juan I. Jottar$^c$}\\

\vskip 0.3in

${}^{a}${\it  Instituto de F\'isica, Pontificia Universidad Cat\'olica de Chile,  \\ Casilla 306, Santiago, Chile} \vskip .5mm
${}^{b}${\it Institute for Theoretical Physics, University of Amsterdam,\\
Science Park 904, Postbus 94485, 1090 GL Amsterdam, The Netherlands} \vskip .5mm
${}^{c}${\it Institut f\"ur Theoretische Physik, ETH Z\"urich, CH-8093 Z\"urich, Switzerland } \vskip .5mm


\end{center}

\vskip 0.45in

\begin{abstract}
\noindent  The gauge sector of three-dimensional higher spin gravities can be formulated as a Chern-Simons theory. In this context, a higher spin black hole corresponds to a flat connection with suitable holonomy (smoothness) conditions which are consistent with the properties of a generalized thermal ensemble.  Building on these ideas, we discuss a definition of black hole extremality which is appropriate to the topological character of $3d$ higher spin theories. Our definition can be phrased in terms of the Jordan class of the holonomy around a non-contractible (angular) cycle, and we show that it is compatible with the zero-temperature limit of smooth black hole solutions. While this notion of extremality does not require supersymmetry, we exemplify its consequences in the context of $sl(3|2)\oplus sl(3|2)$ Chern-Simons theory and show that, as usual, not all extremal solutions preserve supersymmetries. Remarkably, we find in addition that the higher spin setup allows for non-extremal supersymmetric black hole solutions. Furthermore, we discuss our results from the perspective of the holographic duality between $sl(3|2)\oplus sl(3|2)$ Chern-Simons theory and two-dimensional CFTs with $\cW_{(3|2)}$ symmetry, the simplest higher spin extension of the $\cN=2$ super-Virasoro algebra. In particular, we compute $\cW_{(3|2)}$ BPS bounds at the full quantum level, and relate their semiclassical limit to extremal black hole or conical defect solutions in the $3d$ bulk. Along the way, we discuss the role of the spectral flow automorphism and provide a conjecture for the form of the semiclassical BPS bounds in general $\cN=2$ two-dimensional CFTs with extended symmetry algebras.
\end{abstract}

\vfill


\end{titlepage}
}

\newpage

\tableofcontents

\newpage
\section{Introduction}

Higher spin theories provide a framework to explore non-linear and non-local features that are expected to arise in quantum gravity. In a sense, a higher spin theory is characterized by ``too many" gauge symmetries; not only does this feature introduce novel interactions among the fields, but it also calls for a refinement of standard geometrical notions such as causal structure and curvature, which are not invariant under the higher spin symmetries that extend diffeomorphisms. As a consequence, gauge-invariant definitions of concepts such as black hole spacetime become not only desirable, but necessary. In spite of these challenges, holography has provided a useful way to organize our understanding of gravitational higher spin theories and their field theory duals. In particular, the higher spin AdS$_{3}$/CFT$_{2}$ correspondence relates two-dimensional conformal field theories (CFTs) with $\cW$-symmetry algebras and three-dimensional higher spin theories with generalized anti-de Sitter (AdS) boundary conditions, and has proven to be a fruitful arena to explore higher spin holographic dualities and tackle the associated issues. 

An important example of such dualities is the original proposal of Gaberdiel and Gopakumar \cite{Gaberdiel:2010pz,Gaberdiel:2012uj}, which entails a correspondence between $\cW_{N}$ minimal model coset CFTs in the large central charge limit and the interacting Prokushkin-Vasiliev higher spin theory \cite{ Prokushkin:1998vn,Prokushkin:1998bq}. The latter includes matter fields that couple to the $3d$ higher spin fields, but a consistent truncation where the matter representations decouple is possible. In this truncation the pure higher spin sector becomes a three-dimensional Chern-Simons theory based on two copies of the infinite-dimensional Lie algebra known as hs$[\lambda]\,$, which holographically describes the conserved currents in a CFT with $\cW_{\infty}[\lambda]$ as the chiral algebra. A further truncation to $\lambda =  N$ with $N$ an integer is possible, in which the bulk gauge algebra becomes $sl(N)\oplus sl(N)$ Chern-Simons theory, describing the conserved currents of a CFT with $\cW_{N}$ symmetry. From an AdS/CFT perspective, these are rather natural generalizations of the well-known fact that $3d$ Einstein-Hilbert gravity with negative cosmological constant, which displays two copies of the Virasoro algebra as asymptotic symmetries \cite{Brown:1986nw}, can be formulated as an $sl(2)\oplus sl(2)$ Chern-Simons theory \cite{Achucarro:1987vz,Witten:1988hc}.

In the above class of holographic dualities, the semiclassical regime of the CFT maps to classical Chern-Simons theory in the bulk, and the latter provides a powerful framework to address the problems alluded to above. Indeed,  in this setup one can easily capture various local and non-local CFT observables by studying flat connections with suitable boundary conditions. One can, for example, describe a finite-temperature ensemble carrying higher spin charges in the boundary by constructing higher spin black hole solutions in the bulk. Naturally, a gauge-invariant definition of a black hole spacetime is crucial for any further progress along this line.\footnote{ On account of these difficulties, an unambiguous definition of higher spin black holes and their thermodynamics in $4d$ higher spin theories \cite{Fradkin:1986qy,Fradkin:1987ks,Vasiliev:1990en} remains elusive; see \cite{Didenko:2009td} however for an interesting attempt.} Fortunately, the situation is under control for $3d$ higher spin theories, precisely because of the existence of a Chern-Simons formulation. 

In the seminal work of Gutperle and Kraus \cite{Gutperle:2011kf} it was proposed that a suitable definition of $3d$ higher spin black holes consist of a flat connection with trivial holonomy around the Euclidean time cycle.\footnote{ In the context of three-dimensional Einstein-Hilbert gravity formulated as a Chern-Simons theory, such a definition of black hole had appeared long ago in \cite{MaxTesis}.} This is an abstraction of the familiar notion of smoothness of a Euclidean horizon in the metric formulation, and it leads in fact to consistent thermodynamics. Moreover, it was later argued \cite{Ammon:2011nk} that when these smoothness conditions are satisfied, there is precisely one representative in the gauge orbit of a given black hole flat connection whose associated metric displays a smooth horizon in the usual sense. While the original work focused on the $sl(3)\oplus sl(3)$ theory, the definition of higher spin black hole was later extended  to the hs$[\lambda]$ case \cite{Kraus:2011ds} and the corresponding partition function (free energy) was shown to match a perturbative CFT calculation \cite{Gaberdiel:2012yb}. Building on the work in \cite{Banados:2012ue}, which studied the thermodynamics of higher spin black holes from the Euclidean action perspective, general expressions for the higher spin black hole entropy, free energy and associated first law casted solely in terms of the Chern-Simons connections and their holonomies were derived in \cite{deBoer:2013gz}.

 What all the developments described above have in common is that they exploit the topological formulation of the bulk theory and the power of Chern-Simons theory in order to set up and perform calculations that are quite challenging\footnote{ Or even beyond the scope of the existent CFT technology, such as the non-perturbative result for entanglement entropy in $\cW_{3}$ CFTs deformed by sources reported in  \cite{Ammon:2013hba,deBoer:2013vca}.} using solely field-theoretical techniques. In this light, it is natural to ask for a topological definition of extremal higher spin black holes, namely one that is phrased in terms of holonomies of flat connections in Chern-Simons theory, without reference to metric or geometric concepts which are not natural once we go beyond pure gravity. This is the problem we address in the present paper. In particular, we will advocate that the natural definition of extremality in higher spin theories involves the Jordan class of the holonomy around the non-contractible cycle which characterizes three-dimensional black hole topologies. 
 
 More precisely, we will find that the zero-temperature limit of smooth black hole solutions generically results in non-diagonalizable connections, whose structure moreover encode relations between the charges which often saturate higher spin BPS bounds in supersymmetric setups. Quite interestingly, we will find as a by-product that a large enough superalgebra with non-linear symmetry transformations  allows for finite-temperature black hole solutions that carry globally-defined Killing spinors, in stark contrast with the usual gravitational theories where supersymmetry implies extremality. 

The full-fledged higher spin holographic correspondences involving CFTs with infinite-dimensional chiral algebras such as $\cW_{\infty}[\lambda]$ have a rich and complex structure. Moreover, the latter have been recently shown to make an appearance in the tensionless limit of string theory on AdS$_{3}\times S^{3}\times \mathbb{T}^{4}\,$ \cite{Gaberdiel:2014cha,Gaberdiel:2015mra}. The simpler versions of these dualities considered here, which involve only finite-dimensional algebras, provide a useful arena to study and test various aspects of higher spin holographic dualities. An example that has proven particularly fruitful is the correspondence between $sl(N)\oplus sl(N)$ Chern-Simons theory and $\cW_{N}$ CFTs. In this spirit, we will mostly focus on examples involving the pure gravity theory ($N=2$) and the bosonic spin-$3$ higher spin theory ($N=3$), as well as their $\cN=2$ supersymmetric generalizations dual to super-Virasoro and $\cW_{(3|2)}$ CFTs, respectively. While our definition of extremality applies straightforwardly in any finite-$N$ theory, we hope that it will provide guidance in the case of dualities based on infinite-dimensional algebras as well.

The rest of the paper is organized as follows. In section \ref{sec:Definition} we introduce our definition of extremal higher spin black holes and exemplify it in the bosonic $N=2$ (pure gravity) and $N=3$ (spin-3) theories. In section \ref{sec:BHs} we discuss black hole and smooth conical defect solutions in $sl(3|2)\oplus sl(3|2)$ Chern-Simons theory and their supersymmetries, whose number depend on the precise Jordan normal form of the connection. Quite surprisingly, we find a class of finite-temperature black hole solutions which preserves supersymmetry. In  section \ref{sec:BPS} we compute quantum higher spin BPS bounds from the CFT point of view, and relate their semiclassical limit to the holonomies of bulk solutions admitting Killing spinors. In particular, we find that while not all extremal black holes preserve supersymmetries, the ones that do carry charges which fulfill relations that saturate BPS bounds. Furthermore, we provide a conjecture for the generic form of the semiclassical BPS bounds in any $\cN=2$ CFT with an extended symmetry algebra which can be obtained from Drinfeld-Sokolov reduction.  We discuss our findings in section \ref{sec:Disc} and compare them with some previous results in the literature. The appendices contain a brief discussion of the $\cW_{(3|2)}$ algebra as well as other useful formulae and conventions. 

A complete discussion of the $\cN=2$ super-$\cW_{3}$ holographic dictionary will appear in a separate note \cite{BetoJuan}; for the sake of brevity and clarity, we shall henceforth limit ourselves to quoting the corresponding results which are directly relevant for the present discussion.

\section{Extremal higher spin black holes}\label{sec:Definition}
Our aim is targeted towards black hole solutions of three-dimensional Chern-Simons theory. The relevant Euclidean action is
\begin{equation}
I_{CS} = \frac{ik_{cs}}{4\pi}\int_{M}\text{Tr}\Bigl[CS(A)-CS(\bar{A})\Bigr]\,,
\end{equation}
\noindent  where $A$ and $\bar{A}$ are valued in the same algebra (or superalgebra) $\mathfrak{g}\,$, 
\begin{equation}
CS(A) = A\wedge dA + \frac{2}{3}A\wedge A\wedge A
\end{equation}
\noindent is the Chern-Simons form, and $\text{Tr}$ denotes the trace (or supertrace) in the chosen representation. In Euclidean signature the connections are generically complex-valued, with
\be
A^\dagger = -\bar A \,.
\ee

\noindent This condition ensures reality of the action and physical observables. In Lorentzian signature one works instead with two independent connections $A$ and $\bar{A}$, each valued in an appropriate real form of the gauge algebra. In particular, all parameters such as charges and their conjugate potentials are then real in both sectors.

We will as usual exploit the gauge freedom of Chern-Simons theory to ``gauge-away" the radial dependence of the connection\footnote{ In a purely gravitational setup, this possibility can be understood as the familiar fact that in $3d$ the Fefferman-Graham expansion truncates after a finite number of terms in the radial coordinate $\rho$ \cite{Skenderis:1999nb}.} 
\begin{equation}\label{eq:radialgauge}
A(\rho,z,\bar{z}) = b^{-1}(\rho)\Bigl(a(z,\bar{z}) + d\Bigr)b(\rho)\,,\qquad \bar{A}(\rho,z,\bar{z}) = b(\rho)\Bigl(\bar{a}(z,\bar{z}) + d\Bigr)b^{-1}(\rho)\,,
\end{equation}
\noindent and focus on the ``boundary connections" $a(z,\bar{z})$ and $\bar{a}(z,\bar{z})\,$. Here, the boundary coordinates $(z,\bar{z})$ parameterize either the plane, cylinder or torus, or more generally an arbitrary Riemann surface, depending on the topology of the solution under consideration. 

As we will review later on, in the absence of sources one has $a_{\bar{z}} = \bar{a}_{z}=0$ and the remaining components $a_{z}(z)$ and $\bar{a}_{\bar{z}}(\bar{z})$ are respectively holomorphic and anti-holomorphic, so they can be thought of as parameterizing Kac-Moody currents for the algebra $\mathfrak{g}\,$. Further restrictions on the form of these $2d$ flat connections that result in $\cW$-symmetry via Hamiltonian (Drinfeld-Sokolov) reduction of current algebras were discussed long ago in e.g. \cite{Polyakov:1989dm,Bershadsky:1989mf,Bais:1990bs,Lu:1991ux,deBoer:1991jc,DeBoer:1992vm,deBoer:1992sy,deBoer:1993iz} and more recently in \cite{deBoer:1998ip,Henneaux:2010xg,Campoleoni:2010zq,Gaberdiel:2011wb,Campoleoni:2011hg} in the context of the AdS/CFT correspondence, where they were understood as boundary conditions that result in $\cW$-algebras as asymptotic symmetries. We will generically refer to the latter as Drinfeld-Sokolov boundary conditions.

\subsection{Non-extremal higher spin black holes and their thermodynamics: a brief review}\label{sec:finite}
We will now briefly review some key features of non-extremal higher spin black holes and their thermodynamics. Further details can be found in e.g. \cite{Gutperle:2011kf,Ammon:2011nk,Ammon:2012wc,deBoer:2013gz,Bunster:2014mua,deBoer:2014fra}.

 In order to discuss finite-temperature black hole solutions, one compactifies the Euclidean time direction so the $3d$ manifold has the topology of a solid torus. The boundary coordinates $(w,\bar w)$ are then subject to the identifications $w\simeq w + 2\pi \simeq w + 2\pi \tau\,$ where $\tau$ is the modular parameter of the boundary torus, which has volume $\text{Vol}(T^2)=4\pi^{2}\text{Im}(\tau)\,$ in our conventions.\footnote{ Equivalently, one may use coordinates with fixed periodicity $w \simeq w + 2\pi \simeq w+2\pi i\,$, in which case the modular parameter appears explicitly in the connection components.}  In terms of the inverse temperature $\beta$ and the angular velocity of the horizon $\Omega$ one has
\begin{equation}\label{tau and tau bar}
\tau = \frac{i\beta}{2\pi}\bigl(1 +\Omega\bigr)\,,\qquad \bar{\tau} = \frac{i\beta}{2\pi}\bigl(-1+\Omega\bigr)\,,
\end{equation}
\noindent as seen from the canonical relation 
\begin{equation}
2\pi i \tau\left(L_0 - \frac{c}{24}\right) - 2\pi i\bar{\tau}\left(\bar{L}_0 - \frac{c}{24}\right) = -\beta\Bigl(H +\Omega J\Bigr)\,,
\end{equation}
\noindent where $H = L_0 + \bar{L}_0 - \tfrac{c}{12}$ is the CFT Hamiltonian and $J = L_0 - \bar{L}_0$ the angular momentum. The angular potential $\Omega$, and concomitantly the angular momentum $J$, should be continued to purely imaginary values in order to have a real Euclidean section. 

In this language, the BTZ black hole solution \cite{Banados:1992wn,Banados:1992gq} is a constant $sl(2)\oplus sl(2)$ flat connection and reads
\begin{equation}\label{BTZ connection}
a = \left(\begin{array}{cc}
0 &{{\cal L}} \\ 
1& 0
\end{array} \right)dw \,,\qquad \bar{a} = 
-\left(\begin{array}{cc}
0 & 1 \\ 
{\bar {\cal L} } & 0
\end{array} \right)d\bar{w}\,.
\end{equation}
\noindent  Via the usual AdS$_{3}$/CFT$_{2}$ dictionary (see \cite{Kraus:2006wn} for a review), the Chern-Simons level $k$ is given by $k = \ell/(4G_{3})$ in terms of the AdS length $\ell$ and the three-dimensional Newton's constant $G_{3}\,$, and $c = 6k$ is the central charge in the dual CFT \cite{Brown:1986nw}.  Then, $k\mathcal{L} = h- \frac{k}{4}$ and $k\bar{\mathcal{L}}=\bar{h}-\frac{k}{4}$ are seen to correspond to the eigenvalues of the zero modes of the left- and right-moving stress tensor acting on the CFT state dual to the black hole.\footnote{ The shift  is due to the mapping from the cylinder to the plane: $(h,\bar h)$ are the eigenvalues of the Virasoro zero modes $(L_0,\bar L_0)$ on the plane, while the combinations in \eqref{BTZ connection} are related to the eigenvalues of the zero modes $(L_0 -c/24, \bar{L}_0 -c/24)$ on the cylinder/torus for the corresponding state.} In terms of the black hole mass $M$ and angular momentum $\mathcal{J}$ one then has 
\begin{equation}
{h}-{c\over 24}= \frac{1}{2}\bigl(M\ell - \mathcal{J}\bigr)\,,\qquad {\bar h}-{c\over 24}= \frac{1}{2}\bigl(M\ell + \mathcal{J}\bigr)\,.
\end{equation}

As usual, demanding smoothness of the horizon implies a relation between the black hole's charges and potentials, namely
\begin{equation}\label{BTZ smoothness}
\tau = \frac{i}{2}\sqrt{\frac{1}{\cal L}}\,,\qquad \bar{\tau} = -\frac{i}{2}\sqrt{\frac{1}{\bar {\cal L}}}\,,
\end{equation}
\noindent so the black hole is in thermodynamic equilibrium and satisfies the first law of thermodynamics. In Chern-Simons language, these relations imply that the holonomy of the black hole connection around the Euclidean time circle $\mathcal{C}_{E}$ becomes trivial\footnote{ Where by trivial we mean that it belongs to the center of the gauge group \cite{Castro:2011fm}. In the expression below $L_0$ denotes the Cartan element of the $sl(2)$ algebra, and not the zero mode $L_0$ of the CFT stress tensor on the plane. We hope that the meaning is clear from the context, and that no confusion arises from this slight abuse of notation.}
\begin{equation}\label{smoothness}
\mathcal{P}\exp\left(\oint_{\mathcal{C}_E} a\right) = e^{2\pi\left(\tau a_{w} + \bar{\tau}a_{\bar{w}}\right)} = e^{2\pi i L_0} = -\mathds{1}_{2\times 2}\,\,,
\end{equation}
\noindent and similarly in the barred sector, where $L_0$ denotes the Cartan element of $sl(2)\,$. 

 In the original work \cite{Gutperle:2011kf} of Gutperle and Kraus it was proposed that the definition of higher spin black hole consists of promoting the smoothness condition \eqref{smoothness} to the higher spin case, namely to demand
\begin{equation}\label{eigenv conditions}
\text{Eigen}\bigl(a_{\text{contract}}\bigr) =  \text{Eigen}\bigl(\tau a_{w} + \bar{\tau}a_{\bar{w}}\bigr) = \text{Eigen}\bigl(iL_0\bigr)
 \end{equation} 
\noindent in the general case as well. Here, $a_{\text{contract}}$ denotes the component of the connection along the cycle of the boundary torus which becomes contractible in the bulk, and the definition is clearly appropriate to the topological setup. We also stress that \eqref{eigenv conditions} requires the identification of an $sl(2)$ subalgebra embedded in the gauge algebra. Different embeddings result in different theories, with different symmetry algebras (see e.g. \cite{Ammon:2011nk,Castro:2011fm}). 

It is worth pausing at this point to emphasize the general philosophy we follow throughout the remainder of the paper. Upon solving the smoothness conditions in Euclidean signature, we will always continue back to the Lorentzian solution where all the charges and potentials are manifestly real, and smoothness is interpreted as a particular relation between these parameters (which were a-priori independent). The rationale behind this choice is that both the notions of extremality and supersymmetry are properly discussed in Lorentzian signature. Therefore, we will endeavor to cast the conditions defining extremal and/or supersymmetric solutions as further constraints on the aforementioned set of real charges and potentials.

Going back to the general structure of the black hole connections, following the Hamiltonian reduction procedure one finds that sources for the CFT currents are incorporated in the $a_{\bar{w}}$ and $\bar{a}_{w}$ components of the Drinfeld-Sokolov connections, the insight being that the CFT Ward identities in the presence of sources should be equivalent to flatness of the gauge connections. In the pure gravitational ($sl(2)\oplus sl(2)$) case one can choose to incorporate the spin-2 sources in the modular parameter of the boundary torus, in a way that $a_{\bar{w}}=\bar{a}_{w}=0$ still vanish as in the BTZ connection above. However, as soon as one goes beyond $sl(2)$ and incorporates higher spin currents it is in general necessary to turn on the $a_{\bar{w}}$ and $\bar{a}_{w}$ components in order to account for the corresponding sources. The question then becomes whether the currents (or their zero modes, the charges) are incorporated in the holomorphic components $a_{w}$ and $\bar{a}_{\bar{w}}$ as in the absence of sources,  or in $a_{\phi} = a_{w} + a_{\bar{w}}$ and $\bar{a}_{\phi} = \bar{a}_{w} + \bar{a}_{\bar{w}}$ instead. 

From the bulk perspective, it was argued in \cite{Banados:2012ue,Perez:2012cf,Compere:2013nba,Henneaux:2013dra} that choosing to incorporate the charges in the angular components $a_{\phi}$ and $\bar{a}_{\phi}$ was consistent with usual canonical notions in gravitational theories. It was then shown in \cite{deBoer:2014fra}, which we follow here, that the choice $a_{\phi}$ vs. $a_{w}$ (and similarly in the barred sector) for the expectation values (charges) amount to different boundary conditions that map to different partition functions in the CFT side. More precisely, by a careful analysis of Ward identities it was shown in \cite{deBoer:2014fra} that the $a_{\phi}$ choice corresponds to deformations of the CFT Hamiltonian of the form
\begin{equation}\label{Hamiltonian deformation}
H = H_{\text{CFT}} + \oint d\phi\, \sum_{s} \mu_{s}J_{s} + \oint d\phi\, \sum_{s} \bar{\mu}_{s}\bar{J}_{s} \,,
\end{equation}

\noindent while the $a_{w}$ choice corresponds instead to deformations of the CFT action
\begin{equation}\label{action deformations}
I = I_{\text{CFT}} + \int d^{2}w\, \sum_{s}\mu_{s}J_{s} + \int d^{2}w\, \sum_{s}\bar{\mu}_{s}\bar{J}_{s}+\ldots \,.
\end{equation}

\noindent Here $J_{s}$ and $\bar{J}_{s}$ denote conserved currents of weight $(s,0)$ and $(0,s)$, respectively, $\mu_{s}$ and $\bar{\mu}_{s}$ the corresponding sources, and the sums run over the particular spectrum of the theory under consideration. The dots in \eqref{action deformations} signify that, for non-chiral deformations, the action requires corrections to all orders in the sources in order for the associated partition function to realize the symmetry \cite{Schoutens:1990ja}. On the other hand, no such higher order terms are required in the Hamiltonian case \eqref{Hamiltonian deformation} \cite{Mikovic:1991rf,deBoer:2014fra}. It is important to notice that the Legendre transform that connects these two pictures is highly non-trivial for higher spin theories,\footnote{ In particular, the currents and sources in \eqref{Hamiltonian deformation} and \eqref{action deformations} are in fact not the same (as the naive notation could suggest), and are instead related to one another in a non-trivial way.} so a careful choice of boundary conditions is essential.

 In the present paper we will be mostly concerned with black holes which describe ensembles dual to Hamiltonian deformations of the form \eqref{Hamiltonian deformation}, because they have a straightforward interpretation in terms of the canonical CFT partition function 
 \begin{equation}
Z_{\text{can}}\left[\tau,\alpha_{s},\bar{\alpha}_{s}\right]
=
\text{Tr}_{\mathcal{H}}\, \exp 2\pi i\left[\tau\left( L_0-\frac{c}{24}\right) - \bar{\tau}\left( \bar{L}_0-\frac{c}{24}\right)
 + \sum_{s}\left(\alpha_s J_0^{(s)}-\bar{\alpha}_{s}\overline{J}_0^{(s)}\right)\right]
\end{equation}
\noindent on the torus. Here, $J_{0}^{(s)}$ and $\overline{J}_0^{(s)}$ denote the zero modes of the corresponding currents. The thermal sources  $\alpha_s\,$, $\bar{\alpha}_{s}$ are related to the $\mu_{s}$, $\bar{\mu}_{s}$ introduced above by \cite{deBoer:2014fra}
\begin{equation}\label{thermal sources}
\mu_{s} = \frac{i\alpha_{s}}{\text{Im}(\tau)}\,,
 \qquad 
\bar{\mu}_{s}= -\frac{i\bar{\alpha}_{s}}{\text{Im}(\tau)}
 \,.
\end{equation}

\noindent Since we are interested in thermodynamics, or alternatively stationary Euclidean black holes, for the remainder of this section we will restrict ourselves to constant connections on the cylinder/torus, which corresponds to constant sources $\mu_{s}$, $\bar{\mu}_{s}$ and constant charges (the latter being the eigenvalues of the zero-modes $J_0^{(s)}$ and $\overline{J}_0^{(s)}$). 

It is perhaps worth emphasizing that there exist several ways to compute the entropy of higher spin black holes, all giving the same result. This caused some confusion initially and we take this opportunity to compare and clarify the different approaches. In a Hamiltonian slicing of spacetime, the entropy of a black hole is defined as the contribution to the on-shell action coming from the boundary term at the horizon.\footnote{ This definition stems from identifying the on-shell value of the Hamiltonian action $I_{\text{Ham}}=\textrm{bulk}+B_{\infty}-B_+$ with the Helmholtz free energy $\beta F$, and interpreting the boundary term at infinity as the internal energy $\beta E\,$. The factors of $\beta$ come from integration along the compact time direction. Since the bulk contribution vanishes due to the constraints, we find $\beta\left(E-TS\right)=\beta E-B_+\Rightarrow B_+=S$.} Such a term is necessary since the time foliation is singular at $r=r_{+}\,$. In order to make sense of the action and its variation, a small disk is excised (introducing an artificial boundary) where suitable boundary conditions must be imposed. Alternatively, one can evaluate the on-shell action using an angular foliation, which is regular everywhere and no horizon boundary term is required. In either case, additional boundary terms at infinity must be added so as to have a well defined variational principle. By comparing the angular vs. time quantization schemes and paying attention to orientation issues, it was first shown in \cite{Banados:2012ue} that the boundary terms at infinity and at the horizon are related by
\begin{equation}\label{B+}
B_\infty - B_+ = - B_\infty \ \ \ \ \ \Rightarrow \ \ \ \  B_+ = 2 B_\infty = {k_{cs} \over 2\pi} \int \mbox{Tr}\left[A_t A_\phi\right].
\end{equation} 
This can be understood as the Smarr relation between the charges and the entropy.  A direct calculation of the boundary term at the horizon was later performed in \cite{Bunster:2014mua}, yielding concordant results.

The same expression can be arrived at by demanding validity of the first law of thermodynamics, as argued in \cite{Perez:2013xi} for the case of $sl(3)\,$. Moreover, the entropy can also be understood as the on-shell value of the appropriate action functional in a microcanonical ensamble, where the charges at infinity are held fixed. In this context, as first shown in \cite{deBoer:2013gz}, the entropy of a higher spin black hole is given in full generality by
\begin{equation}\label{dBJ entropy formula 1}
 S = -2\pi i k_{cs}\text{Tr}\Bigl[\left(a_{w} + a_{\bar{w}}\right)\left(\tau a_{w} + \bar{\tau}a_{\bar{w}}\right) - \left(\bar{a}_{w} + \bar{a}_{\bar{w}}\right)\left(\tau\bar{a}_{w} + \bar{\tau}\bar{a}_{\bar{w}}\right)\Bigr]\,.
\end{equation}
Evaluated on a black hole solution, \eqref{dBJ entropy formula 1} yields
\begin{equation}
S = -2\pi i \left(2\tau \left(h-{c\over 24}\right)  +\sum_{s} s\,\alpha_{s}Q_{s} \right)+\text{other sector}\,,
\end{equation}
\noindent where $Q_{s}$ denotes the expectation value of the dimension-$s$ charge $J_{0}^{(s)}$. 
More interestingly, one can exploit the holonomy conditions to cast the entropy directly as function of the charges only. Using the smoothness conditions \eqref{eigenv conditions} one finds that \eqref{dBJ entropy formula 1} can be written equivalently as  \cite{deBoer:2013gz}
\begin{equation}\label{dBJ entropy formula 2}
S = 2\pi k_{cs}\text{Tr}\Bigl[\left(\lambda_{\phi} - \bar{\lambda}_{\phi}\right)L_0\Bigr]\,,
\end{equation}
\noindent where $\lambda_{\phi}$ and $\bar{\lambda}_{\phi}$ are diagonal matrices containing the eigenvalues of the angular component of the connection, which carries the expectation values of the charges. It is worth emphasizing that \eqref{dBJ entropy formula 1} and \eqref{dBJ entropy formula 2} are completely general and valid for any algebra and embedding, depending only on the choice \eqref{eigenv conditions} of holonomy condition along the thermal cycle.
%

\subsection{Extremal higher spin black holes}\label{sec:defnext}
In conventional gravitational theories, the notion of extremality is tied to the confluence of two horizons. This feature generically implies that the Hawking temperature of the black hole is zero. We could declare that extremality in higher spin theories is simply defined as a solution at zero temperature. However, our aim is to propose a definition that is along the lines of confluence (degeneration) of the parameters of the solution and that relies only on the topological formulation of the theory, yielding in particular the zero-temperature condition as a consequence.

In this spirit, we propose that a $3d$ extremal higher spin black hole is a solution of Chern-Simons theory corresponding to flat boundary connections $a$ and $\bar{a}$ satisfying the following conditions:
\begin{enumerate}
\item They obey Drinfeld-Sokolov boundary conditions,

\item Their components are constant, and therefore correspond to stationary solutions,

\item They carry charges (expectation values) and chemical potentials (sources), which are manifestly real in the Lorentzian section,

\item The angular component of at least one of $a$ and $\bar{a}$, say $a_{\phi}\,$, is non-diagonalizable.
\end{enumerate}

Naturally, the key point of the definition is the non-diagonalizability of the $a_{\phi}$ component. The rationale behind this requirement is as follows. Suppose both the $a_{\phi}$ and $\bar{a}_{\phi}$ components were diagonalizable. Since the boundary connections are assumed to be constant, by the equations of motion the (Euclidean) time components of the connection commute with the angular components, and can be diagonalized simultaneously with them. It is then possible to solve \eqref{eigenv conditions} and find a non-zero and well-defined temperature and chemical potentials as function of the charges. On the other hand, if at least one of $a_{\phi}$ and $\bar{a}_{\phi}$ is non-diagonalizable then $a_{\rm contract}$ will be non-diagonalizable as well. If we insist upon \eqref{eigenv conditions}, then both features are compatible if we take a zero temperature limit, because the smoothness condition becomes degenerate as well.  This is consistent with the usual notion that the solid torus topology of the finite-temperature black hole should change at extremality. 

We emphasize that it is quite convenient to define extremality in terms of the angular component of the connection, because as explained above the latter carries the charges under canonical boundary conditions, and does not involve the sources. Therefore, the conditions for extremality will be cleanly expressed as relations between the charges carried by the black hole, with no ``contamination" from the sources. This will be particularly important later on when we compare extremality conditions with BPS conditions in the CFT, because the latter are derived directly from the operator algebra and indeed involve the charges only. Furthermore, the non-diagonalizability of the connection can be conveniently encoded in terms of the Jordan class of the angular component of the connection, or equivalently the angular holonomy, and we will do so throughout the paper. Moreover, because the analysis involves the angular component of the connection, the classification of holonomies extends rather straightforwardly to other solutions which are not black holes and do not include sources, such as conical defects. As we will comment in due course, the only change lies in the reality properties of the eigenvalues of the connection.

A final technical note: while for a general connection the degeneration of eigenvalues does not imply non-diagonalizability, the special form of the flat connections dictated by the Drinfeld-Sokolov boundary conditions will guarantee that if two eigenvalues of $a_\phi$ are degenerate, then the connection is non-diagonalizable. From this perspective, we could interpret that equating eigenvalues of $a_\phi$ is in a sense analogous to the confluence of horizons for extremal black holes in general relativity.  

\subsection{Non-supersymmetric examples: extremal BTZ and $sl(3)$ black holes}\label{sec:examples}
We will now exemplify our definition for two simple but important solutions: the BTZ black hole and the $sl_{3}$ higher spin black hole, which are respectively solutions of $sl(2)\oplus sl(2)$ and $sl(3)\oplus sl(3)$ Chern-Simons theory.

Let us start with the BTZ black hole as given in \eqref{BTZ connection}. At finite temperature, using \eqref{dBJ entropy formula 2} and the relation $c=6k$ we easily recover the standard results for the BTZ black hole entropy
\begin{equation}
S = 2\pi \sqrt{\frac{c}{6}\left(h-\frac{c}{24}\right)} + 2\pi \sqrt{\frac{c}{6} \left(\bar h-\frac{c}{24}\right)}\,\,.
\end{equation}

\noindent The angular holonomy of the connection \eqref{BTZ connection} is (up to conjugation)
\begin{equation}
\text{Hol}_{\phi}(a) \sim e^{2\pi a_{\phi}}
=
\left(\begin{array}{cc}
\cosh\left(2\pi\sqrt{{\cal L}}\right) & \sqrt{{\cal L}}\sinh\left(2\pi\sqrt{{\cal L}}\right)  \\ 
\frac{1}{\sqrt{{\cal L}}}\sinh\left(2\pi\sqrt{{\cal L}}\right)
& \cosh\left(2\pi\sqrt{{\cal L}}\right) 
\end{array} \right)\,,
\end{equation}
\noindent and similarly for $\bar{a}_{\phi}\,$; recall that ${\cal L}={h\over k}-{1\over 4}$. For generic values of ${\cal L}$, the angular holonomy has two unequal eigenvalues, given by 
\begin{gather}\label{eigenvalues BTZ holonomy}
\lambda_{1}^{h} = e^{2\pi \sqrt{{\cal L}}}\,,\qquad 
\lambda_{2}^{h} = 
e^{-2\pi \sqrt{{\cal L}}}\,\,.
\end{gather}
 
 \noindent Now, a necessary condition for non-diagonalizability is that both eigenvalues are equal, which implies
\begin{equation}
\text{extremality condition:}\qquad \lambda_{1}^{h}  = \lambda_{2}^{h}  \qquad \Rightarrow \qquad {\cal L}=0\,.
\end{equation}
\noindent From the smoothness conditions \eqref{BTZ smoothness} we immediately see that the non-diagonalizability condition implies that the temperature goes to zero. Moreover, in this limit the holonomy becomes
\begin{equation}
\text{extremal holonomy:}\quad \text{Hol}_{\phi}(a) \sim 
\left(
\begin{array}{cc}
1 & 0 \\ 
2\pi & 1
\end{array} \right).
\end{equation}
\noindent The important observation is that precisely at the extremality point the Jordan class of the holonomy changes: while the finite-temperature holonomy lies on a hyperbolic conjugacy class of $SL(2)$, the extremal black hole holonomy belongs to a parabolic conjugacy class \cite{Banados:1992gq,Martinec:1998wm, Maldacena:1998bw}. 

Let us now move on to the spin-3 black hole of \cite{Gutperle:2011kf}. We will focus on the unbarred sector for concreteness. Using canonical boundary conditions, the boundary connections are given by
\begin{align}
a_{\phi} = a_{w} + a_{\bar{w}} ={}&
\left(
\begin{array}{ccc}
0 & \frac{1}{2}{\cal L} & -\frac{2}{\gamma }W \\ 
1 & 0 & \frac{1}{2}{\cal L} \\ 
0 & 1 & 0
\end{array} 
\right)\,,
\\
ia_{t_{E}}+a_{\phi}  = 2a_{\bar{w}} ={}&
-\frac{\gamma \mu}{2}\left(
\begin{array}{ccc}
-\frac{1}{6}{\cal L} & -\frac{2}{\gamma } W& \frac{1}{4}{\cal L}^2 \\ 
0 & \frac{1}{3}{\cal L} & -\frac{2}{\gamma }W \\ 
1 & 0 & -\frac{1}{6}{\cal L}
\end{array} 
\right)\,,
\label{awbar W3}
\end{align}

\noindent where $k_{cs}{\cal L}$ and $k_{cs}W$ denote, respectively, the expectation values of the zero modes of the stress tensor $T$ and the dimension-3 current $W$ on the cylinder. The normalization constant $\gamma$ takes the value $\gamma^{2} =8/5$, which gives canonical OPE relations on the plane \cite{deBoer:2014fra}. Here $\mu$ denotes the source for the weight-3 current, and we emphasize that we have not added an explicit source for the stress tensor in the connection, because the modular parameter $\tau$ is included explicitly in the coordinate identifications. 

Applying \eqref{dBJ entropy formula 1} to this solution we get
\begin{align}
S ={}&
 -2\pi i k_{cs}\bigl(2\tau {\cal L}+ 3\alpha W \bigr) + \text{other sector}\,,
\end{align}
\noindent where the thermal spin-3 source $\alpha$ is related to the spin-3 chemical potential $\mu$ as in \eqref{thermal sources}:
\begin{equation}
\mu = \frac{2\alpha}{\bar{\tau}-\tau}\,.
\end{equation}
\noindent As pointed out above we can write the entropy  as a function of the charges only. In order to achieve this, we will find it convenient to trade the charges $(\cL,\cW)$ for the eigenvalues of $a_{\phi}$ \cite{deBoer:2013gz,deBoer:2013vca}, which we parameterize as $\text{Eigen}(a_{\phi}) = (\lambda_{1},\lambda_{2},-\lambda_{1}-\lambda_{2})$, so that
\begin{equation}\label{charges to eigenvals}
{\cal L} =\lambda_{1}^{2} + \lambda_{1}\lambda_{2} + \lambda_{2}^{2}\,\,,\qquad 
W = \frac{\gamma}{2}\lambda_{1}\lambda_{2}\left(\lambda_{1} + \lambda_{2}\right)\,,
\end{equation}
\noindent with analogous expressions in the barred sector. In Lorentzian signature the eigenvalues $(\lambda_i,\bar \lambda_i)$ are independent, and real when one chooses the connection to be valued in $sl(3;\mathds{R})\,$.  In Euclidean signature, we have $\lambda_i^*=-\bar\lambda_i$, which implies that ${\cal L}^*=\bar {\cal L}$ and $W^*=-\bar W$.  
Equation \eqref{dBJ entropy formula 2} then gives us immediately the entropy as a function of the charges
\begin{align}\label{eq:enths}
S ={}&
  2\pi k_{cs}\bigl(\lambda_{1}-\lambda_{3}\bigr) +\text{other sector}
 \nonumber\\
  ={}& {2\pi k_{cs}}\bigl(2\lambda_{1}+\lambda_{2}\bigr) +\text{other sector}\,,
\end{align}
\noindent with $\lambda_{1}$ and $\lambda_{2}$ obtained by inverting \eqref{charges to eigenvals} and choosing the branch of the solution that connects smoothly to the BTZ black hole as one turns off the $W$ charge.

Next, in order to obtain the potentials as a function of the charges we solve the smoothness conditions \eqref{smoothness}. This gives
\begin{align}\label{smoothness solution 1}
\tau ={}&
 i \frac{2\lambda_{1}^{2} + 2\lambda_{1}\lambda_{2}-\lambda_{2}^{2}}{\left(\lambda_{1}-\lambda_{2}\right)\left(2\lambda_{1}+\lambda_{2}\right)\left(\lambda_{1}+2\lambda_{2}\right)}\,,
\\ 
 \alpha 
 ={}&
   -{6i\over \gamma}\frac{\lambda_{2}}{\left(\lambda_{1}-\lambda_{2}\right)\left(2\lambda_{1}+\lambda_{2}\right)\left(\lambda_{1}+2\lambda_{2}\right)}\,.
\end{align}

\noindent In Euclidean signature, $(\bar \tau,\bar \alpha)$ are the complex conjugate of the above expressions. When continuing back to Lorentzian signature, the charges and potentials in the two sectors are no longer related to each other by complex conjugation, but are instead each real and independent. Upon performing this continuation, it is convenient to trade the parameters $(\tau,\bar{\tau})$ for the inverse temperature $\beta$ and the (real, Lorentzian) angular velocity $\Omega\,$ via \eqref{tau and tau bar} (which remains true in the higher spin case). We then note that the above relations imply
\begin{align}
\mu ={}&
 6\gamma\bigl(1+\Omega\bigr)\left(\frac{\lambda_{2}}{2\lambda_{1}^{2} + 2\lambda_{1}\lambda_{2}-\lambda_{2}^{2}}\right)
 \\
 \bar{\mu} ={}&
 -6\gamma\bigl(1-\Omega\bigr)\left(\frac{\bar{\lambda}_{2}}{2\bar{\lambda}_{1}^{2} + 2\bar{\lambda}_{1}\bar{\lambda}_{2}-\bar{\lambda}_{2}^{2}}\right).
 \label{smoohtness solution 2}
\end{align}

With these explicit relations we can now implement our definition of extremality. Requiring that $a_{\phi}$ should be non-diagonalizable gives as a necessary condition
\begin{equation}\label{eq:extLW}
\text{extremality condition:}\qquad \lambda_{1} = \lambda_{2} \equiv \lambda \qquad \Rightarrow \qquad  {\cal L}= 3\lambda^{2}\,,\quad W =\gamma \lambda^{3}\,.
\end{equation}
 
\noindent As a consequence, while the finite-temperature angular holonomy is diagonalizable, in the extremal limit we obtain
\begin{equation}\label{eq:hol2}
\text{extremal holonomy:}\quad \text{Hol}_{\phi}(a) \sim 
\left(
\begin{array}{ccc}
e^{-4\pi \lambda} & 0 & 0 \\ 
0 & e^{2\pi \lambda} & 1 \\
0 & 0 & e^{2\pi \lambda}
\end{array} \right)\,,
\end{equation}
\noindent exhibiting the expected non-trivial Jordan normal form. We emphasize that the extremal limit of the spin-3 higher spin black hole was first discussed in \cite{Gutperle:2011kf}: their bound was found as the maximal value of $W$ for a given ${\cal L}$ such that the entropy is real, and it agrees with \eqref{eq:extLW}.

 Turning now our attention to the potentials, from \eqref{smoothness solution 1}-\eqref{smoohtness solution 2} and \eqref{tau and tau bar} we see in particular that in this limit
\begin{equation}\label{W3 extremal potentials}
\text{extremal potentials:}\qquad \beta \to \infty\,,\quad \mu \to 4{\gamma\over \lambda}\,, \quad \Omega\to 1\,, \quad \bar{\mu} \to 0\,,
\end{equation}
\noindent so the temperature is zero as expected. The spin-3 chemical potential $\mu$ remains finite and becomes a simple homogeneous function of the charges, whereas the corresponding thermal source $\alpha$ scales with the inverse temperature and blows up. On the other hand, the barred sector spin-3 potential $\bar{\mu}$ goes to zero because the thermal source $\bar{\alpha}$ remains unconstrained and in particular finite, as no condition is imposed on the barred charges. 

Several comments are now in order.
\begin{enumerate}
\item {\it Jordan decomposition versus zero temperature:} A valid concern is to wonder if our definition of extremality implies zero temperature and vice-versa. From \eqref{smoothness solution 1} it is clear that there are 3 combinations of $\lambda_1$ and $\lambda_2$  that achieve $\beta\to \infty\,$. The additional other branches also give non-trivial Jordan forms, since they just correspond to different pairings of eigenvalues that are degenerate. For this reason, all these cases are captured by \eqref{eq:extLW}: any pairing $\lambda_i=\lambda_j$ with $i\neq j$  implies the  extremality bound ${\cal L}^3=27 \gamma^{-2} W^2$.\footnote{ Different pairings of eigenvalues conflict with the ordering of eigenvalues used in \eqref{eq:enths}, but this is easily fixed by reordering the eigenvalues appropriately.} At least  for $N=2,3$, a non-trivial Jordan decomposition implies zero temperature and vice-versa. And from the heuristic argument in section \ref{sec:defnext}, we expect this to always be the case. 
\item{\it Other Jordan classes:} For $\lambda\equiv \lambda_1=\lambda_2\neq 0$,  $a_\phi$ has only 2 linearly independently eigenvectors. If take first $\lambda_2=0$ and then $\lambda_1=0$, the holonomy of $a_\phi$ belongs to a different Jordan class where there is only one eigenvector; this case corresponds to extremal BTZ within $sl(3)\oplus sl(3)$ Chern-Simons theory.
\item{\it Finite entropy:} We have a continuous family of extremal ${\cal W}_3$ black holes parametrized by $\lambda$, and from \eqref{eq:enths} the contribution of the extremal (unbarred) sector to the total entropy is
\bea
S_{\rm ext}= 2\pi k_{cs} \, \lambda &=&{\pi\over 3}  \sqrt{ {c\over 2}\left(h -\frac{c}{24}\right)}\cr
&=& {\pi\over 2} \left({c\, \qw\over 9\gamma}\right)^{1/3} \,,
\eea
where $c=24 k_{cs}$ and we casted the answer in terms of the charges $(h,\qw)$ on the plane, related to $\left(\mathcal{L},\mathcal{W}\right)$ by $k_{cs}{\cal L}=h -{c\over 24} $ and $k_{cs}{W}=\qw\,$. The answer is clearly finite.  This should be contrasted with extremal BTZ, where the contribution of the extremal sector vanishes. 
It would be interesting to derive such bound and residual entropy in a CFT with ${\cal W}_3$ symmetry.

\item{\it Extremality vs. unitarity:} The extremality condition we have discussed can be thought of as a bound 
\be\label{eq:bulkw3}
{64\over 5 c}\left({h-{c\over 24}}\right)^3\geq 9 \qw^2
\ee
\noindent on the charges of a spin-3 black hole. On the other hand, in a theory with ${\cal W}_3$ symmetry, the unitary bound in the semiclassical limit is \cite{Mizoguchi:1988vk}\footnote{ The quantum (finite-$c$) unitarity bound reported in \cite{Mizoguchi:1988vk} is
\begin{equation*}
{64\over 22 +5c} h^2 \left( h -{1\over 16} -{c\over 32}\right)-9\qw^2 \geq0 ~.
\end{equation*}}
\be\label{eq:bndyw3}
{64\over 5 c} \left( h^3 -{c\over 32} h^2\right)\geq 9 \qw^2~. 
\ee

\noindent It is clear that \eqref{eq:bulkw3} and \eqref{eq:bndyw3} do not agree. However, the ${\cal W}_3$ unitarity bound \eqref{eq:bndyw3} encloses the bulk extremality bound  \eqref{eq:bulkw3}, indicating that all $sl(3)$ black holes are dual to states allowed by unitarity in the dual CFT.

\item{\it Conformal invariance:} In two-derivative theories of gravity in $D=4,5$ all extremal black holes contain an AdS$_2$ factor in its near horizon geometry \cite{Kunduri:2007vf,Kunduri:2013ana}. The enhancement of time translations to conformal transformations is non-trivial and unexpected a priori; moreover, it is key to build microscopic models of extremal black holes. Here we have not investigated this feature explicitly, but we do expect that the connection at the extremal point is invariant a larger set of gauge transformations relative to the non-extremal connection. It would be interesting to quantify these symmetries and understand its role in the dual CFT.  
\end{enumerate}

Having described the general framework to study extremal higher spin black holes, in the remainder of the article we will focus on the interplay between extremality and supersymmetry. Furthermore, we will analyze our results from the perspective of the holographic duality between Chern-Simons supergravities and CFTs with super-$\cW$ symmetry algebras.

\section{Supersymmetric higher spin backgrounds }\label{sec:BHs}
Extremality can be understood as the saturation of certain inequalities involving conserved charges, and it is natural to contrast these inequalities with BPS bounds that appear in supersymmetric setups. It is well known that in two-derivative theories of supergravity these two types of conditions are intimately related: supersymmetry always implies zero temperature and therefore extremality in the context of BPS black holes. The lore behind this is as simple as noticing that on a contractible circle fermions are anti-periodic whereas bosons are periodic, making finite temperature incompatible with symmetries which relate the two kinds of fields.\footnote{ Another way to motivate this relation in a purely gravitational context is via the attractor mechanism \cite{Ferrara:1995ih,Ferrara:1996um,Ferrara:1996dd,Chamseddine:1996pi}, which shows that the BPS equations in $\cN=2$ supergravity in $D=4,5$ have a fixed point which is responsible for the AdS$_2$ factor in the near horizon geometry.} 

In this section we will explore the relation between extremality and supersymmetry for AdS$_3$ higher spin black holes. There is an arbitrarily long list of supersymmetric higher spin theories with AdS$_3$ as its vacuum configuration. Here we will focus on one representative, namely the $sl(3|2)\oplus sl(3|2)$ Chern-Simons theory. This example contains a spin-3 supermultiplet in its spectrum, thus providing new features that have no counterpart in standard ${\cal N}=2$ supergravity. Since the latter theory is included in the higher spin model as a consistent truncation to $sl(2|1)\subset sl(3|2)$, we will also use our results to review some known features of supersymmetric BTZ black holes cast in Chern-Simons language. For completeness, we shall also study supersymmetric smooth conical defects. The discussion in this section focuses on the AdS$_3$ (bulk) properties of the solutions. In section \ref{sec:BPS} we will compare our findings with the CFT dual.

\subsection{$sl(3|2)$ solutions}\label{sec:w32solutions}
We will now study non-perturbative solutions of $sl(3|2)\oplus sl(3|2)$ Chern-Simons supergravity. For simplicity, we will refer to these backgrounds as `$sl(3|2)$ black holes' or `$sl(3|2)$ smooth conical defects.'

Let us begin by summarizing a few facts about the relevant superalgebra. In the principal embedding of $sl(2|1)$ in $sl(3|2)$ \cite{Peng:2012ae,Chen:2013oxa}, the even-graded sector of the superalgebra is decomposed into the $sl(2)$ generators ($L_i$), a spin 0 element ($J$), one spin 1 multiplet ($A_i$), and one spin 2 multiplet ($W_m$). All together, they span the bosonic sub-algebra $sl(3)\oplus sl(2)\oplus u(1)$. The odd-graded elements consist of two spin $1/2$ multiplets  ($H_r$ and $G_r$) and two spin $3/2$ multiplets ($T_s$ and $S_s$). By ``spin" we mean the $sl(2)$ spin $S$, so  within each multiplet the indices range from $-S$ to $S$.  The spin of the corresponding $3d$ bulk field carrying the representation is then $S+1\,$. Now, in order to fully determine the bulk Lorentzian theory, one must additionally specify the real form under consideration in this case. The complex superalgebra $sl(3|2;\mathds{C})$ has several real forms, as listed in e.g. \cite{Frappat:1996pb}, and in this paper we will deal with $su(2,1|1,1)\,$. This choice is intimately linked to the dual $\mathcal{W}_{(3|2)}$ symmetry. In particular, $su(p,3-p|q,2-q)$ is the only real form that has a compact Abelian generator.\footnote{ We thank J. de Boer for bringing this issue to our attention.} We refer to \cite{BetoJuan} for a further discussion of this subtle yet important point. We also encourage the reader to visit appendix \ref{app: sl(3|2)} for a more detailed discussion of the $sl(3|2)$ and $su(2,1|1,1)$ superalgebras, as well as the matrix representation we employ. Several results in this section pend on some of the specifics outlined therein.

Our aim is to characterize a wide class of solutions supported by the even-graded sector of the $su(2,1|1,1)$ superalgebra, which includes black holes and smooth conical defects. Following the discussion in section \ref{sec:finite}, we will incorporate the higher spin sources so as to realize the boundary conditions that are naturally described by a Hamiltonian formulation of the dual CFT$_2\,$. Simply put, after gauge fixing the radial dependence of the connection, the charges will be carried by the the angular component $a_{\phi}$. In particular, we write
\begin{empheq}{alignat=5}\label{eq: sl(3|2) Black Hole connection aphi}
	a_{\phi}&=L_1-\mathcal{L}L_{-1}-iQ_1J-Q_2A_{-1}-iQ_3W_{-2}\,,
\end{empheq}
where $\mathcal{L}$, $Q_1$, $Q_2$ and $Q_3$ are all taken to be constant and real so that (given our realization of the generators in terms of purely real matrices) $a_{\phi}$ lies on the real form $su(2,1|1,1)$.\footnote{ This was implicitly done in \cite{Tan:2012xi} where appropriate factors of $i$ were introduced to obtain the usual Hermiticity relations among the fields, and in particular the right sign of the kinetic terms in the Lagrangian for the $(2,0)$ supergravity truncation.}  Similar expressions hold for $\bar a_\phi$, which we will omit for the rest of the section. 

In \eqref{eq: sl(3|2) Black Hole connection aphi} we have also made a choice of Drinfeld-Sokolov decomposition for the connection. This is motivated by the holographic dictionary used to map these configurations to states in a CFT$_2$ with $\cW_{(3|2)}$ symmetry algebra \cite{BetoJuan}. We will elaborate more on this dictionary below. For the time being, we mention that the parameters appearing in $a_{\phi}$ are related to the zero modes of the $\mathcal{W}$-symmetry generators on the plane as follows:\footnote{ We note that the expression for the CFT stress tensor in terms of the bulk charges differs from that in \cite{Datta:2012km,Datta:2013qja}.}
\begin{equation}
\begin{aligned}\label{eq:mapcharges}
	\mathcal{L}
	&=
	\frac{6}{c}\left(h-\frac{c}{24}-\frac{3}{2c}q^2+\frac{1}{2}\kappa \qv\right)\,,
	\\
	Q_1
	&=
	-\frac{3}{c}q\,,
	\\
	Q_2
	&=
	-\frac{9}{5c}\kappa \qv\,,
	\\
	Q_3
	&=
	\frac{3}{5c}\kappa\left(\qw-\frac{6}{c}q\qv\right)\,.
\end{aligned}
\end{equation}

\noindent In the conventions of appendix \ref{app: W32}, $h$ denotes the zero mode of the stress tensor $T$, $q$ is that of the $U(1)$ current $J$, $\qv$ is the zero mode of the dimension $2$ primary $V$, and $\qw$ corresponds to the dimension $3$ operator $W\,$. The constant $\kappa$ is fixed in terms of the central charge as in \eqref{definition kappa}, and the large-$c$ limit is understood in the above expressions (c.f. \eqref{semiclassical kappa}). Due to the Hermiticity properties of the operators on the plane, we note that $h$, $q$, $\kappa \qv$ and $\kappa \qw$ are real numbers, implying that $\mathcal{L}$, $Q_1$, $Q_2$ and $Q_3$ are also real. 

Much like in the non-supersymmetric examples in section \ref{sec:Definition}, we will find it convenient for our purposes to redefine the charges in terms of the eigenvalues of $a_{\phi}+iQ_1J$, which we label as
\begin{empheq}{alignat=5}\label{eq: eigenvalues of aphi}
	\textrm{eigen}\left(a_{\phi}+iQ_1J\right)&=\left[\lambda_1,-\lambda_1+\lambda_2,-\lambda_2,\frac{1}{2}\lambda_3,-\frac{1}{2}\lambda_3\right]\,.
\end{empheq}
For notational simplicity, we have substracted the $U(1)$ piece from the connection since it commutes with the rest of the generators and is already diagonal. Notice that the matrix $a_{\phi}+iQ_1J$ is traceless and super-traceless, hence the above parametrization. Being block-diagonal, its characteristic polynomial factorizes into a cubic part and quadratic part, namely, 
\begin{empheq}{alignat=5}
	\textrm{det}\bigl(\lambda-a_{\phi}-iQ_1J\bigr)
	&=
	\bigl(\lambda^3-4\left(\mathcal{L}+Q_2\right)\lambda+8iQ_3\bigr)\left(\lambda^2-\mathcal{L}+Q_2\right)
	\cr
	&=
	\left(\lambda-\lambda_1\right)\left(\lambda+\lambda_1-\lambda_2\right)\left(\lambda+\lambda_2\right)\left(\lambda-\frac{1}{2}\lambda_3\right)\left(\lambda+\frac{1}{2}\lambda_3\right)\,.
\end{empheq}
The respective discriminants are
\begin{empheq}{alignat=5}
	\Delta_3&=64\left(4\left(\mathcal{L}+Q_2\right)^3+27Q_3^2\right)
	\cr
	&=\left(2\lambda_1-\lambda_2\right)^2\left(2\lambda_2-\lambda_1\right)^2\left(\lambda_1+\lambda_2\right)^2\,,
\end{empheq}
and
\begin{empheq}{alignat=5}
	\Delta_2&=4\left(\mathcal{L}-Q_2\right)
	\cr
	&=\lambda_3^2\,.
\end{empheq}

Since $\mathcal{L}$, $Q_2$ and $Q_3$ are real, the eigenvalues $\lambda_1$, $-\lambda_1+\lambda_2$ and $-\lambda_2$ are purely imaginary when $\Delta_3<0$, whereas for $\Delta_3>0$ two of the eigenvalues are minus complex conjugates of each other and the third, being minus the sum of the previous two, is purely imaginary. Analogously, we have that $\lambda_3$ is imaginary when $\Delta_2<0$ and real when $\Delta_2>0$. We will later see that black holes and smooth conical defects fall into the following sectors:
\begin{empheq}{alignat=5}
\begin{array}{|c|c|c|}
	\hline
	sl(3|2)\textrm{ solutions} & \Delta_3 & \Delta_2 \\
	\hline
	\textrm{Black holes} & \geq0 & \geq0 \\ 
	\hline
	\textrm{Smooth conical defects} & <0 & <0 \\
	\hline
\end{array}\,.
\end{empheq}
In this paper we will not explore the remaining possibilities $\Delta_3<0$, $\Delta_2\geq0$ and $\Delta_3\geq0$, $\Delta_2<0\,$. As we will discuss momentarily, extremal black holes correspond to the cases where either $\Delta_{3}=0$ or $\Delta_{2}=0\,$. Hence, the classification of the connections in terms of $\Delta_3$ and $\Delta_2$ is a natural generalization of the familiar classification of pure gravity solutions in terms of hyperbolic, parabolic and elliptic conjugacy classes in $SL(2)\,$.

It follows from the above formulae that
\begin{empheq}{alignat=5}
\label{charges as eigenvalues 1}
	\mathcal{L}
	&=
	\frac{1}{8}\left(\lambda_1^2+\lambda_2^2-\lambda_1\lambda_2+\lambda_3^2\right)\,, 
	\cr
	Q_2
	&=
	\frac{1}{8}\left(\lambda_1^2+\lambda_2^2-\lambda_1\lambda_2-\lambda_3^2\right)\,, 
	\\
	Q_3
	&=
	-\frac{i}{8}\left(-\lambda_1+\lambda_2\right)\lambda_1\lambda_2\,.
	\nonumber
\end{empheq}
As expected, the charges are symmetric polynomials in the eigenvalues. We have chosen the relative normalization in \eqref{eq: eigenvalues of aphi} such that the $sl(2|1)$ theory corresponds to $\lambda_1=\lambda_2=\lambda_3\,$, for which
\begin{empheq}{alignat=5}\label{eq:trunceig}
	\mathcal{L}&=\frac{1}{4}\lambda_1^2\,, 
	&\quad 
	Q_2&=0\,, 
	&\quad 
	Q_3&=0\,,
\end{empheq}
and
\begin{empheq}{alignat=5}
	\Delta_3&=4\lambda_1^6\,,
	&\quad 
	\Delta_2&=\lambda_1^2\,.
\end{empheq}

\noindent Notice that there are other values of $\lambda_i$ that give $Q_2=Q_3=0$; these are perfectly admissible and will not be discarded in our discussion. However, they lead to non-vanishing chemical potentials in the higher spin sector (see \eqref{eq: thermal sources sl(3|2) 1} below), which implies that they are not a solution of the $sl(2|1)$ truncation.

As pointed out in section \ref{sec:defnext}, our definition of extremality involves the diagonalizability of the angular component of the connection. It is easy to see that in this example $a_{\phi}$ is diagonalizable if and only if $\Delta_3\neq0$ and $\Delta_2\neq0$, in which case there exists a similarity matrix $V$ that brings it to the form
\begin{empheq}{alignat=5}\label{eq:aphid1}
	V^{-1}a_{\phi}V&=a_{\phi}^D\,,
\end{empheq}
where
\begin{empheq}{align}
	a_{\phi}^D&=\frac{1}{4}\left(\lambda_1+\lambda_2+2\lambda_3\right)L_0+\frac{1}{4}\left(\lambda_1+\lambda_2-2\lambda_3\right)A_0+\frac{3}{4}\left(\lambda_1-\lambda_2\right)W_0-iQ_1J
\end{empheq}
lies in the Cartan subalgebra of $sl(3|2)$, as appropriate. If $a_{\phi}$ is not diagonalizable, and hence {\it extremal} according to our definition, then either $\Delta_3=0$ or $\Delta_2=0$ and the Jordan decomposition becomes
\begin{empheq}{alignat=5}\label{eq: Jordan form}
	V^{-1}a_{\phi}V&=a_{\phi}^D+a_{\phi}^N\,,
\end{empheq}
where $a_{\phi}^D$ is the same diagonal matrix as above and $a_{\phi}^N$ is a nilpotent matrix commuting with $a_{\phi}^D$. The precise form of $a_{\phi}^N$ depends on the class of matrix under consideration, determined by the number of repeated eigenvalues, i.e. the multiplicity of zeros of $\Delta_3$ and $\Delta_2\,$. Generically, it can be written as a matrix with a few non-zero off-diagonal elements, and it is unique up to similarity transformations that leave $a_{\phi}^D$ invariant. We will come back to this point in the next section.

To summarize, a general $sl(3|2)$ Drinfeld-Sokolov connection can reside in any of the ten classes:
\begin{empheq}{alignat*=5}
\begin{array}{ccc}
\begin{array}{|c|c|}
	\hline
	\textrm{Eigenvalues }\left(3\times3\right) & \Delta_3 \\
	\hline
	\lambda_1=\lambda_2=0 & =0 \\ 
	\hline
	\lambda_2=2\lambda_1\neq0 & =0 \\
	\hline
	\lambda_1=2\lambda_2\neq0 & =0 \\
	\hline
	\lambda_1=-\lambda_2\neq0 & =0 \\
	\hline
	2\lambda_2\neq\lambda_1\neq-\lambda_2\neq-2\lambda_1 & \neq0 \\
	\hline
\end{array}
&\bigotimes&
\begin{array}{|c|c|}
	\hline
	\textrm{Eigenvalues }\left(2\times2\right) & \Delta_2 \\
	\hline
	\lambda_3=0 & =0 \\
	\hline
	\lambda_3\neq0 & \neq0 \\
	\hline
\end{array}
\end{array}\,.
\end{empheq}
This classification is further refined by looking at the signs of $\Delta_3$ and $\Delta_2$. Let us highlight some important properties. First, the $U(1)$ charge, $Q_1$, while important for other considerations, does not play a role in the characterization of the Jordan class. Second, the form \eqref{eq: Jordan form} of the connection does not necessarily take values in the real form $su(2,1|1,1)$ of $sl(3|2;\mathds{C})$, just like the diagonal form of a (diagonalizable) real matrix is not necessarily real. This is equivalent to saying that the similarity matrix $V$ that accomplishes \eqref{eq: Jordan form} does not axiomatically belong to the supergroup $SU(2,1|1,1)$. The Jordan form of the connection does, nonetheless, belong to $sl(3|2;\mathds{C})$.  Also, given the reality properties of the eigenvalues in the black hole and smooth conical defect sectors, which we shall derive momentarily, some of the classes will not be relevant for what follows. In particular, we will see that there are only six non-empty classes for black holes solutions while smooth conical defects can only exist when $a_{\phi}$ is diagonalizable. 


\subsubsection{Black holes and their thermodynamics}\label{sec:sl32bhs}
In the absence of a metric formulation, one resorts to the Euclidean description in order to define higher spin black holes via appropiate smoothness conditions on the solid torus \cite{Gutperle:2011kf}. In Chern-Simons language, the continuation from Lorentzian to Euclidean signature is achieved by letting the two connections $A$ and $\bar{A}$ become complex, thus valued in $sl(3|2;\mathds{C})$, with the constraint $\bar{A}=-A^{\dagger}\,$. Consequently, the charges and chemical potentials introduced below become complex numbers. After defining the Euclidean solutions and studying their thermodynamic properties, we will translate back to Lorentzian signature and demand that the gauge fields lie in the correct real form. 

Borrowing from notions well understood for the BTZ black hole, a Euclidean higher spin black hole is defined by a smooth connection on the solid torus that carries charges as well sources (chemical potentials). As mentioned above, the charges have already been incorporated in the highest weight components of $a_{\phi}\,$, so the sources that support the background will be added in the lowest weight components of $ia_{t_E}+a_{\phi}\,$. The Euclidean black hole connection then reads
\begin{align}\label{eq: sl(3|2) Black Hole connection aphi 2}
	a_{\phi}
	&=
	L_1-\mathcal{L}L_{-1}-iQ_1J-Q_2A_{-1}-iQ_3W_{-2}\,,
	\\
	\label{eq: sl(3|2) Black Hole connection atE}
	ia_{t_E}+a_{\phi}&=i\nu_0J+\tilde{\nu}_{-1}\left(A_1-\frac{5}{3}L_1\right)+i\nu_{-2}W_2
	\nonumber\\
	&
	-\left(4Q_3\nu_2+\left(Q_2-\frac{5}{3}\mathcal{L}\right)\tilde{\nu}_{-1}\right)L_{-1}-\left(4Q_3\nu_2+\left(\mathcal{L}-\frac{5}{3}Q_2\right)\tilde{\nu}_{-1}\right)A_{-1}
	\\
	&
	-2i\left(\mathcal{L}+Q_2\right)\mu_{-2}W_0+i\left(\left(\mathcal{L}+Q_2\right)^2\nu_{-2}+\frac{2}{3}Q_3\tilde{\nu}_{-1}\right)W_{-2}\,.
	\nonumber
\end{align}

\noindent The higher weight components of $ia_{t_E}+a_{\phi}$, namely the last two lines in \eqref{eq: sl(3|2) Black Hole connection atE}, are fixed in terms of the charges $\mathcal{L}$, $Q_1$, $Q_2$ and $Q_3$, and sources $\nu_0$, $\tilde{\nu}_1$ and $\nu_{-2}$ by the flatness condition $[a_{t_E},a_{\phi}]=0$. Analogous expressions follow for the other sector $\bar{a}=-a^{\dagger}$.

A few comments are in order. First, black hole solutions correspond to constant (and purely bosonic) connections defined on the boundary cylinder, which after proper identifications becomes a torus. For ease of comparison with CFT variables, however, we have chosen to map the charges appearing in $a_{\phi}$ to the zero-modes of the $\mathcal{W}$-symmetry generators on the plane. This map is given in \eqref{eq:mapcharges}. Secondly, it is important to emphasize that, just as in the bosonic $\cW_{3}$ example \eqref{awbar W3}, the source for the stress tensor has already been incorporated as the modular parameter $\tau$ of the boundary torus and need not be introduced in the connection. That is why we have only turned on a source for the combination $\left(A_{1}-\frac{5}{3}L_{1}\right)$, which implies a chemical potential for the dimension-2 primary $V$ in the CFT, but not the stress tensor \cite{BetoJuan}. Also, the parameters $\nJ$, $\nA$ and $\nW$ are not particularly meaningful, but their relation to the CFT sources can be obtained by analyzing the CFT's Ward identities \cite{BetoJuan}. This gives
\begin{empheq}{alignat=5}\label{potentials 2}
	\nu_0
	&=
	\mu_1+\frac{6}{c}\mu_3\qv\,,
	\cr
	\tilde{\nu}_{-1}
	&=
	-\frac{3\kappa}{10}\left(\tilde{\mu}_2+\frac{6}{c}q\mu_3\right)\,,
	\\
	\nu_{-2}
	&=
	-\frac{3\kappa}{10}\mu_3\,.
	\nonumber
\end{empheq}

\noindent The above redefinitions are such that 
\begin{empheq}{alignat=5}
	k_{cs}\,\textrm{sTr}\bigl[\left(ia_{t_E}+a_{\phi}\right)a_{\phi}\bigr]&=\mu_1q+2\tilde{\mu}_2\qv+3\mu_3\qw\,,
\end{empheq}

\noindent which correctly identifies $(\mu_1,\tilde\mu_2,\mu_3)$ as the potentials conjugate to $(q,\qv,\qw)$, respectively.

Let us now solve the smoothness condition for the Euclidean black hole along the lines of \eqref{eigenv conditions}. As in the bosonic examples, demanding that the holonomy of the connection around the thermal cycle be trivial fixes the chemical potentials in terms of the charges. In supersymmetric theories, however, there is an important new ingredient that is worth highlighting. The center of the bosonic sub-algebra $sl(3)\oplus sl(2)\oplus u(1)\subset sl(3|2)$ is
\begin{empheq}{alignat=5}\label{eq:gammas}
	\Gamma^{\pm}&\equiv\left(
	\begin{array}{cc}
		\mathds{1}_{3\times3} & 0 \\
		0 & \pm\mathds{1}_{2\times2}
	\end{array}
	\right)\,,
\end{empheq}
where the upper and lower blocks correspond to the centers of $sl(3)$ and $sl(2)$, respectively.\footnote{ Technically, there is also an arbitrary $U(1)$ element in the center of the bosonic sub-algebra. However, it does not commute nor anti-commute with the fermionic elements.} Notice though that $\Gamma^-$ anti-commutes with the odd generators of $sl(3|2)$, so the actual center of the superalgebra is just $\Gamma^+=\mathds{1}_{5\times5}\,$. For this reason, one might naively think that the appropriate smoothness condition is to set the holonomy to $\Gamma^+$. This, however, is {\it wrong}. The novel feature here is that the choice of sign reflects upon whether the fermions present in the theory satisfy periodic or anti-periodic boundary conditions, i.e. the spin structure of the manifold. Indeed, it is easy to convince oneself that under a gauge transformation $\Gamma^+$ remains invariant when the fermionic components of the transformation parameter are periodic, while $\Gamma^-$ does so for anti-periodic fields. Since a contractible cycle allows only for the latter possibility, we shall adopt 
\begin{empheq}{alignat=5}
	H_{\mathcal{C}_E}&\equiv\mathcal{P}e^{\oint_{\mathcal{C}_E}a}&\,=\Gamma^-
\end{empheq}
as the {\it correct} holonomy condition for the thermal direction. In addition, $e^{2\pi i L_0}=\Gamma^-$, which agrees with \eqref{eigenv conditions}. Hence, the smoothness requirement translates to
\begin{empheq}{alignat=5}\label{eq:w32smooth}
	V^{-1}\left(\tau a_w+\bar{\tau}a_{\bar{w}}\right)V&=iL_0\,,
\end{empheq}
where $V$ is the same matrix that diagonalizes $a_{\phi}\,$. We could have used a different element of the Cartan sub-algebra to cast $\Gamma^-$, one that involves a suitable combination of $L_0$, $W_0$, $A_0$ and $J$. The choice in \eqref{eq:w32smooth} is the simplest one that accommodates the BTZ black hole. The interpretations of other choices are discussed in \cite{Hijano:2014sqa,deBoer:2014sna}. It is worth pointing out that we can still have periodic or anti-periodic fields in the $\phi$ coordinate, which defines a non-contractible cycle in a black hole topology.

It is now a straightforward task to solve the smoothness conditions, obtaining
\begin{equation}
\begin{aligned}\label{eq: thermal sources sl(3|2) 1}
	\tau&=\frac{i}{3}\left(4\frac{\lambda_1\left(2\lambda_2-\lambda_1\right)+\lambda_2\left(2\lambda_1-\lambda_2\right)}{\left(2\lambda_1-\lambda_2\right)\left(2\lambda_2-\lambda_1\right)\left(\lambda_1+\lambda_2\right)}-\frac{1}{\lambda_3}\right)\,,
	\\
	\gamma_0&=Q_1\tau\,,
	\\
	\gamma_1&=\frac{i}{2}\left(\frac{\lambda_1\left(2\lambda_2-\lambda_1\right)+\lambda_2\left(2\lambda_1-\lambda_2\right)}{\left(2\lambda_1-\lambda_2\right)\left(2\lambda_2-\lambda_1\right)\left(\lambda_1+\lambda_2\right)}-\frac{1}{\lambda_3}\right)\,,
	\\
	\gamma_2&=\frac{3}{2}\frac{-\lambda_1+\lambda_2}{\left(2\lambda_1-\lambda_2\right)\left(2\lambda_2-\lambda_1\right)\left(\lambda_1+\lambda_2\right)}\,,
\end{aligned}
\end{equation}
where we have introduced the thermal sources for the chemical potentials as in \eqref{thermal sources}, i.e.
\begin{empheq}{alignat=5}
	\gamma_0&=\left(\frac{\bar{\tau}-\tau}{2}\right)\nu_0\,,
	&\qquad
	\gamma_1&=\left(\frac{\bar{\tau}-\tau}{2}\right)\tilde{\nu}_{-1}\,,
	&\qquad
	\gamma_2&=\left(\frac{\bar{\tau}-\tau}{2}\right)\nu_{-2}\,.
\end{empheq}

\noindent Naturally, we can also use \eqref{potentials 2} in order to write down the solution to the smoothness conditions for the dual CFT chemical potentials. Notice that $\gamma_1$ and $\gamma_2$ vanish for $\lambda_1=\lambda_2=\lambda_3$ while the modular parameter becomes $\tau=i/(2\sqrt{\mathcal{L}})$, reproducing \eqref{BTZ smoothness} for the BTZ solution.

All of the above considerations take place in Euclidean time. When switching back to Lorentzian signature, we should demand that the chemical potentials are real so that the connection lies in $su(2,1|1,1)$. This forces the eigenvalues to satisfy
\be\label{eq:realeig}
	\lambda_1=\lambda_2^*\,\,,
	\qquad
	\lambda_3=\lambda_3^*\,\,,
\ee
implying that $\Delta_3\geq0$ and $\Delta_2\geq0$, as previously advertised. We also see that the diagonal form \eqref{eq: aphi diagonal} takes values in $su(2,1|1,1)$, a fact which was not obvious a priori. And last, but by no means least, the entropy of an $sl(3|2)$ black hole is 
\begin{empheq}{alignat=5}\label{eq:entropyw32}
	S&=2\pi k_{cs}\left(\lambda_1+\lambda_2-\frac{\lambda_3}{2}\right)+\textrm{other sector}\,,
\end{empheq}
which follows from applying \eqref{dBJ entropy formula 1} to \eqref{eq: sl(3|2) Black Hole connection aphi 2}. It is important to highlight that due to \eqref{eq:realeig} the entropy in each sector is real, as it should be. Also, the $U(1)$ charge does not make an explicit appearance, it only enters indirectly via \eqref{eq:mapcharges}. The reason is rather simple: $Q_1$ is ambigous since its value can be changed by a smooth $U(1)$ gauge transformation, whereas $({\cal L},Q_2,Q_3)$ are invariant under any such symmetry. As we will see in section \ref{sec:BPS}, in the dual CFT this nicely ties to the fact that $({\cal L},Q_2,Q_3)$ are spectral flow invariants (a property that we expect the entropy to respect). 

We take the reality conditions \eqref{eq:realeig} as part of the definition of the Lorentzian black hole solution. Notice that these imply that the four Jordan classes with $\lambda_2=2\lambda_1\neq0$ and $\lambda_1=2\lambda_2\neq0$ are actually empty. This is a consequence of our ordering of the eigenvalues and the choice of smoothness condition in \eqref{eq:w32smooth}; see comment 3 at the end of section \ref{sec:examples}. There are then only six possible Jordan classes for black hole solutions, five of which are extremal:
\begin{empheq}{alignat=5}\nonumber
\begin{array}{|c|c|c|c|c|}
	\hline
	\textrm{\textbf{Class}} & \multicolumn{2}{|c|}{\textrm{Eigenvalues}} & \left(\Delta_3,\Delta_2\right) & \textrm{Extremal?}  \\
	\hline
	\textrm{\textbf{I}} & \lambda_1=\lambda_2=0 & \lambda_3=0 & (=0,=0) & \textrm{Yes}  \\ 
	\hline
	\textrm{\textbf{II}} & \lambda_1=-\lambda_2\neq0 & \lambda_3=0 & (=0,=0) & \textrm{Yes} \\
	\hline
	\textrm{\textbf{III}} & \lambda_1\neq-\lambda_2 & \lambda_3=0 & (>0,=0) & \textrm{Yes} \\
	\hline
	\textrm{\textbf{IV}} & \lambda_1=\lambda_2=0 & \lambda_3\neq0 & (=0,>0) & \textrm{Yes}  \\
	\hline
	\textrm{\textbf{V}} & \lambda_1=-\lambda_2\neq0 & \lambda_3\neq0 & (=0,>0) & \textrm{Yes}  \\
	\hline
	\textrm{\textbf{VI}} & \lambda_1\neq-\lambda_2 & \lambda_3\neq0 & (>0,>0) & \textrm{No}  \\
	\hline
\end{array}\,.
\end{empheq}

The $sl(2|1)$ theory lives in the subsector $\lambda_1=\lambda_2=\lambda_3$ of class \textbf{VI} and in class \textbf{I}. In particular, class {\bf I} describes the extremal charged BTZ solution. For class {\bf VI} we could further restrict $2(\lambda_1+\lambda_2)>\lambda_3>0$. This ensures that the black hole solutions within this class are smoothly connected to BTZ and that the contribution from this sector to the entropy \eqref{eq:entropyw32} is positive. However, there is nothing in principle pathological about solutions with $\lambda_3<0$ given the criteria used here; actually it seems like a more natural choice for class  {\bf IV} and {\bf V} to take $\lambda_3$ negative rather than positive.

We emphasize that the table displayed above only implements our definition extremality, and other smoothness and/or stability conditions might further restrict the $\lambda_i$'s or even eliminate a whole class. For instance we could in addition demand that the entropy \eqref{eq:entropyw32} is positive; note that this includes both sectors so it is one inequality for $\lambda_i$ and $\bar\lambda_i$. There are additional criteria one could use, e.g., thermodynamical stability of the solutions. One could as well study Lorentzian properties which could probe if the solutions has closed timelike curves. These additional properties will not be discussed here and will not impact further conclusions we draw from these solutions; we leave these issues as future directions. 

\subsubsection{Smooth conical defects}\label{sec:smoothcon}
Smooth conical defects are solutions of the Lorentzian theory that have trivial holonomy along the cycle $\phi\sim \phi +2\pi$ \cite{Castro:2011iw}. Triviality of the holonomy takes the same meaning as it did for black holes along the thermal direction, that is,
\begin{empheq}{alignat=5}\label{eq: holonomy phi}
	H_{\phi}&\equiv\mathcal{P}e^{\oint a_{\phi}d\phi}&\,=\Gamma^{\pm}\,.
\end{empheq}
Since the topology of these backgrounds is assumed to be the same as for AdS$_{3}\,$, where the $\phi$ cycle is contractible, smoothness of fermionic fields at the origin would require $H_{\phi}=\Gamma^-\,$. Notwithstanding this consideration, and for the purpose of comparing with the dual CFT in section \ref{sec:BPS}, we will allow for the possibility of periodic fermions, i.e. $H_{\phi}=\Gamma^+$. Of course, this introduces a sigularity at the origin, interpreted as a delta function source, the presence of which we can not justify from the bulk perspective without coupling the theory to matter in a UV complete fashion.

The above holonomy conditions immediately imply that the eigenvalues of $a_{\phi}$ are purely imaginary, putting these solutions in the sector $\Delta_3<0$ and $\Delta_2<0\,$. The most general set of eigenvalues that satisfy \eqref{eq: holonomy phi} reads
\begin{empheq}{alignat=5}\label{eq:lcd}
	\lambda_1=i\left(n_1+\frac{1}{3}n\right)\,,
	\quad
	\lambda_2=i\left(n_2-\frac{1}{3}n\right)\,,
	\quad
	\lambda_3=i\left(2n_3+n\right)\,,
	\quad
	Q_1=\frac{1}{6}n\,,
\end{empheq}
where $n_1$, $n_2$ and $n$ are integers. The parameter $n_3$  labels whether the solution supports periodic or anti-periodic fermions. If $n_3\in \mathds{Z}$ we have that $H_{\phi}=\Gamma^+$, whereas for $n_3\in\mathds{Z}+\frac{1}{2}$ the holonomy becomes $H_{\phi}=\Gamma^-$. At this stage, the bulk Chern-Simons theory gives us no  obvious further restrictions on $(n_i,n)$ other than requiring that the eigenvalues be non-degenerate. However, comparison with the dual ${\cal W}_{(3|2)}$ CFT$_2$ will impose additional constrains on these parameters.
 

\subsection{Supersymmetry}\label{sec:susybulk}
In a supersymmetric Chern-Simons theory, a solution is said to be BPS if there exists a gauge transformation supported by odd elements of the gauge group that leaves the connection invariant. Having surveyed $sl(3|2)$ solutions in detail, our task in this section will be to explore under what conditions these backgrounds are BPS, and identify the precise fermionic symmetries they preserve. We will then compare the supersymmetry constraints with our definition of extremality. Only one sector of the $sl(3|2)\oplus sl(3|2)$ Chern-Simons theory will be considered here. Analogous results follow for the other sector.

Prior work on supersymmetric black holes in higher spin gravity include \cite{Tan:2012xi,Datta:2012km,Datta:2013qja,Chen:2013oxa}. The logic we follow here is most closely related to\cite{Datta:2013qja}, where the role of the odd roots of the superalgebra and the eigenvalues of the connection was emphasized. The main novelty is that we purposefully allow for the possibility that the connection be non-diagonalizable, in accordance with our definition of extremality. Moreover, we use Hamiltonian boundary conditions while most of the literature focuses on a holomorphic description along the lines of \eqref{action deformations}. The advantage of this setup is that comparison with BPS states in the dual $\cW_{(3|2)}$ theory becomes clear and unambiguous. The recent work \cite{Henneaux:2015ywa} does implement boundary conditions analogous to ours, with one important distinction being that the superalgebra considered in  \cite{Henneaux:2015ywa} is $osp(1|4)$ instead of $sl(3|2)$, and consequently the theory displays so-called \textit{hypersymmetry} as opposed to supersymmetry. Further comparisons with \cite{Henneaux:2015ywa} will be discussed in section \ref{sec:Disc}.

Let us now proceed to analyse the conditions for supersymmetry in Chern-Simons theory. Having eliminated the radial dependence, the residual gauge transformations acting on the boundary connection take the familiar form
\be
a \to a' = e^{-\eps}\, a\, e^{\eps} + d\eps\,, 
\ee
where $\epsilon$ is an arbitrary element of the gauge superalgebra, in this case, $su(2,1|1,1)$. As mentioned above, a given background is deemed supersymmetric if $a'=a$ for an odd transformation parameter. Thus, in infinitesimal form, the BPS conditions read
\begin{empheq}{alignat=5}\label{eq:susycs}
	\partial_t\eps+[a_t,\eps]=0\,,
	\quad
	\partial_{\phi}\eps+[a_{\phi},\eps]=0\,.
\end{empheq}
The number of independent solutions to these equations determines the amount of supersymmetry preserved by the background. We will refer to these equations as ``Killing equations''  and label its solutions as ``Killing spinors'', in analogy with standard supergravity nomenclature.

Locally, Killing spinors exist for arbitrary connections and fermionic generators. In fact, since we are only focusing on backgrounds with constant $a_t$ and $a_{\phi}$, the integrability condition $[a_t,a_{\phi}]=0$ allows us to write the general solution to \eqref{eq:susycs} as
\begin{empheq}{alignat=5}\label{eq:killingds}
	\eps(t,\phi)&=e^{-a_tt-a_{\phi}\phi}\,\eps_0\,e^{a_tt+a_{\phi}\phi}\,.
\end{empheq}
Here $\epsilon_0$ is a constant odd element of $su(2,1|1,1)$. The admissible, {\it globally defined} solutions, however, are only those that possess the correct periodicity in $\phi$. Namely, the spinors can be anti-periodic in the Neveu-Schwarz sector or periodic in the Ramond sector.\footnote{ Recall that the $\phi$-cycle is non-contractible for a black hole, hence we can allow for both NS and R boundary conditions.} This imposes constraints on both $a_{\phi}$ and $\eps_0\,$. Most of the discussion in this section focuses on the $\phi$-dependence of $\eps(\phi)\equiv\eps(0,\phi)$. In the last portion we will comment on the time dependence. 

For the purpose of explicitly displaying the global solutions to the Killing equations, let us bring $a_{\phi}$ to its Jordan normal form as in \eqref{eq: Jordan form}. The important properties to remember are that $a_{\phi}^N$ is nilpotent and commutes with $a_{\phi}^D\,$, obviously vanishing for diagonalizable connections. In this decomposition we have\footnote{ Notice  the use of $\eps$ versus $\epsJ$. Apologies to the reader for the inconvenience.}
\begin{empheq}{alignat=5}\label{eq: Killing spinor general}
	\epsJ(\phi)&= e^{-a_{\phi}^D\phi}e^{-a_{\phi}^N\phi}\,\epsJ_0 \, e^{a_{\phi}^D\phi}e^{a_{\phi}^N\phi}\, \,,
\end{empheq}
where
\be
\epsJ(\phi)\equiv V^{-1}\,\eps(\phi)\, V~,\quad \epsJ_0\equiv V^{-1}\,\eps_0\, V\,,
\ee
and $V$ is the constant ($\phi$-independent) matrix defined in \eqref{eq: Jordan form}. We stress again that $a_{\phi}^D$ and  $a_{\phi}^N$, as well as the spinors $\epsJ$ and $\epsJ_0\,$, need not belong to $su(2,1|1,1)\,$. This requisite must be imposed only upon reverting the transformation that takes the connection to its Jordan form. It will also prove convenient to work in the $E_{IJ}$ basis of $sl(3|2)$ generators introduced in appendix \ref{app: sl(3|2)}, where $a_{\phi}^D$ takes the form
\begin{empheq}{align}
	a_{\phi}^D&=\lambda_1\left(E_{11}-E_{22}\right)+\lambda_2\left(E_{22}-E_{33}\right)+\frac{1}{2}\lambda_3\left(E_{44}-E_{55}\right)+iQ_1\left(E_{11}+E_{22}+E_{33}\right)\,,
\end{empheq}
and $a_{\phi}^N$ can be written as a linear combination of $E_{12}$, $E_{13}$, $E_{23}$ and $E_{45}\,$, depending on the Jordan class under consideration. This basis has the advantage that it diagonalizes the adjoint action of the Cartan elements. In particular,
\begin{empheq}{alignat=5}\label{definition odd roots 1}
	[a_{\phi}^D,E_{IJ}]=\omega_{IJ}E_{IJ}\,,
\end{empheq}
where
\begin{empheq}{alignat=5}\label{definition odd roots 2}
	\omega_{IJ}&=\left(a_{\phi}^D\right)_{II}-\left(a_{\phi}^D\right)_{JJ}\,.
\end{empheq}

More generally, as pointed out in \cite{Datta:2013qja} the frequencies $\omega_{IJ}=-\omega_{JI}$ are determined by the roots of the superalgebra and the holonomy of the connection. In order to exhibit this relation in a general and representation-independent way, let $\alpha_{j}$ denote a root of the bulk gauge superalgebra, and let $a_{\phi}^{D}$ denote the diagonal piece of the Jordan normal form of the Drinfeld-Sokolov connection $a_{\phi}$ appropriate to the boundary symmetries being described. Since $a_{\phi}^{D}$ belongs to the Cartan subalgebra $\mathcal{C}$, we can associate an element $\vec{\Lambda}_{\phi} \in \mathcal{C}^{*}$ of the root space with it. Using the isomorphism between $\mathcal{C}$ and the root space $\mathcal{C}^{*}$, we may also associate a Cartan element $H_{j}$ with the root  $\alpha_{j}\,$. Then, using the bilinear form $\langle \cdot\,,\cdot\rangle$ on $\mathcal{C}^{*}$ induced in the usual way by the Killing form, i.e. $\bigl\langle \alpha , \beta\bigr\rangle = \text{Tr}\bigl[\text{adj}\,H_{\alpha}\,\text{adj}\,H_{\beta}\bigr]$, we can write the frequencies \eqref{definition odd roots 1}-\eqref{definition odd roots 2} in a representation-independent way as
\begin{equation}\label{general frequencies}
\omega_{j} = \bigl\langle \vec{\Lambda}_{\phi}\,,\alpha_{j}\bigr\rangle\,.
\end{equation}

\noindent  The precise form of the frequencies $\omega_{j}$ will of course depend on the concrete algebra under consideration and encodes the semiclassical symmetries of the boundary CFT (via the Drinfeld-Sokolov boundary conditions). In the case at hand, since we are interested in fermionic symmetries only, the odd frequencies are given explicitly by
\begin{empheq}{alignat=5}\label{eq: Sec.3 frequencies}
	\omega_{14}&=\lambda_1-\frac{\lambda_3}{2}+iQ_1\,,
	&\qquad
	\omega_{15}&=\lambda_1+\frac{\lambda_3}{2}+iQ_1\,,
	\cr
	\omega_{24}&=-\lambda_1+\lambda_2-\frac{\lambda_3}{2}+iQ_1\,,
	&\qquad
	\omega_{25}&=-\lambda_1+\lambda_2+\frac{\lambda_3}{2}+iQ_1\,,
	\\
	\omega_{34}&=-\lambda_2-\frac{\lambda_3}{2}+iQ_1\,,
	&\qquad
	\omega_{35}&=-\lambda_2+\frac{\lambda_3}{2}+iQ_1\,.
	\nonumber
\end{empheq}

Finally, the constant element $\epsJ_0$ can be expanded into $U(1)$ eigenstates as
\begin{empheq}{alignat=5}\label{eq: epsilon0}
	\epsJ_0&=\epsJ_0^-+\epsJ_0^+\,,
\end{empheq}
where
\be\label{eq: epsilon0+}
	\epsJ_0^-=\sum_{i,\bar{j}}\epsJ_{i\bar{j}}E_{i\bar{j}}\,,\quad
	\epsJ_0^+=\sum_{\bar{i},j}\epsJ_{\bar{i}j}E_{\bar{i}j}\,,
\ee
with $i=\left(1,2,3\right)$ and $\bar{i}=\left(4,5\right)$. There are in total 12 complex parameters $\epsJ_{i\bar{j}}$ and $\epsJ_{\bar{i}j}\,$. However, only half of them are independent because of the reality constraint satisfied by elements of $su(2,1|1,1)$, which ties the two $U(1)$ sectors by complex conjugation. Since the number of real independent coefficients allowed by a background quantifies the amount of preserved supersymmetries, a fully supersymmetric solution will permit a total of 12 real parameters, a $1/2$-BPS one will preserve 6 of them, a $1/3$-BPS background will have 4 free real coefficients, etc. Of course, this counting neglects the other sector $\bar{a}\,$.

We are now in a position to study the conditions under which any given $sl(3|2)$ solution will be invariant under a supersymmetric transformation. First, notice that because $a_{\phi}^N$ is nilpotent, the series expansion of $e^{a_{\phi}^N\phi}$ in \eqref{eq: Killing spinor general} will be truncated at some finite order. To avoid a polynomial $\phi$-dependence in the Killing spinor, which is neither periodic nor anti-periodic, we must require that
\begin{empheq}{alignat=5}\label{eq:ancomm}
	[a_{\phi}^N,\epsJ_0]&=0\,.
\end{empheq}
This condition restricts the number of independent coefficients $\epsJ_{i\bar{j}}$ and $\epsJ_{\bar{i}j}$ appearing in \eqref{eq: epsilon0+}. The remaining $\phi$-dependence of $\epsilon(\phi)$ is controlled by $[a_{\phi}^D,\epsilon_0]\,$.  By means of \eqref{definition odd roots 1} and the Baker-Campbell-Hausdorff formula, we find that \eqref{eq: Killing spinor general} becomes
\begin{empheq}{alignat=5}\label{eq:fullep}
	\epsJ(\phi)&=\epsJ^-(\phi)+\epsJ^+(\phi)\,,
\end{empheq}
with
\be\label{eq:quantawij}
	\epsJ^-(\phi)=\sum_{i,\bar{j}}\epsJ_{i\bar{j}}E_{i\bar{j}}e^{-\omega_{i\bar{j}}\phi}\,,\quad
	\epsJ^+(\phi)=\sum_{\bar{i},j}\epsJ_{\bar{i}j}E_{\bar{i}j}e^{-\omega_{\bar{i}j}\phi}\,.
\ee
These expressions imply that the frequencies $\omega_{i\bar{j}}=-\omega_{\bar{j}i}$ must be quantized into integer or half-integer imaginary values in order for the solution to comply with the required periodicity: 
\begin{empheq}{alignat=5}
\omega_{i\bar{j}}\in\left\{
\begin{array}{ll}
	i\mathds{Z} & \textrm{R sector}
	\\
	i\mathds{Z}+\frac{i}{2} & \textrm{NS sector}
\end{array}
\right.\,.
\end{empheq}
This requirement is in general not fulfilled automatically. The quantization conditions translate into constraints over the charges carried by the background, which may or may not be possible to satisfy, further restricting the number of preserved supersymmetries.

Once one finds the Killing spinor $\epsJ(\phi)$ explicitly, it is a simple matter to undo the similarity transformation and express the solution $\eps(\phi)$ in the form  \eqref{eq:killingds}. We will display our results by writing the supercharges in the language of asymptotic symmetries. As shown in \cite{BetoJuan}, the general fermionic gauge parameter that preserves the Drinfeld-Sokolov form of the connection is
\begin{empheq}{align}\label{eq:asympe}
	\nonumber
	\eps(t,\phi)&=\lH(t,\phi)H_{\frac{1}{2}}+\lG(t,\phi)G_{\frac{1}{2}}+\lT(t,\phi)T_{\frac{3}{2}}+\lS(t,\phi)S_{\frac{3}{2}}
	\\
	&+h_{-\frac{1}{2}}(t,\phi)H_{-\frac{1}{2}}+g_{-\frac{1}{2}}(t,\phi)G_{-\frac{1}{2}}+t_{-\frac{3}{2}}(t,\phi)T_{-\frac{3}{2}}+s_{-\frac{3}{2}}(t,\phi)S_{-\frac{3}{2}}
	\\
	&+t_{-\frac{1}{2}}(t,\phi)T_{-\frac{1}{2}}+s_{-\frac{1}{2}}(t,\phi)S_{-\frac{1}{2}}+t_{\frac{1}{2}}(t,\phi)T_{\frac{1}{2}}+s_{\frac{1}{2}}(t,\phi)S_{\frac{1}{2}}\,,
	\nonumber
\end{empheq}
where the higher weight terms are fixed algebraically in terms of the lowest weight components:
\bea\label{eq:explicitK}
	h_{-\frac{1}{2}}&=&-\partial_{\phi}\lH-iQ_1\lH+2Q_2\lT\,,
	\cr
	g_{-\frac{1}{2}}&=&-\partial_{\phi}\lG+iQ_1\lG-2Q_2\lS\,,
	\cr
	t_{\frac{1}{2}}&=&-\partial_{\phi}\lT-iQ_1\lT\,,
	\cr
	s_{\frac{1}{2}}&=&-\partial_{\phi}\lS+iQ_1\lS\,,
	\cr
	t_{-\frac{1}{2}}&=&\frac{1}{2}\partial^2_{\phi}\lT+iQ_1\partial_{\phi}\lT-\frac{1}{2}\left(3\mathcal{L}+Q_2+Q_1^2\right)\lT\,,
	\\
	s_{-\frac{1}{2}}&=&\frac{1}{2}\partial^2_{\phi}\lS-iQ_1\partial_{\phi}\lS-\frac{1}{2}\left(3\mathcal{L}+Q_2+Q_1^2\right)\lS\,,
	\cr
	t_{-\frac{3}{2}}&=&-\frac{1}{6}\partial_{\phi}^3\lT-\frac{1}{2}iQ_1\partial_{\phi}^2\lT+\frac{1}{3}\partial_{\phi}^2\lH+\frac{1}{2}\left(\frac{7}{3}\mathcal{L}-Q_2+Q_1^2\right)\partial_{\phi}\lT+\frac{2}{3}iQ_1\partial_{\phi}\lH
	\cr
	&&+\frac{1}{2}iQ_1\left(\frac{7}{3}\mathcal{L}-Q_2+\frac{1}{3}Q_1^2\right)\lT-\left(\frac{1}{3}\mathcal{L}+Q_2+\frac{1}{3}Q_1^2\right)\lH\,,
	\cr
	s_{-\frac{3}{2}}&=&-\frac{1}{6}\partial_{\phi}^3\lS+\frac{1}{2}iQ_1\partial_{\phi}^2\lS-\frac{1}{3}\partial_{\phi}^2\lG+\frac{1}{2}\left(\frac{7}{3}\mathcal{L}-Q_2+Q_1^2\right)\partial_{\phi}\lS+\frac{2}{3}iQ_1\partial_{\phi}\lG
	\cr
	&&-\frac{1}{2}iQ_1\left(\frac{7}{3}\mathcal{L}-Q_2+\frac{1}{3}Q_1^2\right)\lS+\left(\frac{1}{3}\mathcal{L}+Q_2+\frac{1}{3}Q_1^2\right)\lG\,.
	\nonumber
\eea
Notice that $\eps\in su(2,1|1,1)$ implies $\lG=i\blH$ and $g_{r}=i\overline{h}_{r}$, as well as $\lS=-i\blT$ and $s_{s}=-i\overline{t}_{s}\,$. Since a Killing spinor corresponds to a particular class of gauge transformations where the bosonic parameters vanish, we will express our findings by specifying $\lH$ and $\lT\,$. One reason why it is worth writing the remaining components explicitly is to illustrate how the fermionic generators are concatenated. For instance, if a global Killing spinor has $\lT=0\,$, it does not necessarily imply that the corresponding background preserves a supercharge lying only within the graviton multiplet, i.e. the $sl(2|1)$ truncation. Indeed, from \eqref{eq:explicitK} it is clear that $\lH$ by itself can induce components in $\eps(\phi)$ that are supported by generators belonging to the higher spin multiplet.

Lastly, we point out that one could, in principle, find the Killing spinors by directly solving the resulting (sixth-order) differential equations for $(\lT,\lH)$ without ever having to resort to the Jordan form of $a_{\phi}\,$. It would still be necessary, however, to distinguish between the different classes of extremal and non-extremal solutions, a task that is far from trivial in the Drinfeld-Sokolov form of the connection, especially when written in terms of the charges instead of eigenvalues. The Jordan form method outlined above is equivalent and it simply presents the $\phi$ dependence in a different manner. 

\subsubsection{Black holes}\label{sec:susyw32bhs}
We will now go through an exhaustive analysis of the above supersymmetry conditions for each Jordan class in the black hole sector; these classes are listed at the end of section \ref{sec:sl32bhs}. Recall that for black hole solutions (and not for smooth conical defects) the diagonal part of the connection \eqref{eq: aphi diagonal} automatically lies in $su(2,1|1,1)$ due to the reality properties of the eigenvalues. In what follows, we will make sure that the nilpotent piece in \eqref{eq: Jordan form} also takes values in this superalgebra and that the similarity transformation that puts $a_{\phi}$ in its Jordan form belongs to the corresponding supergroup. While this is not strictly necessary, it will ensure that the Killing spinors we find always live in the correct real form, regardless of the basis we use to describe them. Since the parameters in \eqref{eq: epsilon0} are then tied by ${\epsJ_0^{\pm}}^{\dagger}=-K\epsJ_0^{\mp}K$ so that $\epsJ_0=\epsJ_0^-+\epsJ_0^+\in su(2,1|1,1)$ (see appendix \ref{app: sl(3|2)}), it suffices to perform the analysis for $\epsJ^-_0$ only.


\subsubsection*{Class I: $\lambda_1=\lambda_2=0$, $\lambda_3=0$}
As our first example, we will show in detail the construction of Killing spinors for solutions in class {\bf I}, which captures the supersymmetric sector of the $sl(2|1)$ truncation that maps to ${\cal N}=2$ supergravity (this truncation is further discussed in section \ref{sec:n2bhs}). We denote these backgrounds as ``BPS charged BTZ black holes''. The other classes follow in an analogous manner.

In this class the charges are given by
\be
\mathcal{L}=0\,,\quad	Q_2=0\,,\quad 	Q_3=0\,,
\ee
leading to a Jordan decomposition \eqref{eq: Jordan form} where
\begin{empheq}{alignat=5}
	a_{\phi}^D&=iQ_1\left(E_{11}+E_{22}+E_{33}\right)\,,
	\quad
	a_{\phi}^N&=-\left(E_{12}+E_{23}+E_{45}\right)\,.
\end{empheq}
The choice for $a_{\phi}^N$ is not unique; any other matrix related to this one by a similarity transformation will yield equivalent results. Using the expansion \eqref{eq: epsilon0+}, we find that
\begin{empheq}{alignat=5}
	[a_{\phi}^N,\epsJ^-_0]&=-\epsJ_{24}E_{14}+\left(\epsJ_{14}-\epsJ_{25}\right)E_{15}-\epsJ_{34}E_{24}+\left(\epsJ_{24}-\epsJ_{35}\right)E_{25}+\epsJ_{34}E_{35}\,.
\end{empheq}
As argued above, this commutator must vanish, which sets
\be
\epsJ_{14}=\epsJ_{25}\,,\quad	\epsJ_{24}=\epsJ_{34}=\epsJ_{35}=0\,.
\ee
This leaves two independent complex coefficients, $\epsJ_{25}$ and $\epsJ_{15}$. Hence, within class {\bf I}, solutions can preserve at most 4 real supercharges. Additionally, we need to ensure that the Killing spinors have the correct periodicity. From \eqref{eq: Sec.3 frequencies} and \eqref{eq:quantawij} we have
\begin{empheq}{alignat=5}
	\epsJ^-(\phi)&=\left(\epsJ_{25}\left(E_{14}+E_{25}\right)+\epsJ_{15}E_{15}\right)e^{-iQ_1\phi}\,.
\end{empheq}
Therefore,
\begin{empheq}{align}
-Q_1=\eta+\frac{1}{2}\in\left\{
\begin{array}{ll}
	\mathds{Z} & \textrm{R sector}
	\\
	\mathds{Z}+\frac{1}{2} & \textrm{NS sector}
\end{array}
\right.\,,
\end{empheq}
i.e. the $U(1)$ charge must be quantized appropriately.  

After undoing the similarity transformation that puts $a_{\phi}$ in its Jordan form and casting the resulting generators as in \eqref{eq:asympe}, we find that the transformation parameters read
\begin{empheq}{alignat=5}
	\lH(\phi)&=\lH(0)e^{i\left(\eta+\frac{1}{2}\right)\phi}\,,
	\quad
	\lT(\phi)&=\lT(0)e^{i\left(\eta+\frac{1}{2}\right)\phi}\,,
\end{empheq}
where we have exchanged the coefficients $\epsJ_{15}$ and $\epsJ_{25}$ for $\lH(0)$ and $\lT(0)$. Since all other components in \eqref{eq:explicitK} vanish, the Killing spinor is simply
\be\label{eq:kclassi}
\eps(\phi)= \lH(\phi)H_{\frac{1}{2}}+i\blH(\phi)G_{\frac{1}{2}}+\lT(\phi)T_{\frac{3}{2}}-i\blT(\phi)S_{\frac{3}{2}}\,.
\ee 


\subsubsection*{Class II: $\lambda_1=-\lambda_2\neq0$, $\lambda_3=0$}
The charges carried by this class are
\be	
\mathcal{L}=\frac{3}{8}\lambda_1^2\,,\quad Q_2=\frac{3}{8}\lambda_1^2\,,\quad Q_3=-\frac{i}{4}\lambda_1^3\,,
\ee
with $\lambda_1$ being purely imaginary. The Jordan form of the connection reads
\begin{empheq}{alignat=5}
	a_{\phi}^D&=\lambda_1\left(E_{11}-2E_{22}+E_{33}\right)+iQ_1\left(E_{11}+E_{22}+E_{33}\right)\,,
	\quad
	a_{\phi}^N&=-\left(\pm iE_{13}+E_{45}\right)\,,
\end{empheq}
where the choice of sign depends on the phase $\lambda_1=\pm i |\lambda_1|$. 
The condition $[a_{\phi}^N,\epsJ^-_0]=0$ then sets
\be	
\epsJ_{14}=\pm i\epsJ_{35}\,,\quad \epsJ_{24}=\epsJ_{34}=0\,.
\ee
The free parameters are $ \epsJ_{i5}$, and from \eqref{eq: Sec.3 frequencies} combined with \eqref{eq:quantawij} we see that the Killing spinors are
\begin{empheq}{alignat=5}
\epsJ^-(\phi)&=\left(\epsJ_{15}E_{15}+\epsJ_{35}\left(E_{35}\pm iE_{14}\right)\right)e^{-i\left(-i\lambda_1+Q_1\right)\phi} +\epsJ_{25}E_{25}e^{-i\left(2i\lambda_1+Q_1\right)\phi}\,.
\end{empheq}

An interesting feature here is that there are two different exponentials for which we need to demand periodicity. Depending on the quantization conditions imposed on the charges, different supersymmetries are preserved. By requiring
\be\label{eq:qc1II}
i\lambda_1-Q_1=\eta+\frac{1}{2}\in\left\{
\begin{array}{ll}
	\mathds{Z} & \textrm{R sector}
	\\
	\mathds{Z}+\frac{1}{2} & \textrm{NS sector}
\end{array}\right.\quad,
\ee
we will preserve 4 supercharges, which in the notation of \eqref{eq:asympe} read
\begin{empheq}{alignat=5}
	\lH(\phi)&=\lH(0)e^{i\left(\eta+\frac{1}{2}\right)\phi}\,,
	&\quad
	\lT(\phi)&=\lT(0)e^{i\left(\eta+\frac{1}{2}\right)\phi}\,.
\end{empheq}
Instead, demanding that
\begin{empheq}{alignat=5}\label{eq:q2II}
-2i\lambda_1-Q_1=\eta+\frac{1}{2}\in\left\{
\begin{array}{ll}
	\mathds{Z} & \textrm{R sector}
	\\
	\mathds{Z}+\frac{1}{2} & \textrm{NS sector}
\end{array}
\right.
\end{empheq}
leads to the two supersymmetries
\begin{empheq}{alignat=5}\label{eq:susyII}
	\lH(\phi)&=\lambda_1\lT(\phi)\,,
	&\quad
	\lT(\phi)&=\lT(0)e^{i\left(\eta+\frac{1}{2}\right)\phi}\,.
\end{empheq}
While for \eqref{eq:qc1II} we can smoothly recover the results for class {\bf I} by taking $\lambda_1=0$, the quantization condition \eqref{eq:q2II} is disconnected from the previous case due to the relations in \eqref{eq:susyII}.  

It is also possible that \eqref{eq:qc1II} and \eqref{eq:q2II} are satisfied simultaneously. This occurs when
\begin{empheq}{alignat=5}\label{eq:mostsusy}
	\lambda_1&=\frac{i}{3}\left(\eta_2-\eta_1\right)
	&\qquad\textrm{and}\qquad
	Q_1&=-\frac{1}{3}\left(2\eta_1+\eta_2\right)-\frac{1}{2}\,,
\end{empheq}

\noindent a scenario which preserves six supersymmetries. This demonstrates explicitly that charged BTZ black holes are {\it not} the most supersymmetric black hole configurations in the higher spin theory, as one might have naively expected. At the level of the entropy, all solutions within classes {\bf I} and {\bf II} have $S=0+\textrm{other sector}$, according to \eqref{eq:entropyw32}. One would generically expect that the most symmetric configuration minimizes the entropy, but this argument would not distinguish between the two classes. 


\subsubsection*{Class III: $\lambda_1\neq-\lambda_2$, $\lambda_3=0$}
The charges carried by this class are
\begin{empheq}{alignat=5}
	\mathcal{L}&=\frac{1}{8}\left(\lambda_1^2+\lambda_2^2-\lambda_1\lambda_2\right)\,,
	\cr
	Q_2&=\frac{1}{8}\left(\lambda_1^2+\lambda_2^2-\lambda_1\lambda_2\right)\,,
	\\
	Q_3&=-\frac{i}{4}\left(-\lambda_1+\lambda_2\right)\lambda_1\lambda_2\,,
	\nonumber
\end{empheq}
with the connection being
\begin{empheq}{alignat=5}
	a_{\phi}^D&=\lambda_1\left(E_{11}-E_{22}\right)+\lambda_2\left(E_{22}-E_{33}\right)+iQ_1\left(E_{11}+E_{22}+E_{33}\right)\,,
	&\quad
	\quad
	a^N_{\phi}&=-E_{45}\,.
\end{empheq}
The condition $[a_{\phi}^N,\epsJ^-_0]=0$ then sets
\be
	\epsJ_{14}=\epsJ_{24}=\epsJ_{34}=0\,,
\ee
leaving $\epsJ_{i5}$ as free parameters. The putative supercharges are then
\begin{empheq}{alignat=5}
	\epsJ^-(\phi)&=\epsJ_{15}E_{15}e^{-\left(\lambda_1+iQ_1\right)\phi}+\epsJ_{25}E_{25}e^{-\left(-\lambda_1+\lambda_2+iQ_1\right)\phi}+\epsJ_{35}E_{35}e^{-\left(-\lambda_2+iQ_1\right)\phi}\,.
\end{empheq}

It is easy to see that $\epsJ_{15}$ and $\epsJ_{35}$ cannot be preserved: by definition $\lambda_1\neq-\lambda_2$ within this class, a fact which coupled to the reality condition $\lambda_1=\lambda_2^*$ does not allow for $\lambda_1$ or $\lambda_2$ to be purely imaginary. Therefore, we have $\epsJ_{15}=\epsJ_{35}=0$ and the quantization condition
\begin{empheq}{alignat=5}
	-i(\lambda_1-\lambda_2)-Q_1&=\eta+\frac{1}{2}\in \left\{
\begin{array}{ll}
	\mathds{Z} & \textrm{R sector}
	\\
	\mathds{Z}+\frac{1}{2} & \textrm{NS sector}
\end{array}
\right. 
\end{empheq}
as requisites for the existence of Killing spinors. All in all this class can only preserve two real supercharges, which are
\begin{empheq}{alignat=5}
	\lH(\phi)&=\frac{1}{2}\left(\lambda_1-\lambda_2\right)\lT(\phi)\,,
	&\quad
	\lT(\phi)&=\lT(0)e^{i\left(\eta+\frac{1}{2}\right)\phi}\,.
\end{empheq}
Notice that the entropy \eqref{eq:entropyw32} is always non-vanishing. Also, setting $\lambda_2=-\lambda_1$ we smoothly recover one of the quantization conditions in class \textbf{II}.

\subsubsection*{Class IV: $\lambda_1=\lambda_2=0$, $\lambda_3\neq0$}
The Jordan form of the connection is
\begin{empheq}{alignat=5}
	a_{\phi}^D&=\frac{1}{2}\lambda_3\left(E_{44}-E_{55}\right)+iQ_1\left(E_{11}+E_{22}+E_{33}\right)\,,
	\quad
	a_{\phi}^N&=-\left(E_{12}+E_{23}\right)\,.
\end{empheq}
This class does not contain any supersymmetric solutions since the exponential dependence of the Killing spinor always has a non-zero real part.
\subsubsection*{Class V: $\lambda_1=-\lambda_2\neq0$, $\lambda_3\neq0$}
In this case we have
\begin{empheq}{alignat=5}
	a_{\phi}^D&=\lambda_1\left(E_{11}-2E_{22}+E_{33}\right)+\frac{1}{2}\lambda_3\left(E_{44}-E_{55}\right)+iQ_1\left(E_{11}+E_{22}+E_{33}\right)\,,
	\quad
	a_{\phi}^N&=\mp iE_{13}\,.
\end{empheq}
Again, this class does not contain any supersymmetric solutions since the exponential dependence of the Killing spinor always has a non-zero real part.
\subsubsection*{Class VI: $\lambda_1\neq-\lambda_2$, $\lambda_3\neq0$}
In this class all charges are generically independent, corresponding to diagonalizable connections whose Jordan form is
\begin{empheq}{alignat=5}
	a_{\phi}^D&=\lambda_1\left(E_{11}-E_{22}\right)+\lambda_2\left(E_{22}-E_{33}\right)+\frac{1}{2}\lambda_3\left(E_{44}-E_{55}\right)+iQ_1\left(E_{11}+E_{22}+E_{33}\right)\,,
	\\
	a_{\phi}^N&=0\,.
\end{empheq}
Note that, according to our definition, solutions in this class are therefore \textit{not} extremal. Since $a_\phi^N$ is trivial, \eqref{eq:ancomm} is automatically satisfied. Still, we need to ensure that $\epsJ^-(\phi)$ in \eqref{eq:quantawij} is single or double-valued along the $\phi$ cycle by adjusting the frequencies $\omega_{i\bar j}$ in \eqref{eq: Sec.3 frequencies}. For convenience, we reproduce them here:
\begin{empheq}{alignat=5}\label{eq:freq2}
	\omega_{14}&=\lambda_1-\frac{\lambda_3}{2}+iQ_1\,,
	&\qquad
	\omega_{15}&=\lambda_1+\frac{\lambda_3}{2}+iQ_1\,,
	\cr
	\omega_{24}&=-\lambda_1+\lambda_2-\frac{\lambda_3}{2}+iQ_1\,,
	&\qquad
	\omega_{25}&=-\lambda_1+\lambda_2+\frac{\lambda_3}{2}+iQ_1\,,
	\\
	\omega_{34}&=-\lambda_2-\frac{\lambda_3}{2}+iQ_1\,,
	&\qquad
	\omega_{35}&=-\lambda_2+\frac{\lambda_3}{2}+iQ_1\,.
	\nonumber
\end{empheq}

Taking into account the reality condition \eqref{eq:realeig} together with $\lambda_1\neq-\lambda_2$ and $\lambda_3\neq0$, it is straightforward to check that $\omega_{24}$ and $\omega_{25}$ cannot be purely imaginary in this class. However, if we set 
\begin{empheq}{alignat=5}\label{eq:cd61}
\lambda_3&=\lambda_1+\lambda_2\quad {\rm and } \quad \frac{i}{2}(\lambda_1-\lambda_2)-Q_1&=\eta+\frac{1}{2}\in \left\{
\begin{array}{ll}
	\mathds{Z} & \textrm{R sector}
	\\
	\mathds{Z}+\frac{1}{2} & \textrm{NS sector}
\end{array}
\right.\,,
\end{empheq}
then $\omega_{14}$ and  $\omega_{35}$ are properly quantized. The coresponding global Killling spinor is 
\begin{empheq}{alignat=5}
	\epsJ^-(\phi)&=\left(\epsJ_{14}E_{14}+\epsJ_{35}E_{35}\right)e^{i\left(\eta+\frac{1}{2}\right)\phi}~,
\end{empheq}
which has 4 real independent parameters.
Alternatively, we can impose 
\begin{empheq}{alignat=5}\label{eq:cd62}
\lambda_3&=-(\lambda_1+\lambda_2)\quad {\rm and } \quad \frac{i}{2}(\lambda_1-\lambda_2)-Q_1&=\eta+\frac{1}{2}\in \left\{
\begin{array}{ll}
	\mathds{Z} & \textrm{R sector}
	\\
	\mathds{Z}+\frac{1}{2} & \textrm{NS sector}
\end{array}
\right.\,,
\end{empheq}
for which $\omega_{15}$ and $\omega_{34}$ are quantized and the fermionic symmetry generator is
\begin{empheq}{alignat=5}
	\epsJ^-(\phi)&=\left(\epsJ_{15}E_{15}+\epsJ_{34}E_{34}\right)e^{i\left(\eta+\frac{1}{2}\right)\phi}~.
\end{empheq}
This solution preserves 4 real supercharges as well. In either case we find 
\begin{empheq}{alignat=5}\label{eq:ftk}
	\lH(\phi)&=\lH(0)e^{i\left(\eta+\frac{1}{2}\right)\phi}\,,
	\quad
	\lT(\phi)&=\lT(0) e^{i\left(\eta+\frac{1}{2}\right)\phi}\,.
\end{empheq}

The conditions \eqref{eq:cd61} and \eqref{eq:cd62} exhaust all possible supersymmetric configurations within this class. Since \eqref{charges as eigenvalues 1} is unaffected by the sign of $\lambda_3$, the corresponding bosonic charges always read
\begin{empheq}{alignat=5}
	\mathcal{L}&=\frac{1}{8}\left(2\lambda_1^2+2\lambda_2^2+\lambda_1\lambda_2\right)\,,
	\nonumber\\
	Q_2&=-\frac{3}{8}\lambda_1\lambda_2\,,
	\\
	Q_3&=-\frac{i}{8}\left(-\lambda_1+\lambda_2\right)\lambda_1\lambda_2\,.
	\nonumber
\end{empheq}
However, from \eqref{eq:entropyw32}, we see that the entropy is sensitive to the choice $\lambda_3=\pm(\lambda_1+\lambda_2)$: 
\begin{empheq}{align}
	\nonumber
	S&=2\pi k_{cs}\left(1\mp {1\over 2}\right) (\lambda_1+\lambda_2)+\textrm{other sector}
	\\
	&=4\pi k_{cs} \left(1\mp {1\over 2}\right) \sqrt{ {\cal L}-Q_2}+\textrm{other sector}\,.
\end{empheq}
The chemical potentials \eqref{eq: thermal sources sl(3|2) 1} are also affected by the sign of $\lambda_3\,$. Recall from the discussion at the end of section \ref{sec:sl32bhs} that $\lambda_3>0$ is slightly preferred since within this branch one could reach the BTZ solution.

There is something disconcerting about our findings. The above analysis shows that setting $\lambda_3=\pm(\lambda_1+\lambda_2)$ allows for supersymmetry within the class of diagonalizable connections, which according to our general definition are \textit{not} extremal. Indeed, since the temperature as defined in \eqref{eq: thermal sources sl(3|2) 1} remains finite, we come to the conclusion that we have successfully constructed {\it globally defined Killing spinors for non-extremal black hole solutions!}. This finding seems to go against the conventional wisdom regarding supersymmetric theories. 
To make the reader (and ourselves) at ease with this undoubtedly peculiar feature, let us highlight some properties of these black hole configurations that should be taken into account before discarding them:
\begin{enumerate}
\item So far, we have focused exclusively on the $\phi$-dependence of the Killing spinors. One could suspect that there exists some incurable illness along the thermal cycle. Naively at least, this does not seem to be the case. It is easy to check, using \eqref{eq:killingds} and the fact that the holonomy of $a_{\rm contract}$ is equal to $\Gamma^-$, that the spinor $\eps(t,\phi)$ is indeed anti-periodic around the contractible direction, as expected for a smooth fermionic field. Recall that $\Gamma^-$  anti-commutes with the odd generators of the algebra. This property is only relevant for class {\bf VI} where the topology of the solutions is a solid tours, whereas for the other classes the topology changes due to the vanishing temperature.
\item The BPS conditions for class {\bf VI} do not allow for solutions in the $sl(2|1)$ truncation. The well-known folklore regarding supersymmetric BTZ black holes remains safe in the higher spin setup.
\item The supercharges preserved by solutions with \eqref{eq:cd61}-\eqref{eq:ftk} are a non-trivial mixture of fermionic generators in the spin-2 and spin-3 multiplets, as reflected by \eqref{eq:explicitK} and \eqref{eq:ftk}. Moreover, the corresponding variations of the charges involve non-linear relations among the bosonic and fermionic fields, which is explicitly seen when studying the asymptotic symmetry group \cite{BetoJuan}. We suspect that these non-linearities are allowing for the solution to balance anti-periodic fermions and periodic bosons on a topology with a contractible cycle. This feature is clearly absent in standard supergravity where BPS conditions always involve relations which are linear in the fermionic fields. 
\item A powerful reason to take these solutions seriously is that we will able to reproduce the BPS bounds \eqref{eq:cd61}-\eqref{eq:cd62} from a calculation of the Kac determinant in a CFT with $\cW_{(3|2)}$ symmetry. We postpone this aspect of the discussion until section \ref{sec:HSBHCFT}.
\end{enumerate}

\subsubsection{Smooth conical defects}
We now move on to the supersymmetry analysis of smooth conical defects, which is significantly simpler than that for black holes. This is mainly because in this case the smoothness condition immediately implies that the connection is diagonalizable. Otherwise it would not have trivial holonomy. We need only look at the periodicity of the Killing spinors dictated by the frequencies \eqref{eq: Sec.3 frequencies}, which upon using \eqref{eq:lcd} read
\begin{empheq}{alignat=5}\label{eq:ccdd1}
	\omega_{14}&=i\left(n_1-n_3\right)\,,
	&\qquad
	\omega_{15}&=i\left(n_1+n_3+n\right)\,,
	\cr
	\omega_{24}&=-i\left(n_1-n_2+n_3+n\right)\,,
	&\qquad
	\omega_{25}&=i\left(-n_1+n_2+n_3\right)\,,
	\\
	\omega_{34}&=-i\left(n_2+n_3\right)\,,
	&\qquad
	\omega_{35}&=i\left(-n_2+n_3+n\right)\,.
	\nonumber
\end{empheq}
Recall that $n_1$, $n_2$ and $n$ are integer numbers, whereas $n_3$ can be an integer of half-integer, determining the periodicity of $\epsilon(\phi)$. Since all the frequencies are automatically quantized by the smoothness requirement, we see that these solutions are always maximally supersymmetric, preserving all 12 real supercharges. The Killing spinors are simply given by \eqref{eq:fullep}-\eqref{eq:quantawij}.

\subsubsection{Summary: supersymmetry versus extremality}\label{sec:summ}
The supersymmetries preserved by $sl(3|2)$ black holes are summarized in table \ref{table: SUSY summary}.
\begin{table}[h]
\begin{empheq}{align*}
\begin{array}{|c|c|c|c|c|c|c|}
	\hline
	\textrm{Class} & \multicolumn{2}{c|}{\textrm{Eigenvalues}} & \textrm{Extremal?} & \textrm{Quantization conditions} & \multicolumn{2}{c|}{\textrm{\# of supersymmetries}} \\\hline
	\textrm{\textbf{I}} & \lambda_1=\lambda_2=0 & \lambda_3=0 & \textrm{Yes} & -Q_1=\eta+\frac{1}{2} & 4 & \phantom{\Big(}\frac{1}{3}\textrm{-BPS}\phantom{\Big)} \\ \hline
	\multirow{2}{*}{\textrm{\textbf{II}}} & \multirow{2}{*}{$\lambda_1=-\lambda_2\neq0$} & \multirow{2}{*}{$\lambda_3=0$} & \multirow{2}{*}{\textrm{Yes}} & i\lambda_1-Q_1=\eta+\frac{1}{2}  & 4  & \phantom{\Big(}\frac{1}{3}\textrm{-BPS}\phantom{\Big)} \\\cline{5-7}
	&  &  &  & -2i\lambda_1-Q_1=\eta+\frac{1}{2} & 2 & \phantom{\Big(}\frac{1}{6}\textrm{-BPS}\phantom{\Big)} \\ \hline
	\textrm{\textbf{III}} & \lambda_1\neq-\lambda_2 & \lambda_3=0 & \textrm{Yes} & -i\left(\lambda_1-\lambda_2\right)-Q_1=\eta+\frac{1}{2} & 2 & \phantom{\Big(}\frac{1}{6}\textrm{-BPS}\phantom{\Big)} \\ \hline
	\textrm{\textbf{IV}} & \lambda_1=\lambda_2=0  & \lambda_3\neq0 & \textrm{Yes} & \textrm{None} & 0  & \phantom{\Big(}\textrm{Not BPS}\phantom{\Big)} \\ \hline
	\textrm{\textbf{V}} & \lambda_1=-\lambda_2\neq0 & \lambda_3\neq0 & \textrm{Yes} & \textrm{None} & 0 & \phantom{\Big(}\textrm{Not BPS}\phantom{\Big)} \\ \hline
	\multirow{2}{*}{\textrm{\textbf{VI}}} & \multirow{2}{*}{$\lambda_1\neq-\lambda_2$} & \multirow{2}{*}{$\lambda_3\neq0$} & \multirow{2}{*}{\textrm{No}} & \phantom{\Big(}\frac{i}{2}\left(\lambda_1-\lambda_2\right)-Q_1=\eta+\frac{1}{2}\,,\phantom{\Big)} & \multirow{2}{*}{$4$} & \multirow{2}{*}{$\frac{1}{3}\textrm{-BPS}$} \\
	&   &   &   & \phantom{\Big(}\lambda_3=\pm\left(\lambda_1+\lambda_2\right)\phantom{\Big)} &   &  \\ \hline
\end{array}\,.
\end{empheq}
\caption{Supersymmetries of  $sl(3|2)$ (or more properly $su(2,1|1,1))$ black holes. The eigenvalues of the connection are parametrized as $\textrm{eigen}\left(a_{\phi}+iQ_1J\right)=\left[\lambda_1,-\lambda_1+\lambda_2,-\lambda_2,\frac{1}{2}\lambda_3,-\frac{1}{2}\lambda_3\right]$, with the reality conditions $\lambda_1^*=\lambda_2$, $\lambda_3^*=\lambda_3$. The charges carried by the solution are given by \eqref{charges as eigenvalues 1}. The quantization parameter $\eta+\frac{1}{2}$ is an integer in the Ramond sector and a half-integer in the Neveu-Schwarz sector.}
\label{table: SUSY summary}
\end{table}
In terms of the eigenvalues of the connection $a_{\phi}\,$, we can group the BPS conditions into two familes, depending on the number of preserved supercharges. The first has
\begin{empheq}{alignat=5}\label{eq:l34}
	\lambda_3&=\pm\left(\lambda_1+\lambda_2\right)
	&\qquad\textrm{and}\qquad
	\frac{i}{2}\left(\lambda_1-\lambda_2\right)-Q_1&=\eta+\frac{1}{2}\,,
\end{empheq}
and implies the existence of four independent Killing spinors. It is accessible only to solutions in classes \textbf{I}, \textbf{II} and \textbf{VI}. Depending on the choice of sign, this is equivalent to demanding
\begin{empheq}{alignat=5}
i\omega_{14}=i\omega_{35}=\eta+\frac{1}{2}\in\left\{
\begin{array}{ll}
	\mathds{Z} & \textrm{R sector}
	\\
	\mathds{Z}+\frac{1}{2} & \textrm{NS sector}
\end{array}
\right.\,,
\end{empheq}
or
\begin{empheq}{alignat=5}
i\omega_{15}=i\omega_{34}=\eta+\frac{1}{2}\in\left\{
\begin{array}{ll}
	\mathds{Z} & \textrm{R sector}
	\\
	\mathds{Z}+\frac{1}{2} & \textrm{NS sector}
\end{array}
\right.\,.
\end{empheq}
The second kind, which gives rise to two supersymmetries, imposes
\begin{empheq}{alignat=5}\label{eq:l32}
	\lambda_3&=0
	&\qquad\textrm{and}\qquad
	-i\left(\lambda_1-\lambda_2\right)-Q_1=\eta+\frac{1}{2}\,,
\end{empheq}
and can occur in classes \textbf{II} and \textbf{III}, which are extremal. Consonantly,
\begin{empheq}{alignat=5}
i\omega_{24}=i\omega_{25}=\eta+\frac{1}{2}\in\left\{
\begin{array}{ll}
	\mathds{Z} & \textrm{R sector}
	\\
	\mathds{Z}+\frac{1}{2} & \textrm{NS sector}
\end{array}
\right.\,.
\end{empheq}
Both families of BPS conditions can intersect in class \textbf{II}, producing and enhancement to six supercharges. Class \textbf{I} also allows in fact for both scenarios, but no enhancement occurs in that case because the two conditions coincide.

The first relation in \eqref{eq:l34} implies that the charges carried by the backgrounds satisfy
\begin{empheq}{alignat=5}\label{eq:ffss}
	4\left(\mathcal{L}+\frac{5}{3}Q_2\right)Q_2^2+9Q_3^2&=0\,.
\end{empheq}
The converse, however, is not true, as can be seen by setting the eigenvalues to the $sl(2|1)$ truncation in class \textbf{VI}. Keeping this caveat in mind, we can relate this BPS condition with extremality by noticing that
\begin{empheq}{alignat=5}\label{eq:ffcc}
	4\left(\mathcal{L}+\frac{5}{3}Q_2\right)Q_2^2+9Q_3^2&=\frac{1}{3}\left(\frac{1}{64}\Delta_3-\Delta_2\left(\mathcal{L}+2Q_2\right)^2\right)\,.
\end{empheq}
It is clear that vanishing of the above combination of charges does not necessarily imply $\Delta_3=0$ or $\Delta_2=0$. Notice also that the second relation in \eqref{eq:l34} can be expressed as
\begin{empheq}{alignat=5}
	-\left(\frac{3}{2}\frac{Q_3}{Q_2}+Q_1\right)&=\eta+\frac{1}{2}\,,
\end{empheq}
where the first term is absent for $Q_2=0$. The other BPS condition, equation \eqref{eq:l32}, is always linked to extremality since $\lambda_3=0$ implies $\Delta_2=0$ and vice versa. In terms of the charges this constraint translates simply to
\begin{empheq}{align}
	\mathcal{L}-Q_2&=0\,.
\end{empheq}

All in all, there are three notable results we would like to highlight. First, there are solutions in class \textbf{II} that preserve more supersymmetries than the charged BTZ black holes in class \textbf{I}. Second, the intricacies of the $sl(3|2)$ algebra allowed us to build non-extremal supersymmetric solutions in class \textbf{VI}. This establishes that extremality is not a necessary condition for supersymmetry as one might have naively suspected. Third, supersymmetric solutions, extremal or not, generically carry residual entropy, which up to a numerical coefficient we find to be
\be
S_{\rm SUSY-BH}\sim 2\pi k_{cs} (\lambda_1+\lambda_2)+\textrm{other sector}\,.
\ee


\subsection{${\cal N}=2$ supergravity truncation}\label{sec:n2bhs}

In the last portion of this section we will take the opportunity to review some aspects of black holes and conical defects in AdS$_{3}$ ${\cal N}=2$ supergravity. Historically, this is one of the first theories to be described in Chern-Simons language \cite{Achucarro:1987vz,Witten:1988hc}. Given that  $sl(2|1)$ is a subalgebra of $sl(3|2)$, and with the intention of avoiding further cluttering, we will simply truncate our results for the $sl(3|2)$ theory. More background on this topic can be found in e.g. \cite{Banados:1998pi,Henneaux:1999ib} and references therein. For a discussion of the theory in metric formulation see e.g. \cite{Izquierdo:1994jz,Balasubramanian:2000rt,Kraus:2006nb}.

In this case the appropriate gauge superalgebra in Lorentzian signature turns out to be $osp(2|2;\mathds{R})\oplus osp(2|2;\mathds{R})$. This is the choice of real form of $sl(2|1;\mathds{C})\oplus sl(2|1;\mathds{C})$ that gives the usual Hermiticity properties for the metric fields. The even-graded sector decomposes into the $sl(2)$ generators ($L_i$) and a spin 0 element ($J$); the bosonic sub-algebra is thus $sl(2)\oplus u(1)$. The odd-graded elements consist of two spin $1/2$ multiplets  ($H_r$ and $G_r$). The non-vanishing commutators of $sl(2|1)\subset sl(3|2)$ can be found in Appendix \ref{app: sl(3|2)}.

As for $sl(3|2)$, one can gauge fix the radial dependence of the connection and define the charges by the highest weight components of $a_{\phi}$. In particular, a black hole connection now reads
\begin{empheq}{alignat=5}\label{eq: sl(2|1) Black Hole connection aphi}
	a_{\phi}&=L_1-\mathcal{L}L_{-1}-iQ_1J\,, \quad  ia_{t_E}+a_{\phi}=i\nu_0J~,
\end{empheq}

\noindent with similar expressions for the components of $\bar{a}\,$. These configurations can be interpreted as states in a theory with ${\cal N}=2$ super-Virasoro symmetry by using the map
\begin{empheq}{alignat=5}\label{eq:mapcharges11}
	\mathcal{L}&=\frac{6}{c}\left(h-\frac{c}{24}-\frac{3}{2c}q^2\right)\,,
	\quad	
	Q_1&=-\frac{3}{c}q\,.
\end{empheq}
Here, $h$ is the zero mode of the stress tensor $T$ on the plane, and $q$ is that of the $U(1)$ current $J$; $\nu_0$ is the source for the $U(1)$ charge. In Euclidean signature, the topology is taken to be that of a solid torus and the source for $T$ is introduced as the modular parameter $\tau\,$ of the boundary two-torus. All parameters are then complex.

The thermodynamics follows as before. Imposing the holonomy condition
\begin{empheq}{alignat=5}
	\mathcal{P}e^{\oint_{\mathcal{C}_E}a}&=\Gamma^-\,,
\end{empheq}
one finds that
 \be\label{eq:tempbtzu1}
 \tau={i\over 2\sqrt{\mathcal{L}}} ~,\qquad  \nu_0= i Q_1{\tau\over  {\rm Im}(\tau)}~.
 \ee
Recall that $\Gamma^-$, given in \eqref{eq:gammas}, is the central element of the bosonic sub-group that is compatible with anti-periodic fermions along the thermal cycle. The entropy carried by the solution is then
\begin{empheq}{align}\label{eq:entbtzu1}
	S&=2\pi \sqrt{\frac{c}{6}\left(h - \frac{c}{24} -\frac{3q^{2}}{2c}\right)} +  2\pi \sqrt{\frac{c}{6}\left(\bar h - \frac{c}{24} -\frac{3{\bar q}^{2}}{2c}\right)}\,.
\end{empheq}
In Lorentzian signature, demanding reality of the entropy restricts ${\cal L} >0$. This equivalent to stating that $a_{\phi}$ has real eigenvalues. Solutions with ${\cal L} <0$ are conical defects, to be expanded on below. In the metric formulation, these backgrounds correspond to BTZ black holes carrying topological $U(1)$ charge (i.e. Abelian Wilson loops). See \cite{Dijkgraaf:2000fq,Kraus:2006nb,Kraus:2006wn} for the explicit solutions.
 
Within this theory there are only two Jordan classes of connections, one where $a_{\phi}$ diagonalizable and another one where it is not. A simple calculation shows that the only way to have degenerate eigenvalues, and therefore a non-diagonalizable matrix, is to set $\cL=0$.\footnote{ Recall that the $sl(2|1)$ truncation is obtained by setting $\lambda_1=\lambda_2=\lambda_3$ so that $\mathcal{L}=\frac{1}{4}\lambda_1^2$ and $Q_2=Q_3=0$ throughout the $sl(3|2)$ analysis.} According to section \ref{sec:defnext}, this defines {\it extremal} charged BTZ black holes. Slightly adapting the machinery developed in section \ref{sec:w32solutions}, the Jordan form of the connection can be cast as
\begin{empheq}{alignat=5}\label{eq:aphid1}
	V^{-1}a_{\phi}V&=a_{\phi}^D+a_{\phi}^N\,,
\end{empheq}
where
\be\label{eq: aphi diagonal}
	a_{\phi}^D=2\sqrt{\cal L}L_0-iQ_1J\,,
	\qquad	a_{\phi}^N=\left\{
\begin{array}{ll}
	0 & \textrm{if}\quad\cL\neq0
	\\
	L_{-1} & \textrm{if}\quad\cL=0
\end{array}
\right.\,.
\ee
Notice that $[a_{\phi}^D,a_{\phi}^N]=0$, as appropriate.

As is clear from \eqref{eq:tempbtzu1}, our notion of extremality is again compatible with the zero temperature limit. Moreover, the contribution from $a_\phi$ to  the entropy in \eqref{eq:entbtzu1} vanishes, while the barred sector remains unchanged. Following the classification exposed in section \ref{sec:sl32bhs}, extremal charged BTZ black holes fall into class {\bf I} and non-extremal charged BTZ solutions belong in (the non-supersymmetric subsector of) class {\bf VI}. In terms of CFT variables the extremality condition reads
\begin{equation}\label{extremal black hole condition}
\text{extremal charged BTZ:}\qquad h =\frac{3q^{2}}{2c} + \frac{c}{24} \,.
\end{equation}

We can also characterize the supersymmetric solutions in the $sl(2|1)$ truncation, in particular, BPS black holes. The details of the analysis were carried out above, the pertinent results being those for class \textbf{I} in section \ref{sec:susyw32bhs}. We found that in order to have a supersymmetric background we need
\be\label{eq:LQsusy}
\mathcal{L}=0\,,\qquad	Q_1\in\left\{
\begin{array}{ll}
	\mathds{Z} & \textrm{R sector}
	\\
	\mathds{Z}+\frac{1}{2} & \textrm{NS sector}
\end{array}
\right.\,.
\ee
The Killing spinor is then
\be\label{eq:kclassn2}
\eps(\phi)= \lH(0)e^{-iQ_1\phi} H_{\frac{1}{2}}+i\blH(0)e^{iQ_1\phi}G_{\frac{1}{2}}\,,
\ee
where $\lH(0)$ is a free complex parameter, which implies that a BPS black hole can preserve 2 supercharges (half-BPS). The introduction of a $U(1)$ charge makes it possible to find supersymmetric solutions in both the NS and R sectors, as seen from the periodicity of the corresponding Killing spinors. This is in contrast to the uncharged case (neutral BTZ), where the Killing spinors carry no dependence on the angular coordinate $\phi$ so BPS black holes lie in the R sector only \cite{Coussaert:1993jp}. In CFT language \eqref{eq:LQsusy} translates to 
\begin{equation}\label{eq:bpscbtz}
\text{BPS charged BTZ }\left\{
\begin{array}{rcl}
\text{R:} & \displaystyle{\left(\frac{6h}{c}\,,\,\frac{6q}{c}\right) = \left(n^{2}+\frac{1}{4}\,,\, 2n\right),} & n \in \mathds{Z} \\ 
&&
\\
\text{NS:} &  \displaystyle{\left(\frac{6h}{c}\,,\,\frac{6q}{c}\right) = \left(r^{2}+\frac{1}{4}\,,\, 2r\right),} & r \in \mathds{Z} +\tfrac{1}{2}
\end{array}  
\right.\,.
\end{equation}
 
It is also interesting to discuss smooth conical defects in ${\cal N}=2$ supergravity. These are a subset of the solutions constructed for the $sl(3|2)$ theory in section \ref{sec:smoothcon}, which have
\begin{empheq}{alignat=5}
	\cL&=-\frac{1}{4}\left(2n_3+3n\right)^2\,,
	&\qquad
	Q_1&=\frac{1}{2}n\,.
\end{empheq}
Since the topology is that of AdS$_{3}$, it must be that $n\in\mathds{Z}$ in order to achieve a trivial holonomy along the contractible cycle $\phi\sim\phi+2\pi$. Additionally, $n_3\in \mathds{Z}$, which allows for periodic fermions (R), or $n_3\in \mathds{Z}+{1\over 2}$, implying anti-periodic boundary conditions (NS). The corresponding connection is always diagonalizable and the solutions are maximally supersymmetric, preserving all four supercharges. The bulk theory imposes no further constraints on $n_3$ and $n$ besides $\mathcal{L}\neq0$.

In this case, the CFT charges of the defects are
\begin{empheq}{alignat=5}\label{eq:nn}
	h=-\frac{c}{6}\left(n_3+n\right)\left(n_3+2n\right)+\frac{c}{24}\,,
	\quad
	q=-\frac{c}{6}n\,.
\end{empheq}
It follows that
\begin{empheq}{alignat=5}\label{eq:boundcon}
	h-\frac{3q^2}{2c}&=-\frac{c}{6}\left(n_3+\frac{3n}{2}+\frac{1}{2}\right)\left(n_3+\frac{3n}{2}-\frac{1}{2}\right)\,,
\end{empheq}
which is the spectral flow invariant combination. As we will review in section \ref{sec:warmup}, the semi-classical unitarity bound of the ${\cal N}=2$ algebra demands that this quantity be positive. This only allows for
\begin{empheq}{alignat=5}
	-\frac{1}{2}\leq n_3+\frac{3n}{2}\leq\frac{1}{2}~.
\end{empheq}
From here we have two options:  $\cL=-\frac{1}{4}$, which corresponds to  global AdS$_3$ supported by a $U(1)$ Chern-Simons field, or $\cL=0$, condition which yields a non-diagonalizable connection with-nontrivial holonomy and must therefore be discarded (it is, in fact, an extremal black hole). The unitary BPS smooth conical defects then have
 \begin{equation}\label{eq:bpscbtz}
\text{BPS smooth conical defects }\left\{
\begin{array}{rcl}
\text{R:} & \displaystyle{\left(\frac{6h}{c}\,,\,\frac{6q}{c}\right) = \left(\left(n+\frac{1}{2}\right)^2\,,\, 2n+1\right),} & n \in \mathds{Z} \\ 
&&
\\
\text{NS:} &  \displaystyle{\left(\frac{6h}{c}\,,\,\frac{6q}{c}\right) = \left(\left(r+\frac{1}{2}\right)^2\,,\, 2r+1\right),} & r \in \mathds{Z} +\tfrac{1}{2}
\end{array}  
\right.\,.
\end{equation}
Here we have set $2n_3+3n=1$ and relabeled $n\rightarrow-(2n+1)$ or $n\rightarrow-(2r+1)$.
 
Before closing this section, some comments regarding the periodicity of Killing spinors are in order. In the black hole case the boundary circle parameterized by the angular coordinate $\phi$ is not contractible in the bulk because of the finite size of the horizon. As a consequence, this cycle supports both periodic and anti-periodic spinors and there exist BPS black holes in both the Neveu-Schwarz and Ramond sectors. For smooth conical defects, on the other hand, the boundary spatial cycle becomes contractible in the bulk, which implies that there is only one admissible spin structure, namely, the one that extends from the circle to the disk. Only anti-periodic Killing spinors are then allowed, making $n_3\in \mathds{Z}+{1\over 2}$ the reasonable choice. For $n_3\in \mathds{Z}$ the bulk solution is singular, which is evident both in metric and Chern-Simons formulations \cite{Izquierdo:1994jz,Balasubramanian:2000rt}. However, if we insist on having periodic fermions along $\phi$, we will get an agreement between the charges of supersymmetric states in the R sector of the dual CFT$_2$ and the charges carried by a smooth conical defect. Since these solutions have a dual interpretation, it is believed that the singularity will be resolved in string theory by either $\alpha'$ corrections or the inclusion of additional directions (KK modes from the three dimensional perspective). These corrections should fatten the contractible cycle and hence allow for periodic fermions.


\section{Higher spin BPS bounds and holography}\label{sec:BPS}

In holography one expects the subset of black hole solutions that admit globally-defined Killing spinors to correspond to states that saturate BPS bounds in the dual CFT.  We will now confirm this expectation in the context of the duality between $sl(3|2)\oplus sl(3|2)$ Chern-Simons theory, whose black hole solutions and corresponding Killing spinors were studied in the previous section, and CFTs with $\cW_{(3|2)}$ symmetry. This setup is of particular interest because the latter theories are based on the simplest higher spin extension of the familiar $\cN=2$ super-Virasoro algebra. In the process we will compute BPS bounds for the $\cW_{(3|2)}$ algebra. Moreover, we will manage to compute these bounds at the full quantum level (i.e. at finite values of the central charge $c$) and show that their semiclassical limit is indeed saturated by the subset of supersymmetry-preserving extremal black hole solutions in classical Chern-Simons supergravity.

Before proceeding with the higher spin case, we will review the problem in the pure $\cN=2$ super-Virasoro case. These considerations will clarify a number of issues, especially in relation to the role of the $U(1)$ charge and the spectral flow automorphism of $\cN=2$ superconformal algebras.

\subsection{Warmup: super-Virasoro BPS bounds and $\mathcal{N}=2$ supergravity backgrounds}\label{sec:warmup}
In our conventions the $\cN=2$ super-conformal algebra is given by \eqref{superVir comms}-\eqref{superVir comms 2}, and we shall assume the standard Hermiticity properties of the generators on the plane
\begin{equation}\label{super Virasoro Hermiticity}
\left(L_{n}\right)^{\dagger} = L_{-n},\qquad \left(J_{n}\right)^{\dagger} = J_{-n}\,,\qquad \left(G^{+}_{r}\right)^{\dagger} = G^{-}_{-r}\,.
\end{equation}

\noindent In the Ramond (R) sector the fermionic generators $G^{\pm}_{r}$ are integer-modded ($r \in \mathds{Z}$), while in the Neveu-Schwarz (NS) sector they are half-integer-modded ($r \in \mathds{Z} + \frac{1}{2}$).

Quite importantly for our purposes, the $\cN=2$ superconformal algebra is invariant under a continuous family of deformations of the generators, the so-called spectral flow automorphism \cite{Schwimmer:1986mf}:
\begin{alignat}{3}\label{spectral flow 1}
L_n &\quad \to \quad & L'_{n} 
={}&
 L_n + \eta J_{n} + \frac{\eta^{2}}{6}c \delta_{n,0}
\\
J_n &\quad \to & J'_{n}  ={}&
 J_n + \frac{c}{3}\eta \delta_{n,0}
\\
G^{\pm}_{r} &\quad \to & G_{r}^{\pm '} 
={}&
 G^{\pm}_{r\pm \eta}\,.
 \label{spectral flow 2}
\end{alignat}

\noindent Here $\eta$ is a continuous parameter; for $\eta \in \mathds{Z} + 1/2$ the flow interpolates between the NS sector and the R sector, while for $\eta \in \mathds{Z}$ it maps the R and NS sector to themselves.

The zero modes $L_0$ and $J_0$ commute, and super-primary states $|h,q\rangle$ are labeled by the eigenvalues $h$ and $q$ of these operators, namely, 
\begin{equation}\label{superVirasoro Cartan}
L_0|h,q\rangle=h|h,q\rangle\,,\qquad J_0|h,q\rangle=q|h,q\rangle\,.
\end{equation}
They further satisfy the usual highest-weight conditions
\begin{alignat}{3}\label{superVirasoro hw conditions 1}
 G^{\pm}_{r}\bigl |h,q\bigr\rangle ={}&
  0\, ,&
  &\qquad\quad &
   r >{}&0
     \\
  L_{n}\bigl |h,q\bigr\rangle ={}& J_{n}\bigl |h,q\bigr\rangle = 0\,,&
    &\qquad\quad &
   n >{}&0\,.
   \label{superVirasoro hw conditions 2}
\end{alignat}
In the NS sector of the Hilbert space of an $\mathcal{N}=2$ SCFT, it is useful to define \textit{(left-)chiral} and \textit{(left-)anti-chiral}\footnote{ Here, ``left" refers to the fact that both definitions involve generators in the holomorphic sector of the algebra, while ``right" would denote generators in the second, anti-holomorphic, copy of the algebra (i.e. the ``barred" sector).} states to be those which, in addition to \eqref{superVirasoro hw conditions 1}-\eqref{superVirasoro hw conditions 2}, satisfy 
\begin{empheq}{align}
	G^{+}_{-\frac{1}{2}}\left|h,q\right\rangle&=0\,,
\end{empheq}
and 
\begin{empheq}{align}
	G^{-}_{-\frac{1}{2}}\left|h,q\right\rangle&=0\,,
\end{empheq}
respectively. Using the mode algebra one easily proves (see e.g. \cite{Blumenhagen:2009zz}) that $\left|h,q\right\rangle$ is an $\mathcal{N}=2$ (anti-)chiral primary if and only if $h = q/2$ ($h=-q/2$). As usual, chiral representations correspond to short supermultiplets, i.e. BPS states in the NS sector.

The full Kac determinant for the $\mathcal{N}=2$ superconformal algebra was given in \cite{Boucher:1986bh}. In the NS sector, the set of unitarity constraints includes a family of BPS bounds which are linear in the $U(1)$ charge $q$,
\begin{equation}\label{Linear NS bound from lit}
\text{NS sector (\textbf{quantum}):}\qquad h \geq r q  + \frac{(c-3)}{24}\left(1-4r^{2}\right)\,,\qquad r \in \mathds{Z} +\tfrac{1}{2}\,,
\end{equation}

\noindent as well as a family of quadratic conditions of the form $h \geq f(c,q^{2})\,$. The latter are not very important for our purposes, as they are not associated with multiplet shortening and supersymmetric states. In other words, they are unitarity bounds, but not BPS bounds. 

The semiclassical limit of \eqref{Linear NS bound from lit} is of particular interest to us, for it is that version of the bounds that we expect to see reflected in the bulk physics. As reviewed in appendix \ref{app:semiclassical}, such limit is realized by scaling $h \to h/\hbar$, $q \to q/\hbar$, $c \to c/\hbar$ and sending $\hbar \to 0$ while keeping the leading terms only. In this way one finds
\begin{align}\label{Relaxed NS bounds}
\text{NS sector (\textbf{semiclassical}):}\qquad 
h \geq{}&
 rq  + \frac{c}{24}\left(1-4r^{2}\right) \,,\qquad r \in \mathds{Z}+\tfrac{1}{2}\,.
\end{align}

\noindent The level-$1/2$ bound, obtained by setting $r=1/2$ in either \eqref{Linear NS bound from lit} or \eqref{Relaxed NS bounds}, corresponds to the usual BPS condition $h \geq \tfrac{|q|}{2}$ saturated by chiral primaries.

It is worth emphasizing that while the quantum bound \eqref{Linear NS bound from lit} is derived by requiring positivity of the norm for all states at a given level $r$, the semiclassical rendering \eqref{Relaxed NS bounds} comes from doing so only for states of the form $G^{\pm}_{-r}|h,q\rangle$ (with $r$ a positive half-integer). In passing, we also mention that the most stringent of the quadratic restrictions $h \geq f(c,q^{2})\,$ becomes simply $h \geq \frac{3q^{2}}{2c}$ in the semiclassical limit. It is interesting to note that this bound can also be obtained from positivity of the norm of level-1 states in the purely bosonic $\text{Virasoro}\oplus U(1)$ Kac-Moody algebra. As mentioned above, $h \geq \frac{3q^{2}}{2c}$ is a semiclassical unitarity condition, but not a proper BPS bound. 

Let us now consider Ramond sector representations. In this sector there is also a family of BPS bounds which are linear in the R-charge $q$ \cite{Boucher:1986bh}, namely
\begin{equation}\label{Linear R bound from lit}
\text{R sector (\textbf{quantum}):}\qquad 
h \geq nq  + \frac{c}{24}\left(1-4n^{2}\right) + \frac{n(n-1)}{2}\,,\qquad n \in \mathds{Z}\,,
\end{equation}

\noindent which are in fact obtained from the corresponding NS sector expressions \eqref{Linear NS bound from lit} by performing a half unit of spectral flow ($\eta = 1/2$) and setting $r = n-1/2\,$. Taking the semiclassical limit as before we get
\begin{align}\label{Relaxed Ramond bounds}
\text{R sector (\textbf{semiclassical}):}\qquad 
h \geq{}&
 nq  + \frac{c}{24}\left(1-4n^{2}\right) \,,\qquad n \in \mathds{Z}\,.
\end{align}

\noindent The level-0 bound, obtained by taking $n=0$ in either \eqref{Linear R bound from lit} or its semiclassical version \eqref{Relaxed Ramond bounds}, yields the usual Ramond sector constraint $h \geq c/24\,$. In particular, it is easy to see that the Ramond ground states $\left| h = \tfrac{c}{24}\,, q\right \rangle$ are related to chiral primaries in the NS sector by spectral flow:
\begin{equation}
\left|h = \frac{q}{2}\, , \, q\right\rangle_{\text{NS}} \xrightarrow[\eta =1/2]{}\left|h' = \frac{c}{24}\, ,\, q'=q - \frac{c}{6}\right\rangle_{\text{R}} \,.
\end{equation}

Due to the presence of fermionic zero modes, in the Ramond case there are in fact two isomorphic irreducible representations of different chirality. One may then fix the ambiguity by demanding e.g. $G_{0}^{+}|h,q\rangle =0$ as part of the definition of the highest weight state. Besides this generic condition, Ramond ground states satisfy $G_{0}^{-}\left| h = \tfrac{c}{24}\,, q\right\rangle = 0\,$ as well. Just as for chiral primaries in the NS sector, the corresponding shortening of the representation is tied to a BPS bound being saturated. One also notes that the level-$n$ semiclassical bound \eqref{Relaxed Ramond bounds} can be obtained by performing $n$ (integer) units of spectral flow on the level-0 bound $h \geq c/24\,$.

We are now in a position to relate the semiclassical BPS bounds to the supersymmetric solutions in the $osp(2|2;\mathds{R})\oplus osp(2|2;\mathds{R})$ supergravity truncation studied in section \ref{sec:n2bhs}. As we saw there, these correspond to extremal BTZ black holes and smooth conical defects whose CFT-translated charges satisfy
 \begin{equation}\label{eq:bpsbtz33}
\text{BPS black holes }\left\{
\begin{array}{rcl}
\text{R:} & \displaystyle{\left(\frac{6h}{c}\,,\,\frac{6q}{c}\right) = \left(n^{2}+\frac{1}{4}\,,\, 2n\right),} & n \in \mathds{Z} \\ 
&&
\\
\text{NS:} &  \displaystyle{\left(\frac{6h}{c}\,,\,\frac{6q}{c}\right) = \left(r^{2}+\frac{1}{4}\,,\, 2r\right),} & r \in \mathds{Z} +\tfrac{1}{2}
\end{array}  
\right.\,.
\end{equation}
and
\begin{equation}\label{eq:bpscon44}
\text{BPS smooth conical defects }\left\{
\begin{array}{rcl}
\text{R:} & \displaystyle{\left(\frac{6h}{c}\,,\,\frac{6 q}{c}\right) = \left(\left(n+\frac{1}{2}\right)^{2}\,,\, 2n+1\right),} & n \in \mathds{Z} \\ 
&&
\\
\text{NS:} &  \displaystyle{\left(\frac{6h}{c}\,,\,\frac{6q}{c}\right) = \left(\left(r+\frac{1}{2}\right)^{2}\,,\, 2r+1\right),} & r \in \mathds{Z} +\tfrac{1}{2}
\end{array}  
\right.\,.
\end{equation}

\noindent Recall that black hole solutions are defined by a natural (extremality) bound relating $h$ and $q$ which comes from the demand that the entropy be real and positive. For conical defects, however, the bulk Chern-Simons theory imposes no obvious restrictions on $h$ and $q$ other than not being a black hole. In anticipation to the upcoming discussion, \eqref{eq:bpscon44} considers only those backgrounds that comply with the CFT unitarity condition $h\geq\frac{3q^2}{2c}$. This excludes the conical surpluses $\mathcal{L}<-\frac{1}{4}\,$, which can certainly be supersymmetric.

\begin{figure}[h]
\centering
\includegraphics[width=15.5cm]{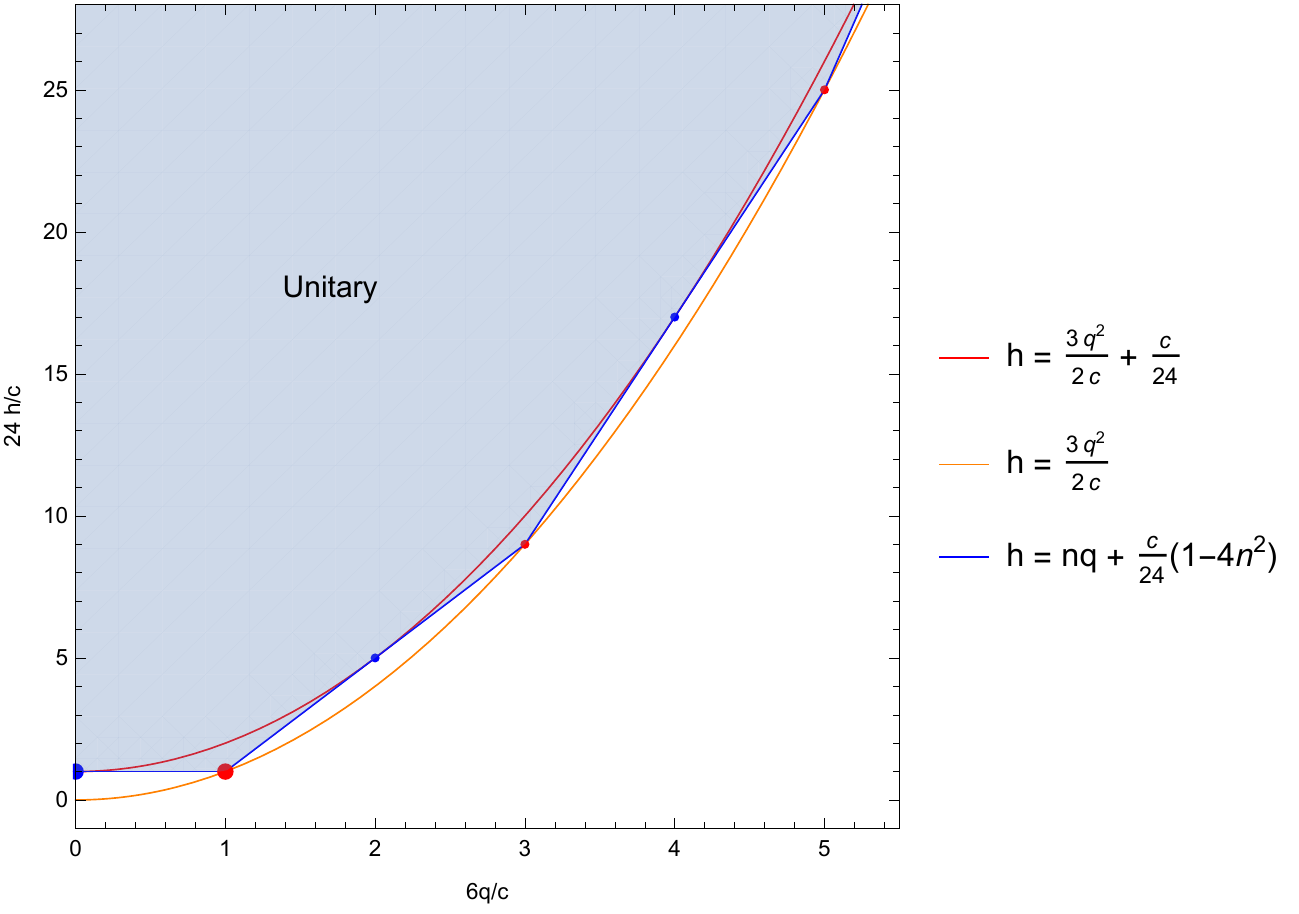}
\caption{Semiclassical unitarity and BPS bounds in the R sector of the $\cN =2$ super-Virasoro algebra. The blue dots correspond to the BPS charged BTZ black holes in \eqref{eq:bpsbtz33}. The red dots denote the BPS smooth conical defects in \eqref{eq:bpscon44}.}
\label{fig:RBounds}
\end{figure}

\begin{figure}[h]
\centering
\includegraphics[width=15.5cm]{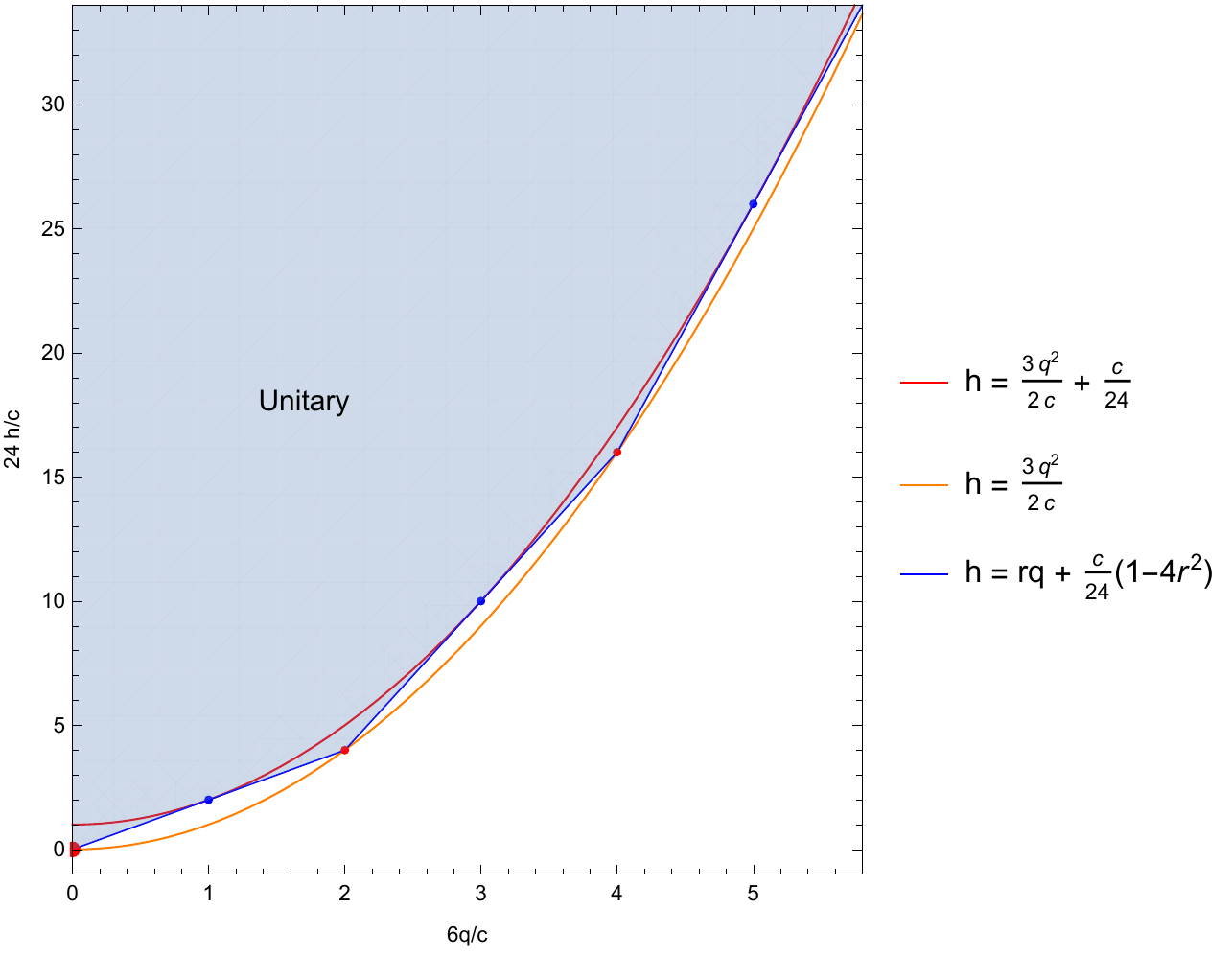}
\caption{Semiclassical unitarity and BPS bounds in the NS sector of the $\cN =2$ super-Virasoro algebra. The blue dots correspond to the BPS charged BTZ black holes in \eqref{eq:bpsbtz33}. The red dots denote the BPS smooth conical defects in \eqref{eq:bpscon44}.}
\label{fig:NSBounds}
\end{figure}

In order to compare bulk versus CFT calculations, in figures \ref{fig:RBounds} (R sector) and \ref{fig:NSBounds} (NS sector) we have plotted the bounds for the conformal weight $h$ as a function of the $U(1)$ charge $q\,$. Since the spectrum and bounds are symmetric under $q\to -q\,$, we have included positive charges only for ease of visualization. The blue straight lines in figure \ref{fig:RBounds} correspond to the semiclassical  R sector BPS bounds \eqref{Relaxed Ramond bounds} for $n=0,1,2,3\,$, while those in figure \ref{fig:NSBounds} represent the NS semiclassical BPS restrictions \eqref{Relaxed NS bounds} for $r=1/2,3/2,5/2\,$. The red parabola is the locus of extremal charged black hole states $h=\frac{3q^2}{2c}+\frac{c}{24}$ ($\mathcal{L}=0$), below which solutions cease to be black holes and become conical defects. The orange parabola corresponds to the level-0 unitarity bound $h=\frac{3q^2}{2c}$ ($\mathcal{L}=-\frac{1}{4}$). Both curves are spectral flow-invariant. The shaded area is the region allowed by unitarity. 

One notices that the quadratic constraint $h  \geq \frac{3q^{2}}{2c}$ is generically not the most stringent one. Therefore, in the semiclassical regime, the unitary domain is determined solely by the BPS bounds. We also note that black hole states, for any value of the temperature, are always allowed by unitarity. As mentioned above, there exist conical defects (surpluses) which lie outside the shaded region and are being omitted from the analysis.

The blue dots in figures \ref{fig:RBounds} and \ref{fig:NSBounds} correspond to the extremal supersymmetric black holes in \eqref{eq:bpsbtz33}. It is clear that they fall on the lines that describe BPS states in the semiclassical limit of the CFT. From figure \ref{fig:RBounds}, it is also evident that all such solutions in the Ramond sector are connected via integer units of spectral flow to the neutral extremal BTZ black hole, which is the uncharged Ramond sector ground state $|h=\frac{c}{24}, q = 0\rangle\,$ and is represented by the large blue dot. The NS black hole configurations in figure \ref{fig:NSBounds} are reached from this state by performing half-integer units of spectral flow.

Supersymmetric smooth conical defects are marked by red points in figures \ref{fig:RBounds} and \ref{fig:NSBounds}. In the R sector, all these states are connected by integer units of spectral flow to the maximally charged Ramond ground states $|h=\frac{c}{24}, q = \pm\frac{c}{6}\rangle\,$, labeled by the large red dot in figure \ref{fig:RBounds}. By the same token, the supersymmetric smooth conical defects in the NS sector are obtained from the NS vacuum $|h=0,q=0\rangle$, dual to global AdS$_{3}$ with vanishing $U(1)$ charge, by performing integer units of spectral flow. This state is the large red dot in figure \ref{fig:NSBounds}. In metric variables all these solutions correspond to a global AdS$_{3}$ metric with a constant Abelian gauge field whose holonomy supports the $U(1)$ charge \cite{Kraus:2006nb,Kraus:2006wn} and controls the periodicity of the Killing spinors.

A more refined analysis shows that the number of null states that appear when the BPS bounds are saturated is exactly the same as the one predicted by the study of Killing spinors in the bulk. We will omit this calculation for the case of $\mathcal{N}=2$ Super-Virasoro as it will be done in detail below in the context of the duality between $sl(3|2)\oplus sl(3|2)$ Chern-Simons theory and $\mathcal{W}_{(3|2)}$ CFTs.

\subsection{Higher spin BPS bounds}

Having reviewed the super-Virasoro case in detail, we will now show that among the extremal higher spin black holes and higher spin smooth conical defects studied in section \ref{sec:BHs}, those admitting Killing spinors correspond to states saturating BPS bounds of the $\cN=2$ super-$\cW_{3}$ algebra. To our knowledge the unitarity and BPS bounds for the $\cW_{(3|2)}$ algebra have not been derived in the literature, so we will start by computing some of these bounds at the full quantum level (finite $c$), and then studying their semiclassical limit. It is worth noticing that the structure of highest weight representations in the R sector of the $\cW_{(3|2)}$ algebra is rather involved due to the increased number of fermionic zero modes with respect to the super-Virasoro case. Consequently, our strategy here will consist on computing the BPS bounds in the NS sector and then obtain their R sector counterparts by performing half-integer units of spectral flow. As we shall see below, the bounds thusly obtained, while by no means exhaustive, will suffice for our purposes.

As before, we will assume the standard Hermiticity properties \eqref{super Virasoro Hermiticity} for the $\mathcal{N}=2$ superconformal generators. The structure of the normal-ordered composites in the OPE algebra then requires the following Hermiticity conditions for the fields in the higher spin multiplet:
\begin{equation}\label{Hermiticity second multiplet}
\left(W_{n}\right)^{\dagger} = \epsilon W_{-n}\,,\qquad \left(V_{n}\right)^{\dagger} = \epsilon V_{-n}\,,\qquad \left(U^{+}_{r}\right)^{\dagger} = \epsilon U^{-}_{-r}\,,
\end{equation}
\noindent with
\begin{equation}\label{epsilon choice}
\epsilon = 
\left\{
\begin{array}{cl}
+1 & \text{if } \kappa \in \mathds{R}\,\,  ( -6 < c < 1 \cup \frac{3}{2} < c <15)  \\ 
-1 & \text{if }  \kappa \text{ is imaginary } (\text{e.g. } c > 15)
\end{array} 
\right.
\end{equation}

\noindent and $\kappa$ defined as in \eqref{definition kappa}. Notice that the rescaled operators $\kappa V_n$, $\kappa W_n$ and $\kappa U^{\pm}_r$ satisfy the usual Hermiticity conditions, and one can rewrite the whole algebra in terms of these operators if desired. This implies in particular that $\kappa V_0$ and $\kappa W_0$ are Hermitian for any value of the central charge.

Constraints on the allowed values of the central charge that follow from unitarity considerations were discussed in \cite{Romans:1991wi}. 
It is important to emphasize that the semiclassical limit implies $c\to \infty$, which lies outside the unitarity window. One might be concerned that this will hinder the holographic interpretation. We will see, however, that the semiclassical limit of the unitarity bounds is exactly saturated by the relevant bulk solutions, and that the geometric (or rather topological) description of the theory in terms of the gravitational dual remains sensible in this limit.\footnote{ Whether this remains true after taking quantum corrections into account is of course a separate issue that goes beyond the scope of the present paper.}

\subsubsection{NS sector highest weight representations of the $\cW_{(3|2)}$ algebra}
Let us now briefly investigate the highest weight representations of the $\cW_{(3|2)}\,$ algebra in the NS sector. The zero modes are given by $L_0$, $J_0$, $V_0$ and $W_0\,$, and both $\{L_0,J_0,V_0\}$ and $\{L_0,J_0,W_0\}$ are sets of mutually commuting operators. However, $V_0$ and $W_0$ do not commute identically, as can be seen from \eqref{VW commutator}\footnote{ As usual for finitely-generated $\cW$-algebras, this feature comes about because of the requirement of closure of the algebra including a finite number of currents only, which forces the appearance of composite operators and the ensuing non-linearity.}
\begin{equation}\label{V and W zero modes}
\left[V_0,W_0\right] = \cC{4}_0\,.
\end{equation}

\noindent   One may then worry that it is not consistent to label the highest weight states by simultaneous eigenvalues of the full set $\{L_0,J_0,V_0,W_0\}\,$, but this expectation is not correct. In order to see this explicitly, we will start by constructing a highest weight representation of the set $\{L_0,J_0,V_0\}$ only.

Consider primary states $\hwr = \left|h,q,\qv\right\rangle$ obeying the usual highest weight conditions
\begin{alignat}{3}
 L_{n}\hwr  ={}& J_{n}\hwr  = V_{n}\hwr 0\,,&
    &\qquad\quad &
   n >{}&0~,
  \cr
 G^{\pm}_{r} \hwr ={}&
  0\, ,&
  &\qquad\quad &
   r >{}&0~.
   \label{hw conditions 2}
\end{alignat}

\noindent The mode algebra then implies\footnote{ Since $\left[G^{\pm}_{r},V_0\right] = \mp U^{\pm}_{r}$, the highest-weight conditions $V_0\hwr = v\hwr$ and $G^{\pm}_{r} \hwr =0$ for $r > 0$ imply  $U^{\pm}_{r} \hwr =0$ for $r > 0\,$. Using the latter condition and taking $\left\{G^{\pm}_{n-t},U^{\mp}_{t}\right\}\hwr =\left( \pm(3n-4t)V_{n} + 2W_{n}\right)\hwr $ with $t=\tfrac{1}{2}$ and $n>0$ yields $U^{\mp}_{\frac{1}{2}}G^{\pm}_{n-\frac{1}{2}}\hwr =2W_{n}\hwr\,$.  Since in the NS sector $G^{\pm}_{n-\frac{1}{2}}\hwr =0 $ for $n>0$, we conclude $W_{n}\hwr = 0$ for $n>0\,$ as well.} 
\begin{equation}
 U^{\pm}_{r}\hwr  =
0
 \quad \text{for}\quad
  r >0\qquad \text{and}\qquad W_{n}\hwr = 0\quad \text{for}\quad  n >0\,.
     \label{hw conditions 3}
\end{equation}

\noindent Next, define the state $\left|\phi_W\right\rangle \equiv W_0\hwr\,$. It is easy to see that it satisfies $L_0\left|\phi_W\right\rangle = h\left|\phi_W\right\rangle$ and $J_0\left|\phi_W\right\rangle = q \left|\phi_W\right\rangle$, as well as $L_{n}\left|\phi_W\right\rangle = J_n\left|\phi_W\right\rangle =0$ for $n>0\,$. Similarly, $G_{r}^{\pm}\left|\phi_W\right\rangle = 0$ for $r > 0$ follows from \eqref{hw conditions 3}. The action of the $V_{n}$ and $W_{n}$ modes on $\left|\phi_W\right\rangle\,$ for $n>0$ is sightly more complex. From  \eqref{WW commutator}-\eqref{VW commutator} we have 
\begin{align}
\left[V_n,W_0\right] 
={}&
 \cC{4}_{n} + 2n\cC{3}_{n} + n(n^2-1)\cC{1}_{n}~,
\cr
\bigl[W_{n}\, ,\, W_{0}\bigr]
  ={}&
  n\cB{4}_{n}+2n(n^{2}-4)\cB{2}_{n}~.
\end{align}
\noindent Since $V_{n>0}$ and $W_{n>0}$ annihilate the highest weight state $\hwr$, these equations translate into
\begin{align}
V_{n>0} \left|\phi_W\right\rangle ={}&
 \left(\cC{4}_{n} + 2n\cC{3}_{n} + n(n^2-1)\cC{1}_{n}\right)\hwr ~,
\cr
W_{n>0} \left|\phi_W\right\rangle ={}&
 \left( n\cB{4}_{n}+2n(n^{2}-4)\cB{2}_{n}\right)\hwr~.
\end{align}
\noindent Furthermore, since $\hwr$ is annihilated by the positive modes of all the currents, it follows that $\cC{1}_{n}\hwr = \cC{3}_{n}\hwr=\cC{4}_{n}\hwr=\cB{2}_{n}\hwr=\cB{4}_{n}\hwr =0$ for all $n>0$, so that
\begin{equation}
V_{n} \left|\phi_W\right\rangle=W_{n} \left|\phi_W\right\rangle =0 \qquad \forall \quad n>0\,.
\end{equation}

\noindent The last step is to check the action of the zero mode $V_0$ on $\left|\phi_W\right\rangle$. The complete mode expansion of $\cC{4}$ in the NS sector is given in \eqref{C4 NS modes}, and one easily verifies $ \cC{4}_0\hwr=0\,$. Hence,
\begin{equation}\label{V0 on W state}
V_0\left|\phi_W\right\rangle = \cC{4}_0\hwr + \qv \left|\phi_W\right\rangle = \qv \left|\phi_W\right\rangle\,.
\end{equation}
Using the above results, combined with $\bigl[G^\pm_r,V_0\bigr]= \mp U^{ \pm}_{r}$, gives $U^{ \pm}_{r}\left|\phi_W\right\rangle=0$ for $r>0$.

Summarizing, we have shown that in the NS sector the state $ \left|\phi_W\right\rangle = W_0\hwr$ carries the same quantum numbers $h,q,\qv$ and satisfies all the same highest weight conditions \eqref{hw conditions 2}-\eqref{hw conditions 3} as $\hwr$ itself. It follows that $\left|\phi_W\right\rangle$  must be proportional to $\hwr$ for the representation to be irreducible. In other words, \textit{if we start with a highest weight representation built from $\{L_0,J_0,V_0\}$ in the NS sector, it will automatically be a highest weight representation of the full set $\{L_0,J_0,V_0,W_0\}$ as well}. 
 
 Given the above analysis, from now on the NS primary $\hwr$ will be taken to be the highest weight state  $\hwr = |h,q,\qv,\qw\bigr\rangle$ satisfying 
\begin{align}\label{NS hw state 1}
L_0 \hwr ={}&
 h\hwr &  J_0\hwr  ={}&
  q\hwr \\
 V_0\hwr 
 ={}&
  \qv\hwr  &
 W_0 \hwr 
 ={}&
  \qw \hwr 
  \label{NS hw state 2}
\end{align}

\noindent with 
\begin{alignat}{3}
\label{NS hw conditions 1}
  L_{n}\hwr  ={}& J_{n}\hwr  =   V_{n}\hwr = W_{n}\hwr  = 0\,,
&
    &\qquad\quad &
   n >{}&0
   \\
    G^{\pm}_{r}\hwr  ={}&   U^{\pm}_{r}\hwr  =
  0\, ,&
  &\qquad\quad &
   r >{}&0\,.
   \label{NS hw conditions 2}
\end{alignat}

\subsubsection{$\cW_{(3|2)}$ BPS bounds and their semiclassical limit}
Having discussed the highest weight representations in the NS sector, we will now compute the basic BPS bound. At level 1/2 we find fermionic descendants
\begin{equation}
\text{level }1/2: \qquad \left | \alpha^{\pm}\right\rangle \equiv G^{\pm}_{-1/2}\hwr\quad\text{and}\quad \left | \beta^{\pm}\right\rangle \equiv U^{\pm}_{-1/2} \hwr \,.
\end{equation}

\noindent Since states with different $U(1)$ charges are orthogonal, we will focus on the charge $q+1$ sector for concreteness. The matrix of inner products at level 1/2 is then
\begin{equation}
K^{(1/2)}=\left(
\begin{array}{cccc}
\langle \alpha^+ | \alpha^+\rangle & \langle \alpha^+ | \beta^+\rangle \\ 
\langle \beta^+ | \alpha^+\rangle & \langle \beta^+ | \beta^+\rangle  
\end{array} 
\right)
=
\left(
\begin{array}{cccc}
2h-q & 2 (\qw-\qv)  \\ 
 2\epsilon(\qw-\qv) & \epsilon\left\langle \cD{4}_0-\cD{3}_0 -2\cD{2}_0 + 2\cD{1}_0\right\rangle 
\end{array} 
\right)\,,
\end{equation}

\noindent where the brackets in the right hand side indicate expectation value in the highest weight state $\hwr\,$ satisfying \eqref{NS hw state 1}-\eqref{NS hw conditions 2} and $\epsilon$ is defined as in \eqref{epsilon choice}. The explicit expressions for the composite operators $\cD{1},\cD{2},\cD{3}, \cD{4}$ as well as their action on a highest weight state are given in appendix \ref{app: Romans composites}. All in all we find that the level-1/2 BPS bound reads
\begin{equation}\label{level 1/2 bound}
\textrm{det}\,K^{(1/2)}=\epsilon\left[\left(2h-q \right) \left\langle \cD{4}_0-\cD{3}_0 -2\cD{2}_0 + 2\cD{1}_0\right\rangle -4\left( \qw-\qv\right)^2\right] \geq 0\,,
\end{equation}

\noindent with
\begin{align}
\left\langle \cD{1}_0\right\rangle
={}&
\frac{q}{4}~,
\cr
\left\langle \cD{2}_0\right\rangle
={}&
\frac{5c-3}{10(c-1)}h + \frac{\kappa}{5}\qv -\frac{3}{10(c-1)}q^{2}~,
\cr
\left\langle \cD{3}_0 \right\rangle
={}&
3\gamma \left(2(5c^{2}+9)qh - 3(4c+3)q^3+\frac{1}{2}(c-3)(13c-6)\frac{q}{3}\right) 
\nonumber\\
&
+ \frac{2\kappa}{5c-12}\Bigl(21 q \qv- (c+6)\qw\Bigr)~,
\cr
\left\langle \cD{4}_0\right\rangle
={}&
6\gamma \Biggl(9c(c-1)h\left(h+\frac{1}{5}\right) + \frac{1}{4}\left(5c^{2} - 51c + 18\right)\left(\frac{7h}{5} - \frac{q}{2}\right)
\nonumber
\cr
&
\qquad
+ 3(4c+3)q^{2}\left(\frac{1}{3}- h\right)
+
\frac{1}{4}(c^{2}-53c + 66)\frac{q^{2}}{5}
\Biggr)
\cr
&
+\frac{6\kappa}{(c+3)(5c-12)}\Biggl(18(c-1)\qv\left(h + \frac{1}{5}\right) + \frac{32}{5}(4c+3)\qv
-2(c-15)q\qw
\Biggr).
\nonumber
\end{align}
It is worth emphasizing that equation \eqref{level 1/2 bound} is the fully quantum (finite-$c$) level-1/2 bound. 

An associated family of quantum BPS bounds at higher Virasoro levels can be obtained by spectral flow,  i.e. replacing
\begin{equation}
\begin{aligned}\label{spec flow 1}
 h 
\to{}&
 h'-\eta q '+ \frac{\eta^{2}}{6}c 
\\
 q \to{}&
 q' -\frac{c}{3}\eta 
\\
 \qv \to{}&
 \qv'
\\
\qw \to{}&
\qw' - 2\eta \qv'
\\
\end{aligned}
\end{equation}

\noindent in \eqref{level 1/2 bound}, with the choice $\eta \in \mathds{Z}$ resulting in an NS bound and $\eta \in \mathds{Z} +\tfrac{1}{2}$ resulting in a Ramond sector bound. We do not expect such a bound to be the most stringent one at the corresponding level, however, because in the quantum regime it is necessary to consider all descendant states at any given level in order to obtain the full set of unitarity constraints. Nevertheless, in analogy with the super-Virasoro case, in the semiclassical limit we anticipate the spectral-flowed bound to capture all the relevant information. As we will see momentarily, this expectation is indeed confirmed via holography: all the bulk solutions admitting Killing spinors saturate the semiclassical spectral-flowed BPS bound which we present below.

In the semiclassical limit described in \ref{app:semiclassical}, the expectation value of the normal-ordered composites becomes
\begin{equation}
\begin{aligned}
\left\langle \cD{1}_0\right\rangle_{\text{semiclassical}}
={}&
\frac{q}{4}~,
\\
\left\langle \cD{2}_0\right\rangle_{\text{semiclassical}}
={}&
\frac{h}{2} + \frac{\kappa}{5}\qv -\frac{3}{10c}q^{2}~,
\\
\left\langle \cD{3}_0 \right\rangle_{\text{semiclassical}}
={}&
\frac{15}{c}q\left(h - \frac{6}{5}\frac{q^{2}}{c}\right) -\frac{2}{5}\kappa\left(w - 21\frac{q}{c}\qv \right)~,
\\
\left\langle \cD{4}_0\right\rangle_{\text{semiclassical}}
={}&
\frac{27}{c}h\left(h-\frac{4q^{2}}{3c}\right) + \frac{3q^{2}}{20c} + \frac{12\kappa}{5c}\left(9h\qv - q\qw\right)~,
\end{aligned}
\end{equation}
\noindent and in particular
\begin{align}\label{UU norm}
\left\langle \cD{4}_0-\cD{3}_0 -2\cD{2}_0 + 2\cD{1}_0\right\rangle_{\text{semiclassical}}
={}&
\frac{27}{2c} (2h-q)\left(h-\frac{4q^{2}}{3c} - \frac{1}{18}\left(q + \frac{2c}{3}\right)\right)
\nonumber\\
&
 + \frac{2\kappa}{5c}\Bigl(c\left(\qw-\qv\right) + 54h\qv - 3q(2\qw+7\qv)\Bigr).
\end{align}

\noindent Consequently, the semiclassical limit of the matrix of inner products at level $1/2$ is
\begin{empheq}{alignat=5}
	K^{(1/2)}_{\text{semiclassical}}&=\left(
	\begin{array}{cc}
		K^{(1/2)}_{1,1} & K^{(1/2)}_{1,2} \\
		K^{(1/2)}_{1,2} &  K^{(1/2)}_{2,2}
	\end{array}
	\right)\,,
\end{empheq}

\noindent where
\begin{equation}
\begin{aligned}
	K^{(1/2)}_{1,1}&=2h-q\,,
	\\
	K^{(1/2)}_{1,2}&=2\left(\qw-\qv\right)\,,
	\\
	K^{(1/2)}_{2,1}&=\epsilon K^{(1/2)}_{1,2}\,,
	\\
	K^{(1/2)}_{2,2}&=\epsilon\left[\frac{27\epsilon}{2c}K^{(1/2)}_{1,1}\left(\frac{1}{2}K^{(1/2)}_{1,1}+\frac{4\kappa}{5}\qv-\frac{1}{27c}\left(c-6q\right)^2\right)+\frac{\kappa}{5c}\left(c-6q\right)K^{(1/2)}_{1,2}\right]\,.
\end{aligned}
\end{equation}

\noindent Performing now $\eta$ units of spectral flow, the matrix elements become
\begin{equation}
\begin{aligned}\label{matrix elements level eta+1/2}
	K^{(\eta +1/2)}_{1,1}&=2h-q\left(1+2\eta\right)+\frac{c}{3}\eta\left(\eta+1\right)\,,
	\\
	K^{(\eta+1/2)}_{1,2}&=2\bigl(\qw-\qv\left(1+2\eta\right)\bigr)\,,
	\\
	K^{(\eta+1/2)}_{2,1}&=\epsilon K^{(\eta+1/2)}_{1,2}\,,
	\\
	K^{(\eta+1/2)}_{2,2}&=\epsilon\Bigg[\frac{27}{2c}K^{(\eta+1/2)}_{1,1}\left(\frac{1}{2}K^{(\eta+1/2)}_{1,1}+\frac{4\kappa}{5}\qv-\frac{1}{27c}\left(c\left(1+2\eta\right)-6q\right)^2\right)
	\\
	&+\frac{\kappa}{5c}\bigl(c\left(1+2\eta\right)-6q\bigr)K^{(\eta+1/2)}_{1,2}\Bigg]\,.
\end{aligned}
\end{equation}

\noindent All in all, the associated level-$(\eta+1/2)$ semiclassical bound reads\footnote{ It is worth emphasizing that this bound can be also obtained by taking the semiclassical limit directly in the spectral-flowed quantum bound obtained by replacing \eqref{spec flow 1} into \eqref{level 1/2 bound}. In this sense, spectral flow and the semiclassical limit can be said to commute.}
\begin{align}\label{main BPS bound}
0 &\leq
 \epsilon\Bigg[\frac{6}{c}\left(h+\frac{c}{6} \eta  (\eta +1)-\Bigl(\eta +\frac{1}{2}\Bigr) q\right)^2 \left[9 h-\Bigl(\eta +\frac{1}{2}\Bigr) q-\frac{12 q^2}{c}+\frac{c}{6}\left(\eta^2+\eta -2\right)\right]
 \nonumber
 \\
&\quad
+\frac{4 \kappa}{5 c}\left(h+\frac{c}{6}  \eta  \left(\eta +1\right)-\Bigl(\eta +\frac{1}{2}\Bigr) q\right) \biggl[\qv \Bigl( 54 h-21q (2 \eta +1)+ 5\eta  (\eta +1)c-c\Bigr)
\\
&
+\qw \Bigl(\left(2\eta +1\right)c-6 q\Bigr)\biggr]-4 \Bigl(\left(2 \eta +1\right)\qv-\qw\Bigr)^2\Bigg]~.
\nonumber
\end{align}

When written in terms of the bosonic zero modes of the CFT generators on the plane, the BPS bound \eqref{main BPS bound} does not look particularly illuminating. Fortunately, we will see that the bulk perspective provides an extremely elegant way of repackaging the information contained in \eqref{main BPS bound} in terms of the holonomy of the Drinfeld-Sokolov Chern-Simons connections. Furthermore, will use the resulting expression to conjecture the generic form of the relevant semiclassical BPS bound in any $\cN =2$ higher spin algebra. To this end, we first recall the holographic dictionary between the CFT zero modes and bulk charges in \eqref{eq:mapcharges}:
\begin{equation}
\begin{aligned}
	h&=\frac{c}{6}\left(\mathcal{L}+\frac{5}{3}Q_2+Q_1^2+\frac{1}{4}\right)\,,
	\cr
	q&=-\frac{c}{3}Q_1\,,
	\cr
	\kappa \qv &=-\frac{5c}{9}Q_2\,,
	\cr
	\kappa \qw &=\frac{5c}{3}\left(Q_3+\frac{2}{3}Q_1Q_2\right)\,,
\end{aligned}
\end{equation}
\noindent which we have further reparameterized in terms of the eigenvalues of the Drinfeld-Sokolov connection (i.e. in terms of its holonomy data) as in \eqref{charges as eigenvalues 1}. In terms of these eigenvalues and the bulk $U(1)$ charge $Q_{1}$, the BPS bound \eqref{main BPS bound} nicely factorizes as
\begin{align}
0 \leq{}&
 \frac{\epsilon c^{2}}{2304}\biggl[\lambda_{3}^{2} + \Bigl(1+2\eta -2i\left(\lambda_{1}+iQ_{1}\right)\Bigr)^{2}\biggr]
 \biggl[\lambda_{3}^{2} + \Bigl(1+2\eta + 2i\bigl(\lambda_{2}-iQ_{1}\bigr)\Bigr)^{2}\biggr]
\nonumber\\
&\quad 
\times
\biggl[\lambda_{3}^{2} + \Bigl(1+2\eta -2i\bigl(\lambda_{2}-\lambda_{1}+iQ_{1}\bigr)\Bigr)^{2}\biggr]\,.
\end{align}

\noindent The final step consists in recognizing that the above expression for the semiclassical BPS bound simplifies even further when written in terms of the frequencies \eqref{eq: Sec.3 frequencies} that control the periodicity of the Killing spinors in the bulk:
\newcommand*\widefboxc[1]{\fbox{\hspace{1em}#1\hspace{1em}}\,,}
\begin{empheq}[]{alignat=5}\label{final expression BPS bound}
	\nonumber
	0&\leq\textrm{det}\,K^{(\eta+1/2)}_{\text{semiclassical}}&\,=\frac{\epsilon c^2}{36}\left[i\omega_{14}-\left(\eta+\frac{1}{2}\right)\right]\left[i\omega_{15}-\left(\eta+\frac{1}{2}\right)\right]\left[i\omega_{24}-\left(\eta+\frac{1}{2}\right)\right] &
	\\
	&&\quad\;\,\times\,\,\left[i\omega_{25}-\left(\eta+\frac{1}{2}\right)\right]\left[i\omega_{34}-\left(\eta+\frac{1}{2}\right)\right]\left[i\omega_{35}-\left(\eta+\frac{1}{2}\right)\right] 
\end{empheq}

\noindent or more tersely
\begin{empheq}[box=\widefboxc]{alignat=5}\label{final expression BPS bound compact}
	0&\leq\textrm{det}\,K^{(\eta+1/2)}_{\text{semiclassical}}&\,=\frac{\epsilon c^2}{36}\prod_{i,\bar{j}}\left(i\omega_{i\bar{j}}-\left(\eta+\frac{1}{2}\right)\right)
\end{empheq}

\noindent where we recall that $\eta \in \mathds{Z}$ leads to a NS bound, while $\eta \in \mathds{Z} +\tfrac{1}{2}$ results in a Ramond sector bound. Equation \eqref{final expression BPS bound compact} is one of our main results: it makes clear, from a CFT perspective, what are the conditions satisfied by bulk solutions saturating BPS bounds, namely the quantization of the frequencies \eqref{eq: Sec.3 frequencies} associated with the holonomy of the Drinfeld-Sokolov boundary connection. We reiterate that the relevance of this quantization had been anticipated in \cite{Datta:2012km,Datta:2013qja} from the point of view of the bulk. Here we have recovered it from a CFT computation, which is reassuring and argues in favor of the consistency of the construction. 

Since the derivation of \eqref{final expression BPS bound} relied solely on the $\cN=2$ structure of the chiral algebra (the spectral flow automorphism in particular) and the properties of the Drinfeld-Sokolov connection, we can provide a conjecture for the form of the general semiclassical BPS bounds in any $\cN=2$ higher spin CFT whose chiral symmetries can be obtained via Hamiltonian reduction of current algebras. Using the notation introduced in \eqref{general frequencies}, quite naturally we expect the generalization of \eqref{final expression BPS bound} to be
\newcommand*\widefboxp[1]{\fbox{\hspace{1em}#1\hspace{1em}}\,.}
\begin{empheq}[box=\widefboxp]{alignat=5}\label{conjectured bound}
	\text{semiclassical BPS bounds:}\qquad \quad 0 \leq -c^{2}\prod_{\alpha_{j}^{\text{odd}} \in \text{ \{odd roots\}}}\left[\bigl\langle \vec{\Lambda}_{\phi}\,,\alpha_{j}^{\text{odd}}\bigr\rangle+i\left(\eta+\frac{1}{2}\right)\right]
\end{empheq}

\noindent where the precise form of the holonomy $\vec{\Lambda}_{\phi}$ and odd roots $\alpha_{j}^{\text{odd}}$ will of course depend on the concrete algebra under consideration and encodes the semiclassical symmetries of the boundary CFT (via Drinfeld-Sokolov reduction).

\subsection{Supersymmetric $sl(3|2)$ backgrounds from a CFT perspective}\label{sec:HSBHCFT}
Above we have shown that the saturation of the $\cW_{(3|2)}$ semiclassical BPS bounds yields exactly the same quantization conditions on the frequencies $\omega_{i\bar{j}}$ as the study of Killing spinors for the higher spin backgrounds introduced in \ref{sec:sl32bhs} \cite{Datta:2012km,Datta:2013qja}. The comparison between the two descriptions can be taken one step further by arguing that for each configuration preserving a supercharge in the bulk there corresponds a null state in the CFT with the same quantum labels. To this end we shall consider the matrix of inner products
\begin{empheq}{alignat=5}\label{matrix level eta+1/2}
	K^{(\eta+1/2)}_{\text{semiclassical}}&=\left(
	\begin{array}{cc}
		K^{(\eta+1/2)}_{1,1} & K^{(\eta+1/2)}_{1,2} \\
		K^{(\eta+1/2)}_{2,1} &  K^{(\eta+1/2)}_{2,2}
	\end{array}
	\right)\,,
\end{empheq}
whose entries are given by \eqref{matrix elements level eta+1/2}, and count the number of non-trivial eigenvectors with eigenvalue zero that appear when its determinant \eqref{final expression BPS bound} vanishes. We emphasize that this matrix does not capture all states at level $\eta+1/2$. It only includes states created by acting with the fermionic generators $G^+_{-\eta-1/2}$ and $U^+_{-\eta-1/2}$ on a highest weight vector. Nevertheless, we will see that analyzing this subsector is sufficient for our purposes.

\subsubsection*{Supersymmetric $sl{(3|2)}$ black holes}
Recall that black hole solutions have $\Delta_3\geq0$ and $\Delta_2\geq0$, property which is implemented by the reality conditions $\lambda_1=\lambda_2^*$ and $\lambda_3=\lambda_3^*$ on the eigenvalues of the connection. From the CFT perspective, this implies that the determinant \eqref{final expression BPS bound} is manifestly semi-positive or semi-negative definite depending on the sign of $\epsilon$. Indeed, realizing that the frequencies in \eqref{eq: Sec.3 frequencies} satisfy
\begin{empheq}{alignat=5}
	\omega_{14}&=-\overline{\omega}_{35}\,,
	&\qquad
	\omega_{15}&=-\overline{\omega}_{34}\,,
	&\qquad
	\omega_{24}&=-\overline{\omega}_{25}
\end{empheq}
identically in this sector, we find that
\begin{empheq}{alignat=5}\label{eq:dk}
	\textrm{det}\,K^{(\eta+1/2)}_{\text{semiclassical}}&=\frac{\epsilon c^2}{36}\left|i\omega_{14}-\left(\eta+\frac{1}{2}\right)\right|^2\left|i\omega_{15}-\left(\eta+\frac{1}{2}\right)\right|^2\left|i\omega_{24}-\left(\eta+\frac{1}{2}\right)\right|^2~.
\end{empheq}
The fact that $\epsilon = -1$ in the semiclassical regime is clearly tied to the theory not being unitary for large values of the central charge $c\,$. However, this issue does not affect the classification of null states, namely the zeroes of the determinant, which is what we matched onto our bulk results. Up to this issue, the fact that the determinant has nicely factorized in the black hole regime suggests that black holes are always allowed in the unitary regime of the dual CFT, and that the remanent of this fact as we push past the unitary regime is the overall sign in the determinant. For the $sl(2|1)$ theory, this is clearly seen in figure \ref{fig:RBounds} and figure \ref{fig:NSBounds} by noticing that the red parabola, which corresponds to the extremality bound $h=\frac{3q^2}{2c}+\frac{c}{24}$ ($\Delta_3=\Delta_2=0$), always lies above the orange parabola representing the unitarity condition $h=\frac{3q^2}{2c}$ and the blue lines that yield BPS bounds.

In the general case, vanishing of the above determinant allows for three possibilities:
\begin{enumerate}
 \item First, the condition $i\omega_{14}=\eta+\frac{1}{2}$ implies
\begin{empheq}{alignat=5}
	\lambda_1+\lambda_2-\lambda_3&=0\,,
	&\qquad
	\frac{i}{2}\left(\lambda_1-\lambda_2\right)-Q_1&=\eta+\frac{1}{2}\,,
\end{empheq}
where we used \eqref{eq: Sec.3 frequencies} to cast $w_{IJ}$ in terms of the eigenvalues of the $sl(3|2)$ connection. It turns out that the matrix of inner products \eqref{matrix level eta+1/2} is identically zero under these conditions. This means that there are 4 supersymmetric states at level $\eta+\frac{1}{2}$, corresponding to the ${1\over 3}$-BPS solutions in class {\bf I}, {\bf II} and {\bf VI}.\footnote{ At each level, the vector space of states has complex dimension $2$. When we count the number of supercharges we mean the number real parameters, hence the doubling. This only includes one $U(1)$-charged sector. However, there is no additional doubling when considering the charge conjugate states.}

\item The second possibility, $i\omega_{15}=\eta+\frac{1}{2}$, is equivalent to the first with $\lambda_3\rightarrow-\lambda_3$ and leads to the same conclusions. 

\item The last alternative is $i\omega_{24}=\eta+\frac{1}{2}$, which requires
\begin{empheq}{alignat=5}
	\lambda_3&=0\,,
	&\qquad	
	-i\left(\lambda_1-\lambda_2\right)-Q_1&=\eta+\frac{1}{2}\,.
\end{empheq}
In this case the inner product matrix reduces to
\begin{empheq}{alignat=5}
	K^{(\eta+1/2)}_{\text{semiclassical}}&=\frac{c}{36}\left(2\lambda_1-\lambda_2\right)\left(2\lambda_2-\lambda_1\right)\left(
	\begin{array}{cc}
		4 & 2\left(\lambda_1-\lambda_2\right) \\
		2\epsilon\left(\lambda_1-\lambda_2\right) & \epsilon\left(\lambda_1-\lambda_2\right)^2
	\end{array}
	\right)\,.
\end{empheq}
It is easy to check that this matrix always has only one non-trivial eigenvector with zero eigenvalue, corresponding to two supersymmetric states. These are the ${1\over 6}$-BPS solutions in classes {\bf II} and {\bf III}.
\end{enumerate}

Finally, notice that conditions 1 and 3 overlap when $i\omega_{14}=\eta_1+\frac{1}{2}$ and $i\omega_{24}=\eta_2+\frac{1}{2}$, or
\begin{empheq}{alignat=5}
	\lambda_1&=\frac{i}{3}\left(\eta_2-\eta_1\right)\,,
	&\qquad
	\lambda_2&=-\lambda_1\,,
	&\qquad
	\lambda_3&=0\,,
	&\qquad
	Q_1&=-\frac{1}{3}\left(2\eta_1+\eta_2\right)-\frac{1}{2}\,.
\end{empheq}
The corresponding matrix \eqref{matrix level eta+1/2} shows the emergence of 4 null states at level $\eta=\eta_1$ and two null states at level $\eta=\eta_2$. This scenario describes the extremal $1\over2$-BPS black holes in class \textbf{II}.
\subsubsection*{Supersymmetric smooth conical defects}
For smooth conical defects, the eigenvalues of the connection $a_{\phi}$ and the corresponding odd frequencies are given in \eqref{eq:lcd} and \eqref{eq:ccdd1}, respectively.
Contrary to what happens for black hole solutions, the determinant \eqref{final expression BPS bound} does not have a definite sign in the conical defect sector $\Delta_3<0$, $\Delta_2<0$. Unitarity should then discard some of the backgrounds. This is already true in the Super Virasoro case, where conical surpluses do not satisfy the constraint $\mathcal{L}\geq-\frac{1}{4}$. For theories with $\mathcal{W}_{(3|2)}$ symmetry, however, this interpretation is further complicated by the overall sign of the semiclassical determinant as we discussed above, so we will avoid making any further claims. Still, we can use our results to track null states in the CFT which one can argue are protected by supersymmetry.

The above configurations obviously make the determinant in \eqref{final expression BPS bound} vanish provided we set $i\omega_{i\bar{j}}=\eta+\frac{1}{2}$. Fixing the parameters $(n_i,n)$, there are six possible values of $\eta$ that make this happen, one for each frequency. Each of these values generically results in a matrix \eqref{matrix level eta+1/2} which has only one non-trivial null eigenvector. There are therefore 12 BPS states, two at each level. For particular choices of $n_i$ and $n$, e.g. the $sl(2|1)$ truncation $n_1=2n_3+\frac{2}{3}n$, $n_2=2n_3+\frac{4}{3}n$, it could be that some of the frequencies $w_{i\bar{j}}$ coincide. This reduces the number of levels where BPS states may appear. It is easy to check, however, that in this case there are still 12 states with vanishing norm, albeit with the degenerate levels displaying 4 instead of 2 of them. Notice that for black holes the reality conditions limit the number of independent frequencies to three, allowing such solutions to carry at most 6 supercharges.

This concludes our search for null vectors in the CFT. The results are in perfect agreement with the analysis of Killing spinors in the $sl(3|2)$ Chern-Simons theory; we can account for all supersymmetric solutions in terms of dual BPS states.

\section{Discussion}\label{sec:Disc}
The purpose of this work was manifold. Our first goal was to provide a definition of extremal black holes in three-dimensional Chern-Simons higher spin gravity that is in harmony with the topological nature of the theory and valid for any gauge algebra, including purely bosonic as well as supersymmetric cases. Secondly, in order to test the proposed notion and some of its consequences, we set out to compare under which conditions extremality implies supersymmetry (and vice-versa) in a theory of higher spin gravity that admits a Chern-Simons formulation. Our third objective was to understand the latter restrictions from the holographic perspective in a theory with $\mathcal{N}=2$ super-$\mathcal{W}_3$ symmetry, the simplest higher spin extension of the familiar $\mathcal{N}=2$ super-Virasoro algebra. This implied, in particular, the necessity to compute certain $\mathcal{W}_{(3|2)}$ BPS bounds which, to the extent of our knowledge, were absent from the literature.

Our main results can be summarized as follows:
\begin{enumerate}
\item A general definition of extremal black hole solutions was given in section \ref{sec:Definition}, which involves as its main ingredient the non-diagonalizability of the angular component of the connection $a_{\phi}$. We argued that extremality, as expressed in terms of non-trivial Jordan classes, is compatible with the notion of zero Hawking temperature of the solution. In particular, this definition was illustrated in theories based on two copies of the $sl(2)$ and $sl(3)$ gauge algebras, as well as their $\mathcal{N}=2$ supersymmetric extensions $sl(2|1)$ and $sl(3|2)$. Furthermore, we identified the appropriate real forms of the algebra for the corresponding Lorentzian theories in each case, namely $osp(2|2;\mathds{R})$ and $su(2,1|1,1)$ respectively. One interesting feature is that, unlike the zero-temperature BTZ solution, in higher spin gravity extremal black holes carry residual entropy in the extremal sector. 

\item To further study the consequences of our definition of extremality, in section \ref{sec:w32solutions} we provided a classification of $sl(3|2)$ backgrounds in terms of the Jordan class of the connection. Here, the discriminants $\Delta_3$ and $\Delta_2$ of the factorized characteristic polynomial of $a_{\phi}$ played a crucial role, allowing us to generalize the notion of hyperbolic, parabolic and elliptic conjugacy classes in $SL(2)$. In particular, we asserted that black holes solutions must have $\Delta_3\geq0$ and $\Delta_2\geq0$ in order for the sources to be real. In contrast, smooth conical defects have $\Delta_3<$ and $\Delta_2<0\,$. The limiting cases $\Delta_3=0$ and $\Delta_2=0$ correspond to extremal black holes.

\item An exhaustive survey of supersymmetric $sl(3|2)$ solutions was carried out in section \ref{sec:susybulk}. The objective was to contrast the conditions imposed by extremality vs. supersymmetry on the charges carried by the background. Most, but not all, supersymmetric solutions in $sl(3|2)\oplus sl(3|2)$ fall within the class of extremal solutions (see point 6 below). In contrast, all supersymmetric solutions in the $sl(2|1)\oplus sl(2|1)$ truncation are extremal. Also, we found that the extremal charged BTZ black holes are not the most supersymmetric black hole solutions in the higher spin theory; that title goes to solutions belonging to a different Jordan class. Supersymmetric conical defects were also analyzed for completeness.

\item We derived novel BPS conditions in theories with $\cW_{(3|2)}$ symmetry, for any value of the central charge. Futhermore, in the semiclassical limit ($\hbar\to 0$, $c\to \infty$) we provided a conjecture for the BPS bounds in a generic $\cW$-algebra with ${\cal N}=2$  supersymmetry.

\item Supersymmetry is generically unaffected by strong coupling regimes, and hence it is natural to ask if the supersymmetric solutions in the bulk can be mapped to BPS states in a CFT with  $\cW_{(3|2)}$ symmetry. In the semiclassical limit  we find perfect agreement between the bulk and boundary BPS conditions.  The non-linearities of the $\cW_{(3|2)}$ algebra are responsible of the non-trivial structure of the bounds, whereas in the bulk the Killing spinor conditions is governed by algebraic properties of $sl(3|2)$ generators. The agreement among the two is non-trivial. 

\item As alluded to above, most notably, we showed that there exist non-extremal solutions in the class of diagonalizable connections that posses 4 independent Killing spinors. This is, within the $sl(3|2)\,$ theory, we managed to construct a smooth higher spin black hole that is both at finite temperature and BPS. We described the features of this solution in section \ref{sec:susyw32bhs}. In addition to its well behaved bulk features, this solution is physical because we can identify an appropriate chiral primary in the CFT that carries the same charges.

\end{enumerate}

Let us now comment on a few implications of our results and compare them with the existing literature. Firstly, the study of extremality and supersymmetry in higher spin gravity made a feature rather evident: supersymmetry does not require extremality. This goes against our intuition in conventional supergravity, nevertheless our results are explicit and well founded. This decoupling between extremality and supersymmetry is evident when we compare them in $sl(3|2)$: this is clear from our BPS conditions require \eqref{eq:ffss} to vanish, whereas extremality is a condition that the discriminants $\Delta_{2,3}$ vanish. 

Our definition of extremality is motivated and inspired by geometrical properties of black holes and in particular BTZ. However, in a holographic context, it is also interesting to compare and contrast unitarity bounds in the CFT versus extremal limits in the bulk. For ${\cal N}=2$ super-Virasoro the unitarity region (blue shaded region in figure \ref{fig:RBounds} and \ref{fig:NSBounds}) shows that the extremal bound for the charged BTZ black hole lies within this region. It would be interesting to make this comparison in higher spin gravity. Taking again the example of $N=3$ studied in section  \ref{sec:examples} as a warm-up: there we found indication that all $\cW_3$ black holes are allowed by the unitarity bound in the CFT (as it happened for ${\cal N}=2$ super-Virasoro). It would be very interesting to study if this always the case: are $\cW_N$ unitarity bounds always compatible with our extremal bounds?. In contrast, for BPS conditions in the boundary and bulk the agreement is {\it exact} in the semiclassical limit.  It is reassuring that supersymmetry is robust and protected in these scenarios. It would be interesting to evaluate unitarity bounds in $\cW_N$ and carry this analysis explicitly. This is a question we leave for future work.

As mentioned throughout various sections, there is an extensive literature on supersymmetric properties of higher spin gravity prior to our work. For instance, the asymptotic symmetries of theories based on $sl(N|N-1)$ were discussed in \cite{Peng:2012ae,Tan:2012xi,Datta:2012km,Hikida:2012eu,Datta:2013qja,Chen:2013oxa} (in particular, the work \cite{Peng:2012ae} discusses both principal and non-principal embeddings). This collection of works contained as well a detailed account of supersymmetric conical defects solutions, and in \cite{Hikida:2012eu} these states were mapped to chiral primaries for the supersymmetric minimal model dualities in \cite{Creutzig:2011fe,Candu:2012tr,Creutzig:2012ar}. The role of the angular holonomy, and its non-trivial Jordan decomposition, was noted back then to be an important key to build Killing spinors. However the discussion was always tied to supersymmetry and not a more general concept of extremality. 

The construction of black holes  in $sl(N|N-1)$ Chern-Simons theory  was as well discussed in prior work, where the emphasis was placed on solutions at finite temperature. This is one of the main differences relative to our work: we treat the sources of the higher spin black hole as a deformation of the CFT Hamiltonian, whereas \cite{Peng:2012ae,Tan:2012xi,Datta:2012km,Datta:2013qja,Chen:2013oxa} utilize a holomorphic deformation of the solution. As it is clear in section \ref{sec:Definition} and \ref{sec:BHs}, the Hamiltonian formulation has the advantage of phrasing  both the extremal and supersymmetric conditions  as conditions among the charges only. This is the construction that natural fits from dual CFT point of view, and our excellent agreement with the CFT is unique from that point of view.   Recently, the concept of hypersymmetry in $osp(1|4)$ Chern Simons theory was studied  in  \cite{Henneaux:2015ywa} with the intention of understanding black holes and their symmetries.  Their definition of extremality relies on requiring that the entropy is real; this is along the lines of the bounds found in \cite{Gutperle:2011kf}. We expect this to agree with our definition, however there is room for ambiguities since the starting point is conceptually rather different. It would be interesting to exhibit the agreement (or lack thereof) explicitly. 

Perhaps a drawback of our choice of examples is that the $\cW_{(3|2)}$ algebra is not unitary in the large-$c$ limit. We have however seen that most of our conclusions are insensitive to this fact, and are expected to hold much more generally. An interesting future direction would be to extend our results to other setups where this problem does not arise, and to study in particular extremal black holes in theories based on infinite-dimensional algebras such as shs$[\lambda]$, as well as other generalizations considered recently in \cite{Gaberdiel:2014cha,Gaberdiel:2015mra}.  The technical difficulty within the shs$[\lambda]$ algebras is that in the bulk it is difficult to impose holonomy conditions; the appeal of course is that these are the relevant structures to study the tensionless limit of string theories and their dual CFT description. Within the class of non-supersymmetric Chern-Simons like theories of gravity, yet another future direction would be study our proposal of extremal black holes in lower spin gravity \cite{Hofman:2014loa} or in non-AdS like theories such as in \cite{Afshar:2012hc,Gary:2014mca}. These interesting problems will be addressed elsewhere.

\vskip 1cm
\centerline{\bf Acknowledgments}
\noindent 
We are specially grateful to Jan de Boer, Matthias Gaberdiel, Diego M. Hofman and Nabil Iqbal for illuminating discussions. It is also a pleasure to thank Marco Baggio, Shouvik Datta, Kevin Ferreira, Micha{\l} Heller, Christoph Keller, Maximilian Kelm, Per Kraus, Cheng Peng, Hai Siong Tan and Carl Vollenweider for helpful conversations and correspondence. J.I.J. would also like to thank the Pontificia Universidad Cat\'olica de Chile, the University of Amsterdam and Perimeter Institute for Theoretical Physics for hospitality while this work was in progress. M.B. is partially supported by FONDECYT Chile, grant \# 1141 221.   A.C. is supported by Nederlandse Organisatie voor Wetenschappelijk Onderzoek (NWO) via a Vidi grant  and in part by  the Delta ITP consortium, a program of the NWO that is funded by the Dutch Ministry of Education, Culture and Science (OCW).  A.F. is supported by CONICYT/PAI ``Apoyo al Retorno" grant 821320022.  The work of J.I.J. is partially supported by the Swiss National Science Foundation and the NCCR SwissMAP.

\appendix

\section{The $sl(3|2)$ superalgebra}\label{app: sl(3|2)}
In this appendix we collect some useful facts and formulae regarding the superalgebra $sl(3|2)$ and its real form $su(2,1|1,1)\,$.
\subsection{Definition and (anti-)commutation relations}\label{app: sl(3|2) definition and commutators} 
The superalgebra $sl(m|n;\mathds{C})$ consists of all complex $(m+n)\times(m+n)$ supermatrices of the form
\begin{empheq}{alignat=5}
	M&=\left(
	\begin{array}{c|c}
		A & B \\\hline
		C & D
	\end{array}
	\right)\,,
\end{empheq}
equipped with the supercommutator
\begin{empheq}{align}
	[M,M'\}&=\left(
	\begin{array}{c|c}
		AA'-A'A+BC'+B'C & AB'-A'B+BD'-B'D \\\hline
		CA'-C'A+DC'-D'C & CB'+C'B+DD'-D'D
	\end{array}
	\right)\,,
\end{empheq}
and satisfying the supertraceless condition
\begin{empheq}{alignat=5}
	\textrm{sTr}(M)&\equiv\textrm{Tr}\left[A\right]-\textrm{Tr}\left[D\right]&\,=0\,.
\end{empheq}
The complex dimension of the superalgebra is $(m+n)^2-1$. Elements with $B=0$ and $C=0$ are called even or bosonic, while those with $A=0$ and $D=0$ are termed odd or fermionic. The even subalgebra is $sl(m;\mathds{C})\oplus sl(n;\mathds{C})\oplus\mathds{C}$. In what follows we deal specifically with $m=3$ and $n=2$. We comment on the real form of interest below.

In the principal embedding of $sl(2|1)$ in $sl(3|2)$ \cite{Peng:2012ae,Chen:2013oxa}, the even-graded sector of the superalgebra is decomposed into the $sl(2)$ generators, $L_i$, one spin 1 multiplet, $A_i$, one spin 2 multiplet, $W_m$, and a spin 0 element, $J$. By spin we mean the $sl(2)$ spin, $S$. Within each multiplet the indices range from $-S$ to $S$, giving a total of $3+3+5+1=12$ bosonic generators. This structure is encoded in the commutation relations
\begin{empheq}{alignat=5}
	[L_i,L_j]&=(i-j)L_{i+j}\,,
	&\qquad
	[L_i,A_j]&=(i-j)A_{i+j}\,,
	&\qquad
	[L_i,W_m]&=(2i-m)W_{i+m}\,.
\end{empheq}
The remaining non-vanishing commutators read
\begin{empheq}{alignat=5}
	[A_i,A_j]&=(i-j)L_{i+j}\,,
	\qquad
	[A_i,W_m]=(2i-m)W_{i+m}\,,
	\\\nonumber
	[W_m,W_n]&=-\frac{1}{6}(m-n)(2m^2+2n^2-mn-8)(L_{m+n}+A_{m+n})\,.
\end{empheq}
Therefore, the bosonic part of the $sl(3|2)$ algebra is $sl(3)\oplus sl(2)\oplus u(1)$, where the $sl(3)$ is generated by $(L_i+A_i)/2$ together with $W_m$, while the $sl(2)$ corresponds to $(L_i-A_i)/2$. The latter factor should not be confused with the ``gravitational'' $sl(2)$ spanned by $L_i$. Of course, the Abelian generator is $J$. In turn, the odd-graded elements consist of two spin $1/2$ multiplets, $H_r$ and $G_r$, and two spin $3/2$ multiplets, $T_s$ and $S_s$;
\begin{empheq}{alignat=5}
	[L_i,G_r]&=\left(\frac{i}{2}-r\right)G_{i+r}\,,
	&\qquad
	[L_i,H_r]&=\left(\frac{i}{2}-r\right)H_{i+r}\,,
	\\\nonumber
	[L_i,S_s]&=\left(\frac{3i}{2}-s\right)S_{i+s}\,,
	&\qquad
	[L_i,T_s]&=\left(\frac{3i}{2}-s\right)T_{i+s}\,.
\end{empheq}
The number of fermionic generators is $2+2+4+4=12$. Their $U(1)$ charge assignments are
\begin{empheq}{alignat=5}
	[J,G_r]&=G_r\,,
	&\qquad
	[J,H_r]&=-H_r\,,
	&\qquad
	[J,S_s]&=S_s\,,
	&\qquad
	[J,T_s]&=-T_s\,.
\end{empheq}
Additionally, they satisfy
\begin{empheq}{alignat=5}
	[A_i,G_r]&=\frac{5}{3}\left(\frac{i}{2}-r\right)G_{i+r}+\frac{4}{3}S_{i+r}\,,
	\qquad
	[A_i,H_r]=\frac{5}{3}\left(\frac{i}{2}-r\right)H_{i+r}-\frac{4}{3}T_{i+r}\,,
	\\\nonumber
	[A_i,S_s]&=\frac{1}{3}\left(\frac{3i}{2}-s\right)S_{i+s}-\frac{1}{3}\left(3i^2-2is+s^2-\frac{9}{4}\right)G_{i+s}\,,
	\\\nonumber
	[A_i,T_s]&=\frac{1}{3}\left(\frac{3i}{2}-s\right)T_{i+s}+\frac{1}{3}\left(3i^2-2is+s^2-\frac{9}{4}\right)H_{i+s}\,,
\end{empheq}
\begin{empheq}{alignat=5}
	[W_m,G_r]&=-\frac{4}{3}\left(\frac{m}{2}-2r\right)S_{m+r}\,,
	\qquad
	[W_m,H_r]=-\frac{4}{3}\left(\frac{m}{2}-2r\right)T_{m+r}\,,
	\\\nonumber
	[W_m,S_s]&=-\frac{1}{3}\left(2s^2-2sm+m^2-\frac{5}{2}\right)S_{m+s}
	\\\nonumber
	&-\frac{1}{6}\left(4s^3-3s^2m+2sm^2-m^3-9s+\frac{19}{4}m\right)G_{m+s}\,,
	\\\nonumber
	[W_m,T_s]&=\frac{1}{3}\left(2s^2-2sm+m^2-\frac{5}{2}\right)T_{m+s}
	\\\nonumber
	&-\frac{1}{6}\left(4s^3-3s^2m+2sm^2-m^3-9s+\frac{19}{4}m\right)H_{m+s}\,,
\end{empheq}
together with the anti-commutation relations
\begin{empheq}{alignat=5}
	\{G_r,H_s\}&=2L_{r+s}+(r-s)J\,,
	\\\nonumber
	\{G_r,T_s\}&=-\frac{3}{2}W_{r+s}+\frac{3}{4}(3r-s)A_{r+s}-\frac{5}{4}(3r-s)L_{r+s}\,,
	\\\nonumber
	\{H_r,S_s\}&=-\frac{3}{2}W_{r+s}-\frac{3}{4}(3r-s)A_{r+s}+\frac{5}{4}(3r-s)L_{r+s}\,,
	\\\nonumber
	\{S_r,T_s\}&=-\frac{3}{4}(r-s)W_{r+s}+\frac{1}{8}\left(3s^2-4rs+3r^2-\frac{9}{2}\right)\left(L_{r+s}-3A_{r+s}\right)
	\\\nonumber
	&-\frac{1}{4}(r-s)\left(r^2+s^2-\frac{5}{2}\right)J\,.
\end{empheq}
Notice that the elements $L_i$, $J$, $H_r$ and $G_r$ generate $sl(2|1)\subset sl(3|2)$, while $osp(2|1)\subset sl(2|1)$ is spanned by $L_i$ and $(H_r+G_r)/\sqrt{2}$.

Another useful basis for $sl(3|2)$, which we use in the analysis of Killing spinors, can be constructed from the twenty-five $5\times5$ matrices
\begin{empheq}{alignat=5}
	\left(e_{IJ}\right)_{KL}&=\delta_{IK}\delta_{JL}\,.
\end{empheq}
It is convenient to split the index $I=\left(1,2,3,4,5\right)$ into $I=\left(i,\bar{i}\right)$, where $i=\left(1,2,3\right)$ and $\bar{i}=\left(4,5\right)$. Then, a basis for the even elements of the superalgebra is given by
\begin{empheq}{alignat=5}
	E_{ij}&=e_{ij}-\delta_{ij}\mathds{1}\,,
	\cr
	E_{\bar{i}\bar{j}}&=e_{\bar{i}\bar{j}}+\delta_{\bar{i}\bar{j}}\mathds{1}\,,
\end{empheq}
while the odd elements are spanned by
\begin{empheq}{alignat=5}
	E_{i\bar{j}}&=e_{i\bar{j}}\,,
	\cr
	E_{\bar{i}j}&=e_{\bar{i}j}\,.
\end{empheq}
Notice that, as expected, the above basis is overcomplete since
\begin{empheq}{alignat=5}
	\sum_iE_{ii}&=-\sum_{\bar{i}}E_{\bar{i}\bar{i}}\,.
\end{empheq}
This matrix actually corresponds to the $U(1)$ generator in the superalgebra. The (anti-)commutation relations in this basis can be found in \cite{Datta:2012km}. 
\subsection{Matrix representation}\label{app: sl(3|2) representation}
For convenience, we have chosen to work in a representation where all matrices are real and satisfy
\begin{empheq}{alignat=5}
	L_i^{\dagger}=(-1)^iL_{-i}\,,
	&\qquad
	A_i^{\dagger}=(-1)^iA_{-i}\,,
	&\qquad
	W_m^{\dagger}=(-1)^mW_{-m}\,,
\end{empheq}
and
\begin{empheq}{alignat=5}
	H_r^{\dagger}&=(-1)^{r+\frac{1}{2}}G_{-r}\,,
	&\qquad
	T_s^{\dagger}&=(-1)^{s+\frac{1}{2}}S_{-s}\,.
\end{empheq}
The generators in this basis are \cite{Chen:2013oxa}
\begin{empheq}{alignat=5}
	L_1&=\left(
	\begin{array}{ccc|cc}
		0 & 0 & 0 & 0 & 0 \\
		\sqrt{2} & 0 & 0 & 0 & 0 \\
		0 & \sqrt{2} & 0 & 0 & 0 \\\hline
		0 & 0 & 0 & 0 & 0 \\
		0 & 0 & 0 & 1 & 0 \\
	\end{array}
	\right)\,,
	&\qquad
	L_0&=\left(
	\begin{array}{ccccc}
		1 & 0 & 0 & 0 & 0 \\
		0 & 0 & 0 & 0 & 0 \\
		0 & 0 & -1 & 0 & 0 \\
		0 & 0 & 0 & \frac{1}{2} & 0 \\
		0 & 0 & 0 & 0 & -\frac{1}{2} \\
	\end{array}
	\right)\,,
\end{empheq}
\begin{empheq}{alignat=5}
	A_1&=\left(
	\begin{array}{ccccc}
		0 & 0 & 0 & 0 & 0 \\
		\sqrt{2} & 0 & 0 & 0 & 0 \\
		0 & \sqrt{2} & 0 & 0 & 0 \\
		0 & 0 & 0 & 0 & 0 \\
		0 & 0 & 0 & -1 & 0 \\
	\end{array}
	\right)\,,
	&\qquad
	A_0&=\left(
	\begin{array}{ccccc}
		1 & 0 & 0 & 0 & 0 \\
		0 & 0 & 0 & 0 & 0 \\
		0 & 0 & -1 & 0 & 0 \\
		0 & 0 & 0 & -\frac{1}{2} & 0 \\
		0 & 0 & 0 & 0 & \frac{1}{2} \\
	\end{array}
	\right)\,,
\end{empheq}
\begin{empheq}{alignat=5}
	W_2&=\left(
	\begin{array}{ccccc}
		0 & 0 & 0 & 0 & 0 \\
		0 & 0 & 0 & 0 & 0 \\
		4 & 0 & 0 & 0 & 0 \\
		0 & 0 & 0 & 0 & 0 \\
		0 & 0 & 0 & 0 & 0 \\
	\end{array}
	\right)\,,
	&\qquad
	W_1&=\left(
	\begin{array}{ccccc}
		0 & 0 & 0 & 0 & 0 \\
		\sqrt{2} & 0 & 0 & 0 & 0 \\
		0 & -\sqrt{2} & 0 & 0 & 0 \\
		0 & 0 & 0 & 0 & 0 \\
		0 & 0 & 0 & 0 & 0 \\
	\end{array}
	\right)\,,
\end{empheq}
\begin{empheq}{alignat=5}
	W_0&=\left(
	\begin{array}{ccccc}
		\frac{2}{3} & 0 & 0 & 0 & 0 \\
		0 & -\frac{4}{3} & 0 & 0 & 0 \\
		0 & 0 & \frac{2}{3} & 0 & 0 \\
		0 & 0 & 0 & 0 & 0 \\
		0 & 0 & 0 & 0 & 0 \\
	\end{array}
	\right)\,,
	&\qquad
	J&=\left(
	\begin{array}{ccccc}
		2 & 0 & 0 & 0 & 0 \\
		0 & 2 & 0 & 0 & 0 \\
		0 & 0 & 2 & 0 & 0 \\
		0 & 0 & 0 & 3 & 0 \\
		0 & 0 & 0 & 0 & 3 \\
	\end{array}
	\right)\,,
\end{empheq}
\begin{empheq}{alignat=5}
	G_{\frac{1}{2}}&=\left(
	\begin{array}{ccc|cc}
		0 & 0 & 0 & 0 & 0 \\
		0 & 0 & 0 & 0 & 0 \\
		0 & 0 & 0 & 0 & 0 \\\hline
		2 & 0 & 0 & 0 & 0 \\
		0 & \sqrt{2} & 0 & 0 & 0 \\
	\end{array}
	\right)\,,
	&\qquad
	H_{\frac{1}{2}}&=\left(
	\begin{array}{ccccc}
		0 & 0 & 0 & 0 & 0 \\
		0 & 0 & 0 & \sqrt{2} & 0 \\
		0 & 0 & 0 & 0 & 2 \\
		0 & 0 & 0 & 0 & 0 \\
		0 & 0 & 0 & 0 & 0 \\
	\end{array}
	\right)\,,
\end{empheq}
\begin{empheq}{alignat=5}
	S_{\frac{3}{2}}&=\left(
	\begin{array}{ccccc}
		0 & 0 & 0 & 0 & 0 \\
		0 & 0 & 0 & 0 & 0 \\
		0 & 0 & 0 & 0 & 0 \\
		0 & 0 & 0 & 0 & 0 \\
		-3 & 0 & 0 & 0 & 0 \\
	\end{array}
	\right)\,,
	&\qquad
	S_{\frac{1}{2}}&=\left(
	\begin{array}{ccccc}
		0 & 0 & 0 & 0 & 0 \\
		0 & 0 & 0 & 0 & 0 \\
		0 & 0 & 0 & 0 & 0 \\
		-1 & 0 & 0 & 0 & 0 \\
		0 & \sqrt{2} & 0 & 0 & 0 \\
	\end{array}
	\right)\,,
\end{empheq}
\begin{empheq}{alignat=5}
	T_{\frac{3}{2}}&=\left(
	\begin{array}{ccccc}
		0 & 0 & 0 & 0 & 0 \\
		0 & 0 & 0 & 0 & 0 \\
		0 & 0 & 0 & -3 & 0 \\
		0 & 0 & 0 & 0 & 0 \\
		0 & 0 & 0 & 0 & 0 \\
	\end{array}
	\right)\,,
	&\qquad
	T_{\frac{1}{2}}&=\left(
	\begin{array}{ccccc}
		0 & 0 & 0 & 0 & 0 \\
		0 & 0 & 0 & -\sqrt{2} & 0 \\
		0 & 0 & 0 & 0 & 1 \\
		0 & 0 & 0 & 0 & 0 \\
		0 & 0 & 0 & 0 & 0 \\
	\end{array}
	\right)\,.
\end{empheq}

\subsection{The real form $su(2,1|1,1)$}\label{app: sl(3|2) real form} 
As listed in \cite{Frappat:1996pb}, the real forms associated with $sl(3|2;\mathds{C})$ are:
\begin{empheq}{alignat=5}
	&sl(3|2;\mathds{R})\supset sl(3;\mathds{R})\oplus sl(2;\mathds{R})\oplus\mathds{R}\,,
	\cr
	&sl(3|2;\mathds{H})\supset su^*(3)\oplus su^*(2)\oplus\mathds{R}\,,
	\cr
	&su(p,3-p|q,2-q)\supset su(p,3-p)\oplus su(q,2-q)\oplus i\mathds{R}\,.
\end{empheq}
In the present work are interested in the last possibility with $p=2$ and $q=1$, the main reason being that it is this choice of bulk superalgebra that makes natural contact with the boundary $\mathcal{W}_{(3|2)}$ theory \cite{BetoJuan}. In particular, notice that $su(p,3-p|q,2-q)$ is the only real form with a compact $u(1)$ generator.

The superalgebra $su(2,1|1,1)\supset su(2,1)\oplus su(1,1)\oplus i\mathds{R}$ is defined as the set of supertraceless $5\times5$ supermatrices $M$ satisfying
\begin{empheq}{alignat=5}
	M^{\dagger}K+KM&=0\,,
\end{empheq}
where $K$ is a non-degenerate Hermitian form of signature $(2,1|1,1)$. One can check that in our representation of $sl(3|2)$ the generators
\begin{empheq}{alignat=5}
	L_i,\quad A_i,\quad iW_m,\quad iJ,\,
\end{empheq}
and
\begin{empheq}{alignat=5}
	e^{i\pi/4}\left(H_r+G_r\right),\quad e^{i\pi/4}i\left(H_r-G_r\right),\quad e^{3i\pi/4}\left(T_s+S_s\right),\quad e^{3i\pi/4}i\left(T_s-S_s\right)\,,
\end{empheq}
satisfy the above property with
\begin{empheq}{alignat=5}
	K&=\left(
	\begin{array}{ccccc}
		0 & 0 & -1 & 0 & 0 \\
		0 & 1 & 0 & 0 & 0 \\
		-1 & 0 & 0 & 0 & 0 \\
		0 & 0 & 0 & 0 & i \\
		0 & 0 & 0 & -i & 0 \\
	\end{array}
	\right)\,.
\end{empheq}
Notice that $K$ has the correct eigenvalues. Therefore, these particular combinations of generators, with the above pre-factors included, form a basis for the \emph{real} superalgebra $su(2,1|1,1)$.

In the analysis of Killing spinors in the $sl(3|2)$ theory we have decomposed the fermionic parameter as
\begin{empheq}{align}
	\epsilon&=\epsilon^-+\epsilon^+\,,
\end{empheq}
where $\epsilon^{\pm}$ are $U(1)$ eigenstates. Demanding that this matrix belong to $su(2,1|1,1)$ implies that
\begin{empheq}{alignat=5}
	\epsilon^{\dagger}K+K\epsilon&=0
	&\qquad\Leftrightarrow\qquad
	{\epsilon^{\pm}}^{\dagger}&=-K\epsilon^{\mp}K\,.
\end{empheq}


\section{The $\cW_{(3|2)}$ algebra}\label{app: W32}

In what follows we will briefly discuss some basic aspects of the $\cW_{(3|2)}$ algebra, and collect some useful formulae. The material below follows \cite{Romans:1991wi,Candu:2012tr} closely. As it is common usage in the literature, we shall use the terms ``spin" and 	``conformal dimension/weight" interchangeably.

\subsection{Commutator algebra and spectral flow}

Before diving into the higher spin algebra, let us recall the structure of the $\cN=2$ super-Virasoro algebra. In addition to the stress tensor $T$, this algebra contains a weight-1 $U(1)$ current $J$ and two weight-3/2 fermionic currents $G^+$ and $G^{-}$ with $U(1)$ charges $+1$ and $-1$, respectively. The corresponding commutators are given by
\begin{align}\label{superVir comms}
\bigl[L_m,L_n\bigr]
 ={}&
  (m-n)L_{m+n} +\tfrac{c}{12}m(m^2-1)\delta_{m+n,0}
  &
 \\
\bigl[L_m,J_n\bigr]
={}&
 -n J_{m+n}
 &
  \bigl[L_m,G^\pm_r\bigr]
  ={}&
   \left(\tfrac{m}{2}-r\right) G^\pm_{m+r}
   \\
\bigl[J_m,J_n\bigr]
={}&
\tfrac{c}{3}m\delta_{m+n,0}
& 
 \bigl[J_m,G^\pm_r\bigr]
  ={}&
  \pm G^\pm_{m+r}
  \\
\bigl\{G^+_r,G^-_s\bigr\}
 ={}&
2L_{r+s}+(r-s)J_{r+s} +\tfrac{c}{3}\left(r^2-\tfrac{1}{4}\right)\delta_{r+s,0}
&
\bigl\{G^\pm_r,G^\pm_s\bigr\}
={}&
0\ .
\label{superVir comms 2}
\end{align}

In addition to the super-Virasoro operators, the $\cW_{(3|2)}$ algebra contains an additional $\cN=2$ multiplet generated by a dimension-$2$ superconformal primary. We shall adopt the notation $\left\{V,U^{+},U^{-},W\right\}$ for the currents in this multiplet. $V$ has conformal dimension $2$ and $U(1)$ charge zero, $U^{\pm}$ have weight $5/2$ and $U(1)$ charge $\pm 1$, and $W$ has dimension $3$ and $U(1)$ charge zero. The commutators between the super-Virasoro currents and the higher spin multiplet fields are
\begin{align}\label{superVirW2comms}
\bigl[L_m,V_n\bigr] 
={}&
(m-n)V_{m+n}
&
\bigl[L_m,U^{\pm}_r\bigr]
={}&
\left(\tfrac{3}{2}m-r\right) U^{\pm}_{m+r}
\\
\bigl[L_m,W_n\bigr]
={}&
(2 m-n)W_{m+n}
&
  \bigl[J_m,V_n\bigr]
  ={}&
  0
   \\
\bigl[J_m,W_n\bigr]
={}&
 2m\, V_{m+n}
 &
\bigl[J_m,U^{ \pm}_r\bigr]
={}&
\pm U^{ \pm}_{m+r} 
\\
\bigl\{G^\pm_r,U^{\mp}_t\bigr\}
={}&
 \pm (3r-t) V_{r+t}+2 W_{r+t}
 &
\bigl[G^\pm_r,V_n\bigr]
={}&
 \mp U^{ \pm}_{r+n}\label{}
 \\
 \bigl[G^\pm_r, W_n\bigr]
 ={}&
   \bigl(2r-\tfrac{1}{2}n\bigr) U^{ \pm}_{r+n}
  &
\big\{G^\pm_r,U^{ \pm}_t\bigr\}
={}&
 0\ .
\end{align}

\noindent in agreement with the $\cN=2$ supersymmetric structure.

Finally, the commutators of the higher spin multiplet operators with themselves were given by Romans in \cite{Romans:1991wi}
\begin{align}
 \bigl[V_{m}\, ,\, V_{n}\bigr]
  ={}& 
 (m-n)\cA{2}_{m+n}+\tfrac{c}{12} m(m^2-1)\delta_{m+n,0}
  \\
  \nonumber\\
 \bigl[W_{m}\, ,\, W_{n}\bigr]
  ={}&
   \tfrac{c}{48} m(m^2-1)(m^2-4)\delta_{m+n,0}+(m-n)\cB{4}_{m+n}
   \\
&
+(m-n)(2m^2-mn+2n^2-8)\cB{2}_{m+n}
   \label{WW commutator}
 \\
   \nonumber\\
   \label{VW commutator}
 \bigl[V_{m}\, ,\, W_{n}\bigr]
 ={}&
  \cC{4}_{m+n}+ (2m-n)\cC{3}_{m+n}
+(6m^2-3mn+n^2-4)\underbrace{\cC{2}_{m+n}}_{=0}+m(m^2-1)\cC{1}_{m+n} 
\\
  \nonumber\\
 \bigl\{U^{+}_{r}\, ,\, U^{-}_{s}\bigr\}
 ={}&
  \cD{4}_{r+s}  +(r-s)\cD{3}_{r+s}+\left(3r^2-4rs+3s^2-\tfrac{9}{2}\right)\cD{2}_{r+s}
  \\
&
 +(r-s)\left(r^2+s^2-\tfrac{5}{2}\right)\cD{1}_{r+s}
+ \tfrac{c}{12}\left( r^2-\tfrac{1}{4}\right)\left( r^2-\tfrac{9}{4}\right)\delta_{r+s,0}
 \\
   \nonumber\\
\bigl\{U^{\pm}_{r}\, ,\, U^{\pm}_{s}\bigr\}
={}&
 \left(\cE{4}{\pm}\right)_{r+s}
  \\
 \nonumber\\
\bigl[V_{m}\, ,\, U^{\pm}_{r}\bigr]
={}&
 \left(\cPhi{7/2}{\pm}\right)_{m+r} +\left( \tfrac{3}{2}m - r\right) \left(\cPhi{5/2}{\pm}\right)_{m+r}
+ \left( 3m^2 - 2mr +r^2-\tfrac{9}{4}\right)\left( \cPhi{3/2}{\pm}\right)_{m+r}
 \end{align}
 \begin{align}
\bigl[U^{\pm}_{r}\, ,\, W_{m}\bigr]
 ={}&
  \left(\cPsi{9/2}{\pm}\right)_{r+m} + \left( 2r-\tfrac{3}{2}m \right)\left(\cPsi{7/2}{\pm}\right)_{r+m}
+ \left( 2r^2 - 2rm +m^2 -\tfrac{5}{2}\right) \left(\cPsi{5/2}{\pm}\right)_{r+m}
\\
&
+ \left( 4r^3 -3r^2m +2rm^2 - m^3 - 9r +\tfrac{19}{4}m  \right)\left(\cPsi{3/2}{\pm}\right)_{r+m}\ .
\end{align}

 \noindent Here $\cA{s}$, $\cB{s}$, $\cC{s}$, $\cD{s}$, $\cE{s}{\pm}$, $\cPhi{s}{\pm}$, $\cPsi{s}{\pm}$ are normal-ordered composite operators built out of primary and quasi-primary operators, with their precise form fixed by Jacobi identities (the explicit expressions have been given in \cite{Romans:1991wi}), and the self-coupling $\kappa$ of the higher spin multiplet with itself is fixed in terms of the central charge as 
\begin{equation}\label{definition kappa}
\kappa =\pm  \frac{(c+3)(5c-12)}{\sqrt{2(c+6)(c-1)(2c-3)(15-c)}} \,.
\end{equation}

\noindent As pointed out in \cite{Romans:1991wi}, the sign ambiguity in $\kappa$ corresponds merely to the freedom of simultaneously flipping the sign of all fields in the higher spin multiplet. Note also that $\kappa$ is real only for $-6 <c<1$ or $\frac{3}{2} < c < 15\,$, and in particular purely imaginary as $c \to \infty\,$. For $c >15$ $\left\{V,U^{+},U^{-},W\right\}$ are anti-Hermitian while the remaining currents are Hermitian. Alternatively, the full algebra can be written in terms of operators with standard Hermiticity properties by rescaling the higher spin multiplet currents by $\kappa$.

As noted in \cite{Romans:1991wi} the redefinition 
\begin{align}
 L'_{n} 
={}&
 L_n + \eta J_{n} + \frac{\eta^{2}}{6}c \delta_{n,0}
\\
 J'_{n}  ={}&
 J_n + \frac{c}{3}\eta \delta_{n,0}
\\
 G_{r}^{\pm '} 
={}&
 G^{\pm}_{r\pm \eta}
 \\
 V'_{n} ={}&
  V_{n}
 \\
 U^{\pm '}_{r} ={}&
 U^{\pm}_{r\pm \eta}
\\
W'_{n} ={}&
 W_{n} + 2\eta V_{n}
\end{align}

\noindent is an automorphism of the full $\mathcal{W}_{(3|2)}$ algebra for any $\eta\,$. We have exploited this property extensively in our calculations.

\subsection{The semiclassical limit}\label{app:semiclassical}

Our discussion has been fully quantum so far. However, in order to compare with the results from the bulk calculations in the main text, we need to consider the semiclassical limit of the $\cW_{(3|2)}$ algebra whose commutation relations we have given above. Roughly speaking this entails a ``large-$c$" limit, but the correct procedure is a bit more subtle than a naive expansion in $1/c\,$. The appropriate limiting procedure has been discussed in \cite{Candu:2013uya},\footnote{ We thank Carl Vollenweider for helpful discussions on this matter.} which we follow here. 

In the present context, we are instructed to first rescale the currents (denoted collectively by $J_{s}(z)$) and central charge as
\begin{equation}
J_{s}(z) =\hbar^{-1}\tilde{J}_{s}(z)\,,\qquad c =\hbar^{-1}\tilde{c}\,.
\end{equation}

\noindent Expanding now in $\hbar\to 0\,$ (keeping the rescaled currents and central charge fixed) these rescalings imply that the r.h.s. of the OPEs and commutation relations are linear in $\hbar\,$, with corrections of order $\mathcal{O}(\hbar^{2})\,$. In particular, the semiclassical OPE algebra (which translates into Poisson brackets) is obtained from the quantum OPE algebra (which translates into commutators) by taking the limit\footnote{ In terms of a free-field realization of the currents, the semiclassical limit amounts to dropping terms containing more than a single Wick contraction.}
\begin{equation}
\left.\tilde{J}_{s}(z)\tilde{J}_{s'}(w)\right|_{\text{semiclass}} \equiv \lim_{\hbar\to 0}\frac{1}{\hbar}\tilde{J}_{s}(z)\tilde{J}_{s'}(w)\,.
\end{equation}

\noindent Furthermore, the leading term in the $\hbar$ expansion of the composite fields $\cA{s}$, $\cB{s}$, $\cC{s}$, $\cD{s}$, $\cE{s}{\pm}$, $\cPhi{s}{\pm}$, $\cPsi{s}{\pm}$ is of order $\mathcal{O}(\hbar^{-1})$ and it precisely agrees with the corresponding expressions obtained from the bulk analysis \cite{BetoJuan} of asymptotic symmetries.

Finally, we notice that under the above procedure the parameter $\kappa$ defined in \eqref{definition kappa} becomes
\begin{equation}\label{semiclassical kappa}
\kappa \xrightarrow[\text{semi-classical}]{} \pm \frac{5i}{2}\,.
\end{equation}

\subsection{Normal-ordered composite operators}\label{app: Romans composites}

In order to compute unitarity and BPS bounds for the $\cW_{(3|2)}$ algebra one requires the action of the modes of various normal-ordered composite operators on highest weight vectors. As in the main text, we consider NS highest weight states $\hwr$ satisfying 
\begin{align}\label{hw conditions 1 app}
L_0 \hwr ={}&
 h \hwr 
  \\
  J_0 \hwr
    ={}&
  q \hwr
  \\
 V_0 \hwr 
 ={}&
  \qv \hwr 
  \\
  W_0 \hwr
  ={}&
  \qw \hwr
  \label{hw conditions 2 app}
\end{align}

\noindent and\footnote{ Note that $\left[G^{\pm}_{r},V_0\right] = \mp U^{\pm}_{r}$. Therefore, the highest-weight conditions $V_0\hwr = v\hwr$ and $G^{\pm}_{r} \hwr =0$ ($r > 0$) imply  $U^{\pm}_{r} \hwr =0$ ($r > 0$) as well.}
\begin{alignat}{3}
 L_{n}\hwr  ={}& J_{n}\hwr  = V_{n}\hwr 0\,,&
    &\qquad\quad &
   n >{}&0
  \\
 G^{\pm}_{r} \hwr ={}&
  0\, ,&
  &\qquad\quad &
   r >{}&0
   \\
   U^{\pm}_{r}\hwr  ={}&
  0\, ,&
  &\qquad\quad &
   r >{}&0
   \label{hw conditions 3 app}
\end{alignat}

The definition of normal-ordering we use follows \cite{Romans:1991wi,Thielemans:1994er} and reads
\begin{equation}
\left(:AB:\right)_{n} \equiv \sum_{p \leq -\Delta_{A}}A_{p}B_{n-p} + \left(-1\right)^{AB}\sum_{p > -\Delta_{A}}B_{n-p}A_{p}\,,
\end{equation}

\noindent where $\Delta_{A}$ is the dimension of $A$ and $(-1)^{AB}$ is $-1$ if both $A$ and $B$ are fermionic, and $+1$ otherwise. Products of more than two fields are defined recursively, grouping them as follows:
\begin{equation}
:A_{1}A_{2}\ldots A_{i-1}A_{i}: \equiv \left(:A_{1}\left(:A_{2}\left(\ldots \left(:A_{i-1}A_{i}:\right)\ldots \right):\right):\right).
\end{equation}

\noindent We will also need a formula for the modes of derivatives of operators:
\begin{equation}\label{modes of derivative}
\left(\partial A\right)_{n} = -\left(n +\Delta_{A}\right)A_{n}\,.
\end{equation}

\noindent Some additional useful relations for normal-ordered products can be found in \cite{Thielemans:1994er}, such as
\begin{equation}\label{NO identity 1}
:BA: \,= \left(-1\right)^{|A||B|}\left(:AB: + \sum_{\ell\geq 1}\frac{(-1)^{\ell}}{\ell !}\partial^{(\ell)}\left[AB\right]_{\ell}\right),
\end{equation}

\noindent where $\left[AB\right]_{\ell}$ is the coefficient of the $(z-w)^{-\ell}$ term (i.e. the $\ell$-th pole) in the $AB$ OPE. We can e.g. apply this formula to show
\begin{equation}
:\partial^{p}J\, J: = :J\partial^{p}J:
\end{equation}

\noindent which follows from \eqref{NO identity 1} and the fact that the $JJ$ OPE consists of just an anomalous (field-independent) term. For the remainder of this section, all composite fields are assumed to be normal-ordered, and we use a square bracket to denote combinations that are primary or quasi-primary operators.

The explicit expression for the composite field $\cC{4}$ is \cite{Romans:1991wi}
\begin{equation}\label{C4 Romans}
\cC{4} = \frac{2}{c-1}\left[J\partial T - 2\partial J \,T\right] + \frac{\kappa}{c+3}\left(2\left[J\partial V - 2\partial J \,V\right]-3\left[G^{+}U^{-} + G^{-}U^{+} - \frac{4}{3}\partial W\right]\right)
\end{equation}

\noindent With the help of the above formulae, in the NS sector we get the mode expansion
\begin{align}\label{C4 NS modes}
\left(\cC{4}\right)_{n}
={}&
\frac{2}{c-1}\left[-nL_{n}J_{0}+\sum_{p \geq 1} \left(3p-n\right)L_{n-p}J_{p}+\sum_{p \leq -1}\left(3p -n\right)J_{p} L_{n-p} \right]
\nonumber\\
&
+\frac{2\kappa}{c+3}\left[-nV_{n}J_{0}+\sum_{p \geq 1} \left(3p-n\right)V_{n-p}J_{p}+\sum_{p \leq -1}\left(3p -n\right)J_{p} V_{n-p} \right]
\\
& 
- \frac{3\kappa}{c+3}\left[ \frac{4n}{3}W_{n} 
+\sum_{p \leq -1/2}\left(G^{+}_{p}U^{-}_{n-p} + G^{-}_{p}U^{+}_{n-p}\right) - \sum_{p \geq 1/2}\left(U^{-}_{n-p}G_{p}^{+}+U^{+}_{n-p}G^{-}_{p}\right)\right]
\nonumber
\end{align}

\noindent and we conclude the important result
\begin{equation}
\cC{4}_0\hwr_{\text{\tiny{NS}}}
=
0\qquad \text{and} \qquad  \cC{4}_{n}\hwr_{\text{\tiny{NS}}} = 0\qquad \text{for }n>0\,.
\end{equation}

Other composites whose explicit action on highest weight state we have used in the main text are (all composite operators below are assumed to be normal-ordered)
\begin{align}
\cD{1} 
={}&
 \frac{1}{4}J
\\
\cD{2} 
={}&
\frac{1}{10(c-1)}\Bigl((5c-3)\left[T\right] - 3\left[J^{2}\right]\Bigr) + \frac{\kappa}{5}\left[V\right]
\\
\cD{3}
 ={}&
3\gamma \left(2(5c^{2}+9)\left[JT\right] - 3(4c+3)\left[J^{3}\right]+\frac{1}{2}(c-3)(13c-6)\left[G^{+}G^{-}-\partial T - \frac{1}{3}\partial^{2}J\right]\right) 
\nonumber\\
&
+ \frac{2\kappa}{5c-12}\Bigl(21 \left[JV\right] - (c+6)W\Bigr)
\\
\cD{4} ={}&
6\gamma \Biggl\{9c(c-1)\left[T^{2} - \frac{3}{10}\partial^{2}T\right] 
+ 3(4c+3)\left(\left[JG^{+}G^{-} - J\partial T - \frac{1}{3}J\partial^{2}J\right] - \left[J^{2}T\right]\right)
\\
&\hphantom{6\gamma \Biggl\{}
+ \frac{1}{4}\left(5c^{2} - 51c + 18\right)\left[\partial G^{+}\, G^{-} - G^{+}\partial G^{-} + \frac{2}{5}\partial^{2}T + \frac{1}{6}\partial^{3}J\right] 
\\
&\hphantom{6\gamma \Biggl\{}
+
\frac{1}{4}(c^{2}-53c + 66)\left[J\partial^{2} J - \frac{3}{10}\partial^{2}\left(J^{2}\right)\right]
\Biggr\}
\\
&
+\frac{6\kappa}{(c+3)(5c-12)}\Biggl\{18(c-1)\left[TV - \frac{3}{10}\partial^{2}V\right] 
-2(c-15)\left[JW\right]
\\
&\hphantom{+\frac{6\kappa}{(c+3)(5c-12)}\Biggl\{}
+ (4c+3)\left[G^{-}U^{+} - G^{+}U^{-} + \frac{2}{5}\partial^{2}V\right]
\Biggr\}
\end{align}

\noindent where $\gamma = \frac{1}{(c-1)(c+6)(2c-3)}\,$. By explicit computation we find
\begin{align}
\cD{1}_{0}\hwr_{\text{\tiny{NS}}} 
={}&
\frac{q}{4}\hwr_{\text{\tiny{NS}}} 
\\
\cD{2}_{0}\hwr_{\text{\tiny{NS}}} 
={}&
\left(\frac{5c-3}{10(c-1)}h + \frac{\kappa}{5}\qv -\frac{3}{10(c-1)}q^{2}\right)\hwr_{\text{\tiny{NS}}}
\\
\cD{3}_{0}\hwr_{\text{\tiny{NS}}} 
={}&
3\gamma \left(2(5c^{2}+9)qh - 3(4c+3)q^3+\frac{1}{2}(c-3)(13c-6)\frac{q}{3}\right) \hwr_{\text{\tiny{NS}}}
\nonumber\\
&
+ \frac{2\kappa}{5c-12}\Bigl(21 q \qv- (c+6)\qw\Bigr)\hwr_{\text{\tiny{NS}}}
\\
\cD{4}_{0}\hwr_{\text{\tiny{NS}}} 
={}&
6\gamma \Biggl(9c(c-1)h\left(h+\frac{1}{5}\right) + \frac{1}{4}\left(5c^{2} - 51c + 18\right)\left(\frac{7h}{5} - \frac{q}{2}\right)
\nonumber
\\
&
\qquad
+ 3(4c+3)q^{2}\left(\frac{1}{3}- h\right)
+
\frac{1}{4}(c^{2}-53c + 66)\frac{q^{2}}{5}
\Biggr)\hwr_{\text{\tiny{NS}}}
\\
&
+\frac{6\kappa }{(c+3)(5c-12)}\Biggl(18(c-1)\qv\left(h + \frac{1}{5}\right) + \frac{32}{5}(4c+3)\qv
-2(c-15)q\qw
\Biggr)\hwr_{\text{\tiny{NS}}}
\nonumber
\end{align}

\noindent and

\begin{equation}
 \cD{1}_{n}\hwr_{\text{\tiny{NS}}}=\cD{2}_{n}\hwr_{\text{\tiny{NS}}}=\cD{3}_{n}\hwr_{\text{\tiny{NS}}}=\cD{4}_{n}\hwr_{\text{\tiny{NS}}} = 0\,,\quad \text{for }n>0
\end{equation}

%

\providecommand{\href}[2]{#2}\begingroup\raggedright\endgroup

\end{document}